\documentclass[a4paper,11pt]{article}
\pdfoutput=1 
\usepackage{jheppub} 

\usepackage{caption} 
\usepackage[]{color}

\usepackage{commath}
\renewcommand{\thesection}{\arabic{section}}

\title{\boldmath Rotating Hairy Black Holes in AdS$_5\times$S$^5$}

 \author{Julija Markevi\v{c}i\={u}t\.e}
 \affiliation{Department of Applied Mathematics and Theoretical Physics, \\
 University of Cambridge, \\ Cambridge, CB3 0WA, UK}
 \emailAdd{j.markeviciute@damtp.cam.ac.uk}
  
 \abstract{We present a numerical study of fully non-linear, rotating and charged hairy black hole solutions in five-dimensional anti-de Sitter space, which originate from a consistent truncation of $\mathcal{N}=8$ supergravity, and can be consistently embedded in type IIB supergravity with AdS$_5\times$S$^5$ asymptotics. The hairy black holes have one scalar field charged under a $U(1)$ gauge field, and branch from the near-extremal Cveti\v{c}, L\"{u} and Pope solutions. We give numerical evidence that the hairy solutions  exist arbitrarily close to the BPS bound for all charges, and saturate it in the $T\rightarrow 0$ and $T\rightarrow\infty$ limits. We give further evidence for the conjecture of Markevi\v{c}i\={u}t\.e and Santos, that on the BPS bound, the rotating hairy black holes form a two-parameter family of solutions with finite entropy, and can be regarded as a one-parameter extension of the supersymmetric Gutowski and Reall black hole. We analyse the approach to the supersymmetric limit and explore the full phase diagram. In the planar horizon limit we find a two parameter family of rotating hairy black brane solutions which cannot be obtained via a Lorentz boost. The field theory dual exhibits a spontaneously generated current. The results of this paper suggest rich and intricate structure of hairy black hole solutions in AdS$_5\times$S$^5$, and highlight their importance in understanding the thermodynamics of $\mathcal{N}=4$ SYM.
}
\begin{document} 
\maketitle
\flushbottom
\newpage
\raggedbottom
 
\section{\label{sec:introduction}Introduction}
 
The AdS/CFT correspondence \cite{Maldacena:1997re,Gubser:1998bc,Witten:1998qj,Aharony:1999ti} asserts that there is a ``holographic'' duality between gravity in \textit{d}-dimensional anti-de Sitter (AdS) spacetime and a conformal field theory (CFT) with fewer dimensions. This is a weak-strong duality, allowing us to study strongly coupled field theories using classical supergravity. At the heart of this programme is Maldacena's original correspondence, which relates $\mathcal{N}=4$ super Yang-Mills (SYM) with gauge group $SU(N)$, and type IIB string theory with AdS$_5\times$S$^5$ asymptotics~\cite{Maldacena:1997re}. By studying the black hole solution space in the supergravity limit we can learn about the thermal states of the strongly coupled dual CFT at energies of order $N^2$.

Five-dimensional gauged $\mathcal{N}=8$ $SO(6)$ supergravity arises as a consistent truncation of type IIB SUGRA on S$^5$, where the isometry group of S$^5$ corresponds to global R-symmetry of the dual field theory~\cite{deWit:2013ija,Godazgar:2013pfa,Godazgar:2013nma,Godazgar:2013dma,Godazgar:2013oba,Lee:2014mla,Baguet:2015sma}. The bosonic sector has 42 scalars, which are organised in $SO(6)$ irreducible representations with masses $m^2L^2=-4,-3,0$, where $L$ is the radius of curvature of AdS$_5$. A further truncation is given by~\cite{Cvetic:2000nc}, where the 20 scalars are in the multiplet that transforms in the $\textbf{20}'$ sector of the $SO(6)$. This is still a very complicated model, and in order to make this problem numerically tractable we will employ a further truncation given by \cite{Bhattacharyya:2010yg}, with the three diagonal $U(1)$ charges in the maximal $U^3$ subgroup of $SO(6)$ set equal. It retains a complex charged scalar field $\phi$ with charge $e=2$, charged under a $U(1)$ gauge field, and has mass $m_\phi^2 L^2=-4$ which saturates the five-dimensional Breitenl\"ohner-Freedman (BF) bound~\cite{Breitenlohner:1982jf}, required to ensure the perturbative stability of AdS.

We will be primarily interested in solutions with global AdS$_5$ asymptotics, \textit{i.e.} the black hole solutions that are dual to the states of the CFT residing on $\mathbb{R}_t\times S^3$. We will also study the large horizon radius limit, in which the black holes are (locally) described by solutions with planar horizon topology. When the scalar field vanishes, the action reduces to that of the bosonic action for $D=5$ minimal gauged supergravity. In this limit, rotating, charged black hole solutions are known in analytic form \cite{Cvetic:2004hs,Cvetic:2004ny,Cvetic:2005zi,Chong:2005da,Chong:2005hr,Chong:2006zx,Wu:2011gq}, and admit a regular supersymmetric black hole limit, the Gutowski and Reall supersymmetric black hole~\cite{Gutowski:2004ez,Gutowski:2004yv,Kunduri:2006ek}. These solutions can be uplifted to type IIB string theory on AdS$_5\times$S$^5$ \cite{Chamblin:1999tk,Gauntlett:2004cm}. Supersymmetric solutions in this theory necessarily rotate, thus prompting the interest in generalising the results of~\cite{Bhattacharyya:2010yg,Markeviciute:2016ivy} to include rotation.

There are at least two instabilities affecting the Reissner-Nordstr\"om type solutions in AdS-Maxwell-Scalar theories whereby the charged black hole acquires hair. The near extremal black holes are unstable against the tachyonic condensation of a scalar field, if the effective mass of the scalar field due to the coupling of the scalar to the Maxwell field violates the effective near horizon Breitenl\"ohner-Freedman bound~\cite{Breitenlohner:1982jf}
\cite{Gubser:2008px,Hartnoll:2008vx,Hartnoll:2008kx}. In fact neither the scalar field nor the black hole need be charged \cite{Dias:2010ma}, but the charge exacerbates the instability. 
 
A bottom-up approach was analysed in \cite{Basu:2010uz} where the massless Abelian Higgs model was consdered in AdS$_5$. The small (compared to the radius of AdS), near extremal charged black holes were found to be unstable against forming scalar field condensate, and the endpoint hairy black holes were constructed perturbatively. The instability in this case was driven by the charged superradiance in AdS\footnote{The sufficiently charged scalar field scattering off the charged black hole becomes amplified \textit{ad infinitum} due to the reflective nature of the conformal boundary, thereby leading to the instability (\cite{Basu:2010uz} and references therein).}, and it was found to affect the small charged black holes with scalar field charges $e$ satisfying $e \mu \geq \Delta_0$, where the energy of the lowest sclar mode is given by $\Delta_0=4$ for $m_\phi^2=0$ (setting $L=1$). If we work in units where the chemical potential $\mu$ of a small RNAdS black hole is bounded above by $\mu\leq 1$, we find that the marginal scalar field charge for the instability is $e_c=4$. 

In \cite{Dias:2011tj} these solutions were constructed numerically and the hairy black hole phase diagram was found to depend qualitatively on the massless scalar field charge~$e$, the magnitude of which influences the two kinds of instabilities. The tachyonic instability was found to set in for lower charges than the superradiance. For charges $e>e_c$ where the two regimes coexist, the hairy black holes were found at all charges. They reduce to the smooth, ground state soliton branch in the $T\rightarrow\infty$ limit for $Q<Q_\mathrm{max}(e)$, and for $Q>Q_\mathrm{max}(e)$ reduce to some extremal, possibly singular, configurations. For $e>e_c$, the smooth soliton exists for all charges $Q$, while $e\leq e_c$ solitons approach the ``Chandrasekhar'' bound in a spiraling fashion\footnote{The soliton solution space exhibits interesting structure and there also exist other branches of solutions \cite{Gentle:2011kv}.}, where they become singular. However, they are never limiting solutions for the hairy black holes. At $e=e_c$, $Q_\mathrm{max}(e)=0$ and the lowest mass limit is a $T=0$ extremal limit, which does not coincide with the soliton. Rotating, charged hairy black holes were constructed numerically by~\cite{Brihaye:2011fj}, in a model with a minimally coupled charged massive scalar field $m_\phi^2=-3$, and for different values of $e$.


The model which we consider in this manuscript was first studied by~\cite{Bhattacharyya:2010yg}, and can be embedded in string theory. The scalar field is non-minimally coupled to gravity and has $m_\phi^2=-4$, $e=2$ and $\Delta=2$, therefore this particular truncation is a marginal case for the superradiant instability. In \cite{Bhattacharyya:2010yg} small hairy black holes were constructed in a matched asymptotic expansion, along with smooth supersymmetric soliton solutions. The latter were found to be the $T=0$ limit of the hairy black holes. There are three soliton families which play an important role in the hairy black hole solution space. They were constructed perturbatively and numerically by employing the BPS equations~\cite{Chong:2004ce,Liu:2007rv,Liu:2007xj,Chen:2007du,Bhattacharyya:2010yg}. The smooth soliton is continuously connected to the vacuum AdS$_5$, in the sense that in the small charge limit it reduces to the global AdS perturbed by the ground state excitation of the scalar field.

Fully non-linear solutions were constructed numerically in~\cite{Markeviciute:2016ivy}, where we found that the static hairy black holes exist for all charges. We summarize the phase diagram of~\cite{Markeviciute:2016ivy} in section~\ref{sec:summary}. The authors of~\cite{Bhattacharyya:2010yg} have proposed a number of intriguing possibilities for the rotating black hole solution space in this model, and theorized the existence of regular hairy supersymmetric black holes based on a non-interacting thermodynamic model. In the present paper we will find numerical evidence supporting this idea, and focus on exploring the moduli space near the BPS bound.

Perhaps one of the main achievements of string theory is understanding the Bekenstein-Hawking entropy of supersymmetric black holes in terms of microstate counting, both in flat and asymptotically AdS spaces~\cite{Strominger:1996sh,Strominger:1997eq,Benini:2015eyy,Benini:2016rke}. However, this has not yet been accomplished in Maldacena's original correspondence, even though a lot has been understood about the dual CFT.

From the ten-dimensional perspective, the supersymmetric black holes with three ${\text{R-charges}}$ and independent angular momenta preserve 2 supercharges, i.e. are dual to 1/16 BPS states. We will consider a truncation in which these black holes have self dual spin and three equal magnitude charges, and thus are described by one free parameter~\cite{Gutowski:2004ez}; the BPS relation is given by $M=J_1+J_2+\sum_i Q_i=2J+3Q$. However, the general $1/16$ BPS state in the dual field theory within our symmetry class depends on two fugacities, which poses a problem for the stringy microstate counting \cite{Kinney:2005ej,Berkooz:2006wc}. One possible resolution of this discrepancy is that a more general family of supersymmetric black holes might exist~\cite{Kunduri:2007qy}.

In \cite{Markeviciute:2018yal} we presented our first results for the rotating system; we reported a limit in which we can find hairy rotating black holes with a finite entropy $S>0$ arbitrarily close to the BPS bound, with no divergence in the scalar field. These black holes have finite curvature invariants, and the Christoffel symbols remain square-integrable as $T\rightarrow 0$. The near extremal hairy black holes display parallely propagated tidal force singularities which are Tipler-weak, so that the black holes admit a consistent propagation of infalling classical objects~\cite{Markeviciute:2018yal}. We argue that the limiting $T=0$ solutions constitute a one parameter generalization of the supersymmetric Gutowski-Reall black hole, and therefore play an important role in resolving the long-standing problem of black hole microstate counting in type IIB supergravity. We propose that the missing parameter is the horizon scalar hair, and provide further evidence for this conjecture.

In the present manuscript we present full analysis of the rotating hairy black holes, and construct the phase diagram which is governed by three parameters $\{Q,M,J\}$. This paper is organised as follows. In section~\ref{sec:summary} we briefly recall the results of~\cite{Markeviciute:2016ivy}, and present the phase diagram for the rotating, charged hairy black holes. We comment on the possible behaviour of the solutions in the regimes which were inaccessible to our numerical scheme. In section~\ref{sec:setup} we detail the setup and ansatz for the solutions, and give the equations of motion explicitly. Next, in section~\ref{sec:Hairy}, we describe the numerical approach and the construction of the solutions, followed by the numerical results in section~\ref{sec:Phase}. We obtain the onset of the superradiance via linear analysis, and we show that it agrees with the full non-linear results. We discuss the parameter space and provide different ways of tracing out the phase diagram, and we thoroughly study the near extremal behaviour of the hairy solutions. In section~\ref{sec:therm} we analyse the thermodynamics, and section~\ref{sec:planar} is devoted to the planar limit of our solutions. We obtain a two parameter family of ``rotating'' hairy branes which are not Lorentz boosted, and analyse their properties. We close with discussion and prospects for future work in section~\ref{sec:discussion}.

\section{\label{sec:summary}Summary of the phase diagram}

\begin{figure}[!htpb]
\centering
    \begin{minipage}[t]{0.44\textwidth}
    \includegraphics[width=\textwidth]{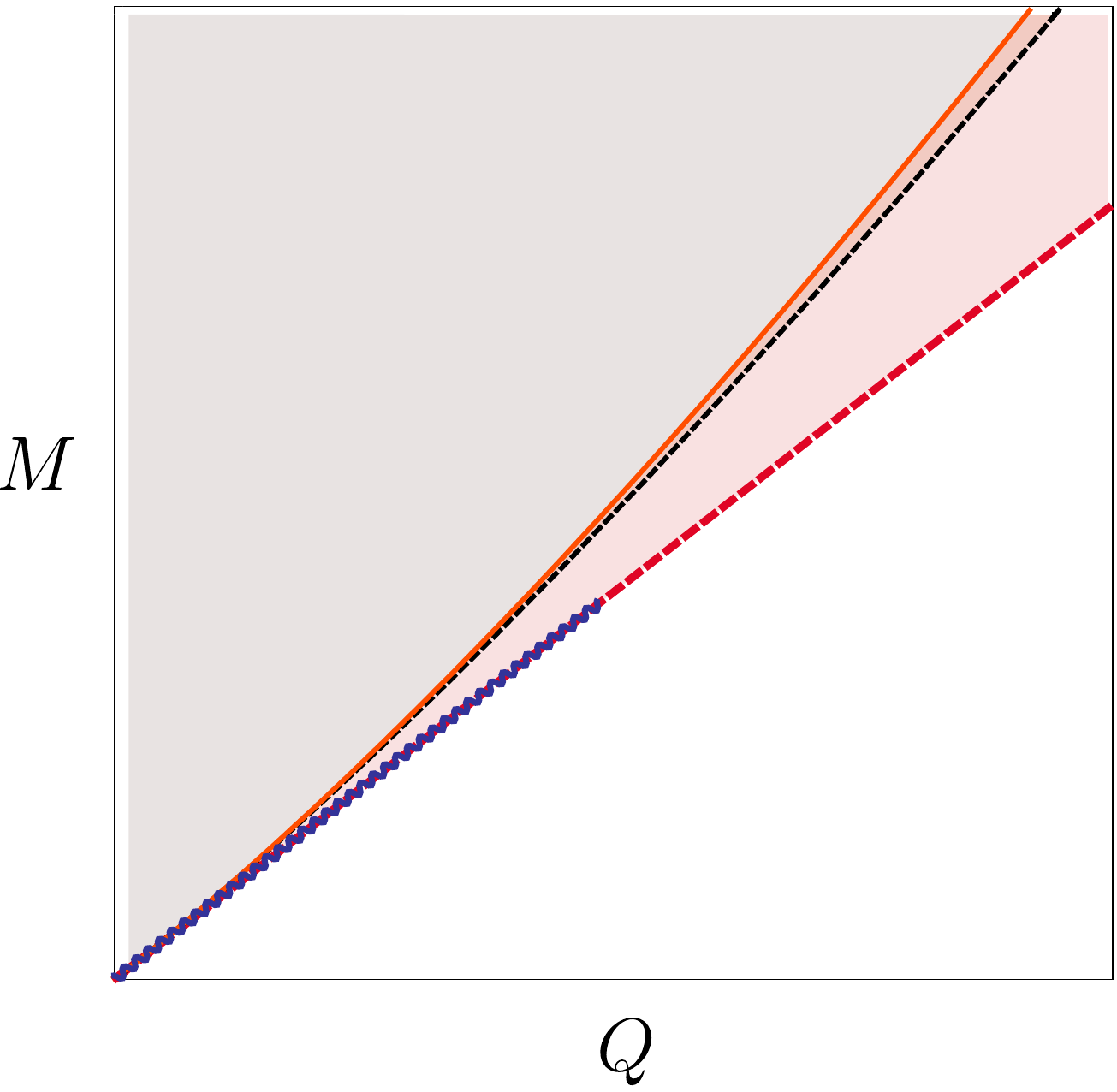}
  \end{minipage}
  \hfill
      \begin{minipage}[t]{0.43\textwidth}
    \includegraphics[width=\textwidth]{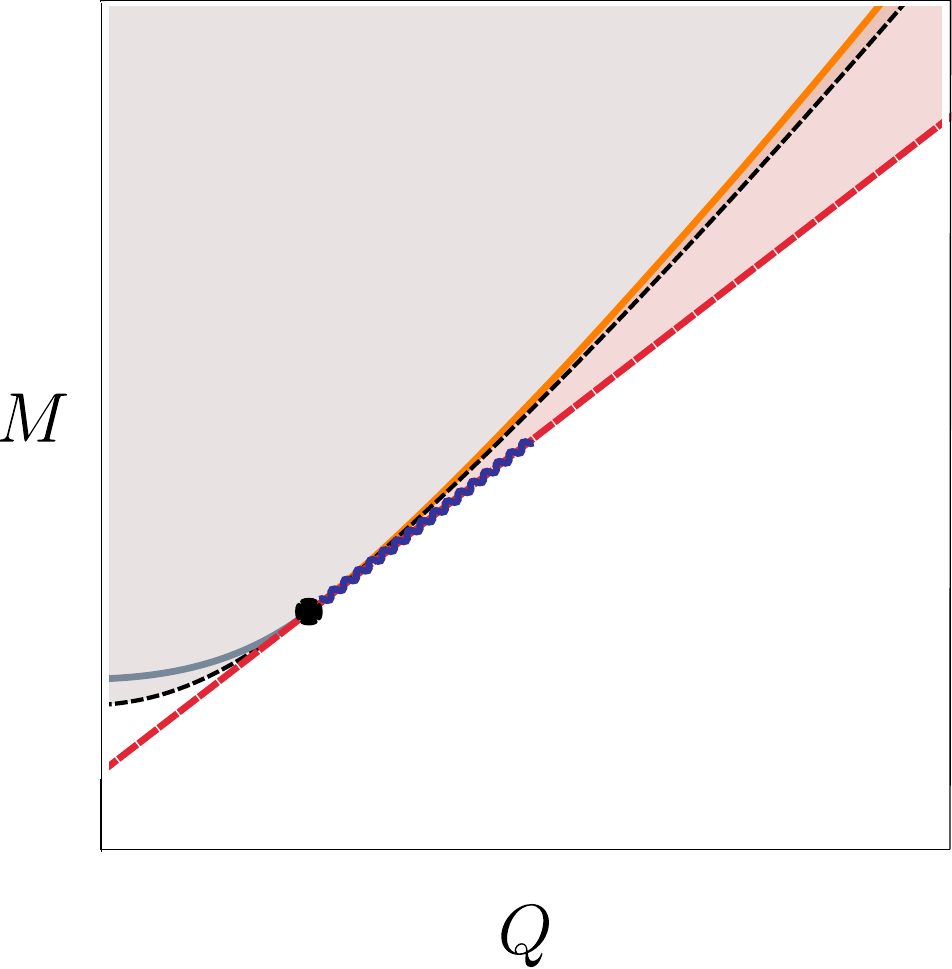}
  \end{minipage}
  \caption{\label{fig:figures} \textit{Left:} Microcanonical phase diagram of charged, non-rotating hairy black holes~\cite{Bhattacharyya:2010yg,Markeviciute:2016ivy}. The Reissner-Nordstr\"om black holes exist above their extremal limit (black dashed line), and the hairy black holes exist below the bold orange line, which shows the onset of the superradiant instability. These solutions extend down to the BPS bound, where they reduce to the smooth soliton (the purple wavy line) in the limit $T\rightarrow 0$, while keeping the value of the scalar field at the horizon $\varepsilon_H$ fixed. \textit{Right:} Microcanonical phase diagram of charged, rotating hairy black holes at a constant angular momentum $J$. The hairy black holes exist between the merger line (orange) and the BPS bound (dashed red), in the region shaded in light red. The black disk is the supersymmetric Gutowski-Reall black hole~\cite{Gutowski:2004ez}, which lies at the intersection of the extremal and the BPS planes. The purple wavy line shows a line of the conjectured supersymmetric hairy black holes, which terminates at a finite charge $Q_\mathrm{max}(J)$, where the black holes become singular.}
\end{figure}

We briefly summarize the results of the non-rotating case presented in~\cite{Markeviciute:2016ivy} in Fig.~\ref{fig:figures}~(\textit{left}). The Reisner--Nordstr\"om black holes (RNAdS) are bounded below by the $T=0$ extremality curve (dashed black), and the hairy black holes exist between the merger line which indicates the appearance of a marginal mode (solid orange) and the BPS bound given by ${M=3Q}$ (dashed red). There are three families of solitonic solutions which are relevant to the hairy black hole phase diagram\footnote{There is at least one more known family of solitons which does not connect to black hole solutions.}. The smooth solitons, shown by the purple wavy line, are perturbatively connected to the pure AdS, and as we increase the central scalar field density to infinity this branch terminates at the unique singular solution with $Q_c\simeq0.2613$. We will see that this solutions is quite special even in the $J>0$ phase space. Here, the smooth and the singular soliton branches merge; the singular soliton family has no upper bound on the charge and admits a scaling limit. The smooth BPS solitons can be regarded as the zero temperature limit of the hairy black hole families with a constant horizon scalar $\varepsilon_H$~\cite{Bhattacharyya:2010yg,Markeviciute:2016ivy}, and the singular solitons represent the $\varepsilon_H\rightarrow\infty$, $T\rightarrow\infty$ limit of the hairy black holes. In the planar limit, we have a one parameter class of hairy black branes that connect to the large charge limit of the singular soliton branch as $\varepsilon_H\rightarrow\infty$.

The present manuscript extends our previous work by considering rotating solutions, which include a one parameter family of supersymmetric Gutowski-Reall black holes. Charged, rotating solutions in this theory have analytic expressions, and were first presented by Cveti\v{c}, L\"u and Pope (CLP). The hairy black holes extend the non-rotating solution space up to the BPS bound, and on the BPS bound smoothly connect to the soliton solutions. We find that when rotation is included, the picture is surprisingly similar, and the hairy black holes interpolate the solution space between the extremal plane and the BPS bound for all charges\footnote{We find that only black holes with horizon angular velocity $\Omega_H<1$ are unstable to forming scalar hair, and all hairy black holes in this truncation have $\Omega_H\leq 1$.}. We find rotating hairy black holes arbitrarily close to the supersymmetric limit, and, for charges sufficiently close to the Gutowski-Reall black hole, the hairy solutions maintain finite entropy and are free from scalar curvature singularities~\cite{Markeviciute:2018yal}.

The results of this paper are summarised in Fig.~\ref{fig:figures}~(\textit{right}), where we present the microcanonical phase diagram of the hairy solutions at a fixed angular momentum $J$. The hairy black holes exist between the solid orange line which marks the onset of the superradiant instability, and the red dashed curve showing the BPS bound, which is given by ${M=3Q+2J}$. The extremality for the charged rotating black holes is given by the dashed black curve. The solid grey line shows the charged rotating black holes with $\Omega_H=1$, and the black holes below this line have $\Omega_H>1$~\footnote{All CLP black holes above this line have $\Omega_H<1$.}. The black disk shows the supersymmetric Gutowski-Reall black hole, and the purple wavy line shows the conjectured supersymmetric rotating hairy black holes~\cite{Markeviciute:2018yal}, which exist up to some maximal charge $Q_\mathrm{max}(J)$, where the entropy $S\rightarrow 0$. At that point a special solution might exist at which all isotherms intersect. For larger charges the BPS solutions are singular, and can be regarded as a $T\rightarrow\infty$ limit of the regular hairy black hole solutions. As we are constrained by the scope of our numerical methods, we can only conjecture the possible limiting behaviour of these solutions. In the following chapters, we will aim to give a convincing numerical evidence which supports this picture. For a further discussion on the extent of the evidence we refer the reader to the concluding section~\ref{sec:discussion}.

Finally, the large charge rotating hairy solutions admit a scaling limit. While the planar CLP black holes can be obtained from the RNAdS holes via a simple boost, we argue that the hairy rotating branes cannot be boosts of the non-rotating counterparts due to a non-vanishing magnetic field on the horizon. The two parameter family of solutions has interesting properties, and displays a phase transition in the canonical ensemble. Such black holes exhibit retrograde condensation, as found in the non-rotating case, and in the large temperature limit they approach the singular soliton of~\cite{Bhattacharyya:2010yg}.

\clearpage
\section{\label{sec:setup}Setup}
\subsection{The action and field equations}

The action given by the consistent truncation of~\cite{Bhattacharyya:2010yg} is

\begin{multline}
\label{eq:action}
 S=\frac{1}{16\pi G_5}\int \mathrm{d}^5x\sqrt{g}\left\{R[g]+12-\frac{3}{4}F_{\mu\nu}F^{\mu\nu}-\frac{3}{8}\left[(D_\mu\phi)(D^\mu\phi)^\dagger-\frac{\nabla_\mu\phi \phi^\dagger\,\nabla^\mu\phi \phi^\dagger}{4(4+\phi \phi^\dagger)}-4\phi \phi^\dagger\right]\right\}
 \\
 -\frac{1}{16\pi G_5}\int F\wedge F\wedge A,
\end{multline} 
with $F_{\mu\nu}=2\partial_{[\mu}A_{\nu]}$, $D_\mu\phi=\nabla_\mu\phi-i\,e A_\mu\phi$, the radius of five-dimensional AdS is set to $L=1$, and we work in units where the $d=5$ Newton's constant is given by $G_5=\pi/(2N^2)$. The complex scalar field $\phi$ has mass $m_\phi^2=-4$ saturating the BF bound, and the electric charge $e=2$ which is prescribed by the consistent truncation. In the AdS/CFT dictionary the operator which is dual to the complex scalar field has conformal dimension $m_\phi^2=\Delta(\Delta-4)=-4$, \textit{i.e.} $\Delta=2$.

The equations of motion read 
\begin{subequations}
\begin{align}
\label{eq:eeq}
&G_{\mu\nu}-6 g_{\mu\nu}=\frac{3}{2}T_{\mu\nu}^{\mathrm{em}}+\frac{3}{8}T_{\mu\nu}^{\mathrm{mat}}\\
\label{eq:maxwell}
&\nabla_\rho F^\rho{}_\mu=\frac{1}{4}\varepsilon_{\mu\lambda\nu\rho\sigma}\,F^{\lambda\nu}F^{\rho\sigma}-\frac{i}{4}\left[\phi(D_\mu\phi)^\dagger-\phi^\dagger D_\mu\phi\right]\\
\label{eq:scalar}
&D_\mu D^\mu\phi+\phi\left[\frac{(\nabla_\mu\phi \phi^\dagger)(\nabla^\mu\phi \phi^\dagger)}{4(4+\phi \phi^\dagger)^2}-\frac{\nabla_\mu \nabla^\mu \phi \phi^\dagger}{2(4+\phi \phi^\dagger)}+4\right]=0,
\end{align}
\label{eq:fieldeq}
\end{subequations} 
and the energy-momentum tensor is given by
\begin{align*}
T_{\mu\nu}^{\mathrm{em}}&=F_\mu{}^\lambda F_{\nu\lambda}-\frac{1}{4}g_{\mu\nu}\,F^2,\\
T_{\mu\nu}^{\mathrm{mat}}&=\frac{1}{2}\left[D_\mu\phi\,(D_\nu\phi)^\dagger+D_\nu\phi\,(D_\mu\phi)^\dagger\right]-\frac{1}{2}g_{\mu\nu}(D_\lambda\phi)(D^\lambda\phi)^\dagger+2g_{\mu\nu}\, \phi \phi^\dagger\\
&\qquad\qquad\qquad\qquad-\frac{1}{4(4+\phi \phi^\dagger)}\left[(\nabla_\mu\phi \phi^\dagger)(\nabla_\nu\phi \phi^\dagger)-\frac{1}{2}g_{\mu\nu}(\nabla_\lambda\phi \phi^\dagger)(\nabla^\lambda\phi \phi^\dagger)\right].
\end{align*}

We are interested in stationary, asymptotically anti-de Sitter black holes, with spherical horizon topology. The most general such solution can have two independent angular momenta, but we will consider a doubly spinning solution for which the magnitude of angular momenta along the two rotation directions is the same. Keeping the gauge arbitrary, we can write down the following general cohomogeneity-one ansatz~\cite{Kunduri:2006qa},
\begin{equation}
 \mathrm{d}s^2=-f(r)\,\mathrm{d}t^2+g(r)\,\mathrm{d}r^2+\Sigma(r)^2 \left[h(r)\left(\mathrm{d}\psi+\frac{1}{2}\cos{\theta}\mathrm{d}\phi-\Omega(r)\,\mathrm{d}t\right)^2+\frac{1}{4}\mathrm{d}\Omega^2_2\right]\,.
 \label{eq:ansatzr}
\end{equation}
When considering the AdS limit of the ansatz, we can see that the metric on the round three sphere is written as a Hopf fibration over the unit $\mathbb{CP}^1$ space, with the corresponding K\"ahler potential\footnote{In the sense that the K\"ahler form for $\mathbb{CP}^1$ is given by $J=\frac{1}{2}\mathrm{d}\mathcal{A}$.} ${\mathcal{A}=\mathcal{A}_a\mathrm{d}x^a=\frac{1}{2}\cos{\theta}\mathrm{d}\phi}$. It is given by $\mathrm{d}\Omega^2_3=(\mathrm{d}\psi+\mathcal{A})^2+\mathrm{d}\tilde{\Omega}^2_2$, with the standard Fubini-Study metric ${\mathrm{d}\tilde{\Omega}^2_2=\frac{1}{4}\left(\mathrm{d}\theta^2+\sin^2{\theta}\,\mathrm{d}\phi^2\right)}$. The fiber is parametrised by the coordinate $\psi$ with a period $2\pi$~\footnote{The Hopf fibration requires that the coordinate parametrising the $\mathbb{S}^1$ fiber has $\psi\rightarrow\psi+\pi$ when $\phi\rightarrow\phi+2\pi$. The two orthogonal planes of rotation are at $\theta=0$ and $\theta=\pi$.}, where $\theta, \phi$ are the standard polar coordinates on $S^2$. The level surfaces of the ansatz are given by homogeneously squashed 3-spheres. 

 The gauge field ansatz which respects the symmetries of~(\ref{eq:ansatzr}) is given by
 \begin{equation}
 A=A_t(r)\mathrm{d}t+A_\psi (r)(\mathrm{d}\psi+\mathcal{A}),
 \label{eq:gauge}
 \end{equation}
and we also take ${\phi=\phi^\dagger=\phi(r)}$, as the phase of the scalar field is removed by a $U(1)$ gauge transformation. The ansatz~(\ref{eq:ansatzr}) has a residual gauge freedom 
\begin{equation}
\psi\rightarrow\psi +\alpha t,\qquad\Omega\rightarrow\Omega+\alpha,
\label{eq:gaugeres}
\end{equation}
\noindent where $\alpha$ is a constant. The gauge in which the frame is static at asymptotic infinity simplifies the thermodynamic analysis \cite{Hawking:1998kw,Hawking:1999dp,Gibbons:2004ai}, thus we will make use of~(\ref{eq:gaugeres}) to set the angular velocity $\Omega(r)\rightarrow 0$ as $r\rightarrow \infty$.

The Einstein, Maxwell and scalar equations~(\ref{eq:fieldeq}) comprise a system of seven non-linear differential equations, two of first order and five of second order. Using the ansatz \eqref{eq:ansatzr} the equations can be compactly written as
\begin{align}
\label{eq:eom}
\begin{split}
f''&+f'\left(\Xi-\frac{h'}{h}-\frac{f'}{f}-\frac{5 \Sigma '}{\Sigma }\right)-f\left(\Pi-\frac{3 \phi '^2}{\phi ^2+4}\right)-5 \left(A'+\Omega A_\psi'\right)^2-2 h \Sigma ^2\Omega'^2=0,
\\
g'\phantom{'}&+2g \left(\Xi-\frac{f'}{2f}-\frac{h'}{2h}-\frac{2\Sigma '}{\Sigma }\right)=0,
\\
h''&+h' \left(\Xi+\frac{\Sigma'}{\Sigma}-\frac{h'}{h}\right)+h^2 \left(\frac{\Sigma ^2\Omega '^2}{f}-\frac{8 g}{\Sigma ^2}\right)+\frac{4 g h}{\Sigma ^4} \left(2 \Sigma ^2-3 A_\psi^2\right)+\frac{3}{\Sigma ^2}\left(A_\psi^2 g \phi ^2+A_\psi'^2\right)=0,
\\
\Omega ''&+\Omega ' \left(\Xi-\frac{f'}{f}+\frac{h'}{h}+\frac{3 \Sigma '}{\Sigma }\right)-\frac{3 \Omega}{h\Sigma ^2}\left(A_\psi^2 g \phi ^2+A_\psi'^2\right)-\frac{3}{h \Sigma ^2}\left(A' A_\psi'+A A_\psi g \phi ^2\right)=0,
\\
A''&-\frac{1}{2} A' \left(\frac{g'}{g}-\frac{h'}{h}+\frac{f'}{f}-\frac{6 \Sigma '}{\Sigma }-X\right)-A g \phi ^2
\\
&+A_\psi' \left[\frac{4 A_\psi\sqrt{f g h}}{h\Sigma^3 }-\Omega  \left(\frac{f'}{f}-\frac{h'}{h}-\frac{2\Sigma '}{\Sigma }-\frac{1}{2}X\right)+\Omega '\right]+\frac{4 A_\psi g h \Omega }{\Sigma ^2}=0,
\\  
A_\psi''&-\frac{1}{2}A_\psi' \left(-\frac{f'}{f}+\frac{g'}{g}+\frac{h'}{h}-\frac{2 \Sigma '}{\Sigma }+X\right)-A_\psi g \left(\frac{4 h}{\Sigma ^2}+\phi ^2\right)-\frac{1}{2\Omega} A' X=0,
\\
\phi''&+\frac{1}{2} \phi ' \left(\frac{f'}{f}-\frac{g'}{g}+\frac{h'}{h}+\frac{6 \Sigma '}{\Sigma }\right)+g \phi  \left(\phi ^2+4\right) \left[\frac{(A+A_\psi \Omega )^2}{f}-\frac{A_\psi^2}{h \Sigma ^2}+1\right]-\frac{\phi  \phi'^2}{\phi ^2+4}=0,
\end{split}
\end{align}
with an additional constraint equation. Here
\begin{align*}
\begin{split}
\Xi&=\dfrac{A_\psi'^2}{2 h \Sigma  \Sigma'}-\dfrac{\Sigma }{2 f \Sigma '}\left(A'+\Omega A_\psi'\right)^2-\dfrac{\Sigma ''}{\Sigma '}+\dfrac{g}{2 \Sigma ^3 \Sigma '} \left[\Sigma ^2 \left(-4 h+\Sigma ^2 \left(\phi ^2+8\right)+8\right)-8A_\psi^2\right],
\end{split}
\end{align*}
\begin{align}
\begin{split}
\Pi&=\frac{1}{h \Sigma ^2}\left[3 A_\psi^2 g \phi ^2-2A_\psi'^2+4 \Sigma ' \left(\Sigma  h'+3 h \Sigma '\right)\right]-\frac{2 g }{\Sigma ^4}\left[\Sigma ^2 \left(-2 h+\Sigma ^2 \left(\phi ^2+8\right)+8\right)-8 A_\psi^2\right],\\
X&=\frac{2\Omega}{\Sigma f}\left[h\Sigma^3 \Omega'-4 A_\psi \sqrt{fgh}\right],
\end{split}
\end{align}
\noindent and the $'$ denotes the differentiation with respect to the coordinate $r$. It is possible to obtain a first order differential equation for $f(r)$ and subsequently eliminate it altogether, but the equations get increasingly complicated.

Known solutions to~(\ref{eq:eom}) have been obtained in the gauge where $\Sigma(r)=r$, and we will use this radial gauge throughout unless otherwise stated. In the radial gauge we are looking for solutions asymptoting to AdS$_5$~\cite{Ashtekar:1984zz,Henneaux:1985tv,Henningson:1998gx,deHaro:2000vlm}, \textit{i.e.} obeying
\begin{align}
\begin{split}
f(r)&=r^2+1+\frac{C_f}{r^2}+\mathcal{O}(r^{-4}),\quad g(r)=\frac{1}{r^2}-\frac{1}{r^4}+\mathcal{O}(r^{-6}),\quad h(r)=1+\frac{C_h}{r^4}+\mathcal{O}(r^{-6}),\\
\Omega(r)&=\frac{2J}{r^4}+\mathcal{O}(r^{-5}),\quad A_t(r)=\mu_\infty+\frac{2Q}{r^2}+\mathcal{O}(r^{-6}),\quad A_\psi(r)=\mathcal{O}(r^{-2}),\\
\phi(r)&=\frac{C_\phi}{r^2}+\mathcal{O}(r^{-4})\,.
\end{split}
\label{eq:expansion}
\end{align}
\noindent Here $\mu_\infty$ is the electrostatic potential at the boundary, $Q$ is the black hole charge, $J$ is the black hole angular momentum, and the black hole mass is given by $M=\frac{1}{4}\left(C_h-3C_f\right)$, where the constants $C_h$ and $C_f$ are to be extracted from the large $r$ behaviour of the functions. We note that here, and throughout the paper, the conserved charges of the system are rescaled by $N^2$. Further discussion on the thermodynamics will be given in subsection~\ref{subsubsec:qua}. The constant $C_\phi$ is the expectation value of the operator dual to the scalar field, $C_\phi=\langle \mathcal{O}_\phi\rangle$; here we assume the standard quantisation and the operator is not sourced.

The black hole horizon $r_+$ is defined as the largest root where $f(r_+)=0$, and all other functions are regular. We also identify $\Omega_H=\Omega(r_+)$ as the horizon angular velocity.

\subsection{Charged, equally-rotating black holes}

\begin{figure}[t]
\centering
  \begin{minipage}[t]{0.5\textwidth}
    \includegraphics[width=\textwidth]{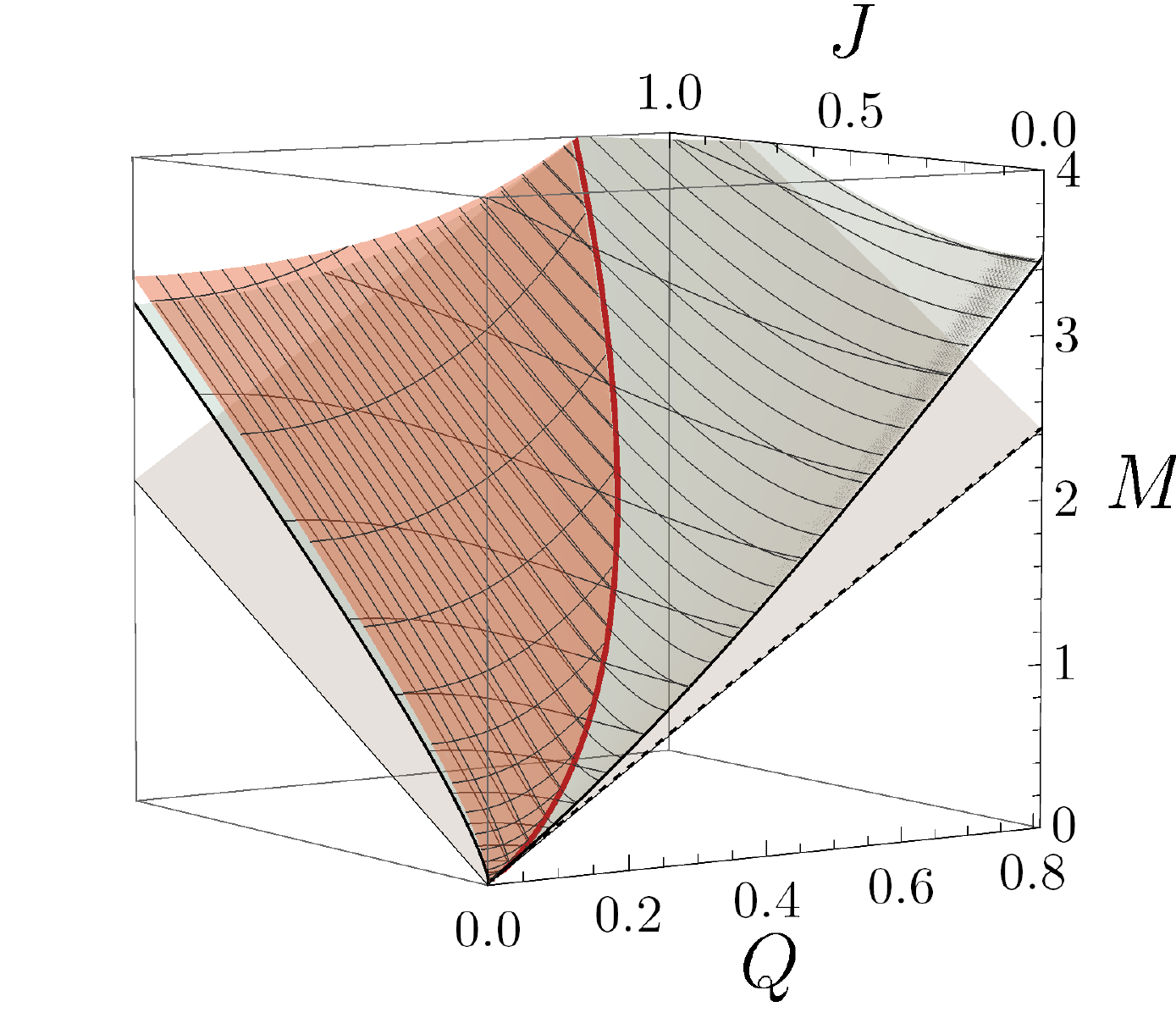}
  \end{minipage}
\hfill
  \begin{minipage}[t]{0.4\textwidth}
    \includegraphics[width=\textwidth]{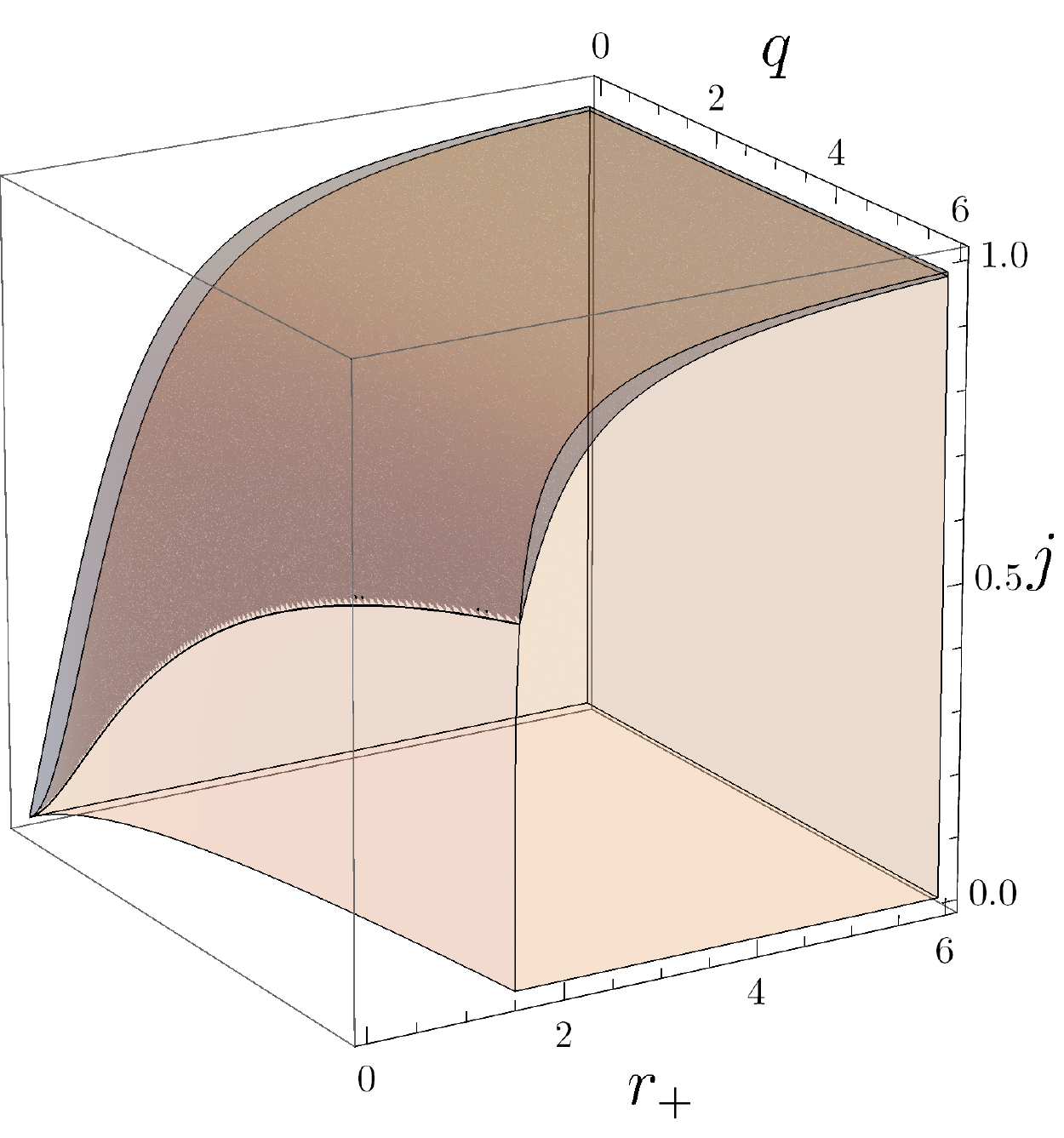}
  \end{minipage} 
\caption[]{\textit{Left}: Phase diagram for the regular CLP black hole solutions, showing possible $Q$, $M$ and $J$ values. The grey plane shows the BPS limit $M=2 J+3 Q$, the blue plane is the extremal limit with $T=0$. The red line shows Real-Gutowski holes, which are extremal BPS solutions, and lie on the intersection of the extremal and BPS planes. The orange plane shows the $\Omega_H=1$ limit for regular black holes with $T>0$. Such black holes exist above the extremal plane, and regular solutions with $\Omega_H>1$ exist only between the orange and blue planes. For $J=0$, we recover RNAdS solutions, and for $Q=0$ we recover MPAdS holes. \textit{Right}: Allowed parameter space for regular CLP black holes. It is bounded by the extremal surface, which also guarantees a non-negative entropy. The grey region on top shows the moduli space with $\Omega_H \geq 1$. The Gutowski-Reall black holes are the curve on the front left face, where the gray region meets the orange region.}
    \label{fig:CLP}
\end{figure}  

General five-dimensional, charged and equally rotating black hole solutions with a cosmological constant were first presented by Cveti\v c, L\"u and Pope (CLP) \cite{Cvetic:2004hs}, in the bosonic sector of minimal gauged supergravity, which is the $\phi=0$ limit of the action~(\ref{eq:action}). The black holes are governed by three parameters $\{q,j,m\}$, which are related to the conserved charges $\{Q,J,M\}$ \cite{Madden:2004ym}. In terms of our ansatz~(\ref{eq:ansatzr}), the solutions in a static frame at infinity are given as 
\begin{align}
\label{eq:CLP}
\begin{split}
f(r)&=F(r)/h(r),\qquad g(r)=1/F(r),\qquad h(r)=j^2\left(\frac{m}{r^4}-\frac{q^2}{r^6}\right)+1,\\
\Omega(r)&=\frac{j}{ h(r)}\left(\frac{m-q}{r^4}-\frac{q^2}{r^6}\right),\qquad A(r)=q/r^2,\qquad A_\psi(r)= -j q/r^2,
\end{split}
\end{align}
\noindent where
\begin{equation}
\begin{split}
F(r)&=\frac{1}{r^4}\left[ q^2\left(1-\frac{j^2}{L^2}\right)+j^2m\right]-\frac{1}{r^2}\left[m\left(1-\frac{j^2}{L^2}\right)-2q\right]+\frac{r^2}{L^2}+1,
\end{split}
\end{equation}
\noindent and we keep the AdS radius $L$ for clarity, in this subsection only. The thermodynamic quantities temperature $T$, entropy $S$, chemical potential $\mu$, electric charge $Q$, horizon angular velocity $\Omega_H$, angular momentum $J$ and mass $M$ are given by

\begin{align}
\label{eq:CLPT}
\begin{split}
&T=\frac{F'(r_+)}{4\pi \sqrt{h(r_+)}}=\frac{r_+^6-j^2 \left(2 L^2 m+m r_+^2-2 q^2\right)+L^2 \left(r_+^2(m -2 q)-2 q^2\right)}{2 \pi  L^2 r_+^2 \sqrt{j^2 \left(mr_+^2-q^2\right)+r_+^6}}\,,\\
&S=\pi  \sqrt{j^2 \left(m r_+^2-q^2\right)+r_+^6},\qquad \mu=q\frac{r_+^2-j^2 \left(1+r_+^2/L^2\right)}{j^2 q+r_+^4}\,,\\
&Q=\frac{1}{2}q,\qquad \Omega_\mathrm{H}=j\frac{q+r_+^2(1+r_+^2/L^2)}{j^2 q+r_+^4}\,,\\
&J=\frac{1}{2} j (m-q),\qquad M=\frac{1}{4} \left[m \left(3+\frac{j^2}{L^2}\right)-6 q\right]\,,
\end{split}
\end{align}
\noindent where the parameter $m$ can be written as
\begin{equation}
m=\frac{\left(q+r_+^2\right)^2-j^2 q^2/l^2+r_+^6/L^2}{r_+^2-j^2 \left(1+r_+^2/L^2\right)},
\end{equation}

\noindent and parameter $r_+$ is determined by requiring that $F(r_+)=0$. The BPS bound is given by~\cite{2001JHEP...02..031K}
\begin{equation}
M=2 J+3 Q.
\end{equation}
All extremal CLP's have $\Omega_H\leq 1$ (and $\mu\geq 1$), where it is equal to one only on the supersymmetric bound.

The solution space of the regular CLP black holes with thermodynamics described by~(\ref{eq:CLPT}) is shown in~Fig.~\ref{fig:CLP} (\textit{left}), together with the allowed parameter range in~Fig.~\ref{fig:CLP} (\textit{right}). We note that the presence of the Chern-Simons term breaks charge reversal invariance $Q\rightarrow -Q$, and in this paper we will only consider $Q>0$, and hence $\mu>0$ solutions.

The non-rotating solutions are the Reissner-Nordstr\"om black holes (RNAdS) and can be recovered by taking $j=0$, $m=R ^2 \left(R ^2+(\mu +1)^2\right)$ and $q=\mu R^2$. For completeness we note down the solution in terms of~(\ref{eq:ansatzr})
\begin{align}
\label{eq:RN} 	
\begin{split}
&f(r)=1+\frac{r^2}{L^2}-\frac{(R^2/L^2+\mu^2+1)R^2}{r^2}+\frac{\mu^2 R^4}{r^4},\\
&g(r)=\frac{1}{f(r)},\quad A(r)=\mu\left(1-\frac{R^2}{r^2}\right),\quad A_\psi(r)=0,\quad \Omega(r)=0, \quad\phi(r)=0.
\end{split}
\end{align}
The thermodynamic quantities are given as
\begin{align}
\begin{split}
M&=\frac{3}{4}R^2\left(1+R^2/L^2+\mu^2\right),\qquad\qquad Q=\frac{1}{2}\mu R^2, \qquad\qquad S=\pi R^3,\\ T&=\frac{1}{2\pi R}\left(1+2R^2/L^2-\mu^2 \right),
\end{split}
\end{align}
\noindent and the black holes are extremal if $\mu=\sqrt{1+2R^2/L^2}$. Another well known solution is the uncharged limit of~(\ref{eq:CLP}), the five-dimensional equally-rotating Myers-Perry-AdS (MPAdS) black hole~\cite{1986AnPhy.172..304M,Hawking:1998kw}.

The supersymmetric limit of the CLP black holes~\cite{Gutowski:2004yv} is a one parameter family, and can be obtained by setting
\begin{equation}
q=\left(1+\frac{r_+^2}{2L^2}\right)r_+^2,\qquad j=\left(1+\frac{r_+^2}{2L^2}\right)^{-1}\frac{r_+^2}{2}, \qquad m=4\left(1+\frac{r_+^2}{2L^2}\right)^2r_+^2.
\end{equation}
\noindent These solutions are both extremal and supersymmetric; they have $\mu=1$ and $\Omega_H=1$.
 
\section{\label{sec:Hairy}Hairy Black holes}

\subsection{\label{subsec:Numerical}Numerical method}

\begin{figure}[t]
\centering
  \begin{minipage}[t]{0.48\textwidth}
    \includegraphics[width=\textwidth]{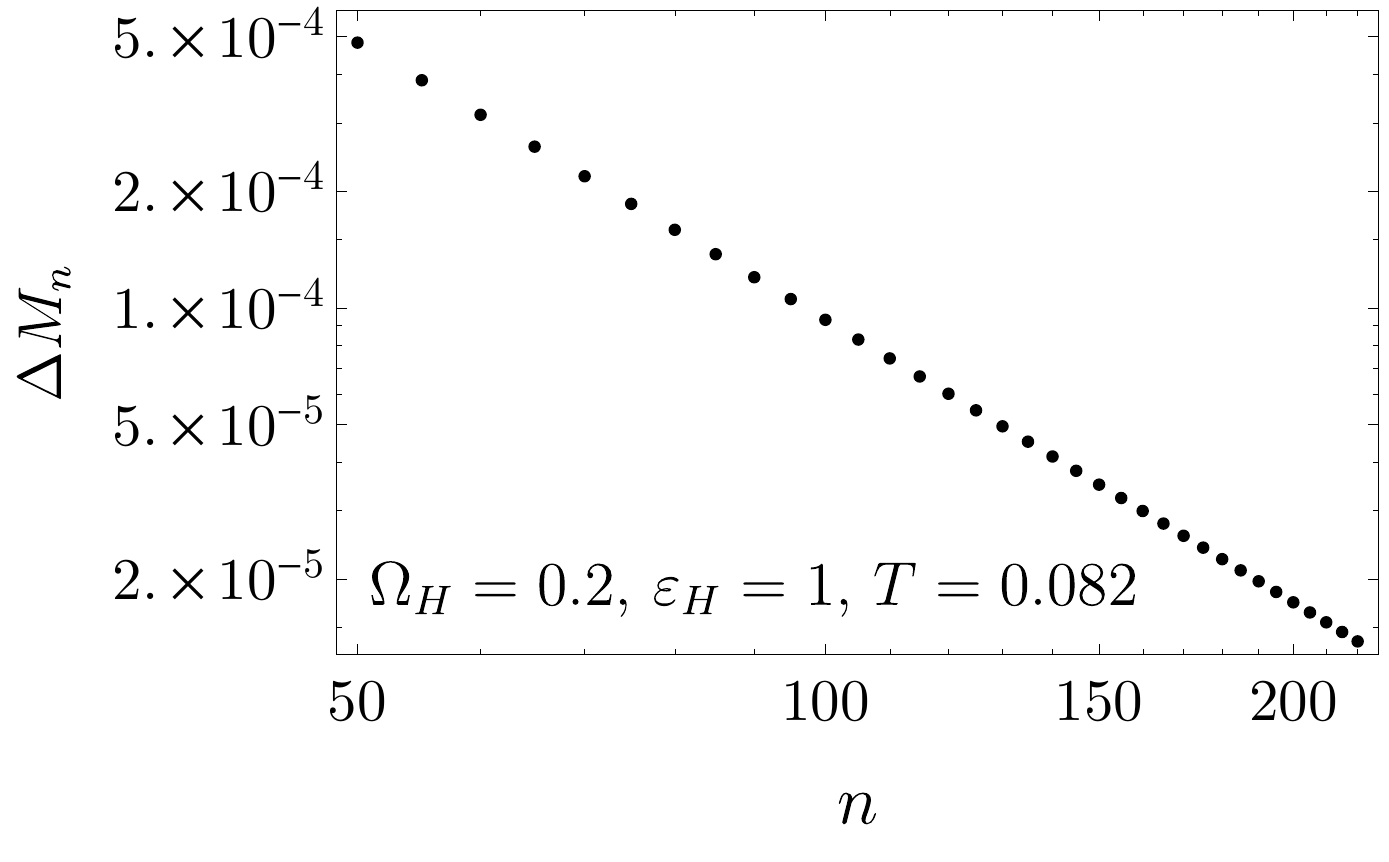}
  \end{minipage}
  \hfill
    \begin{minipage}[t]{0.48\textwidth}
    \includegraphics[width=\textwidth]{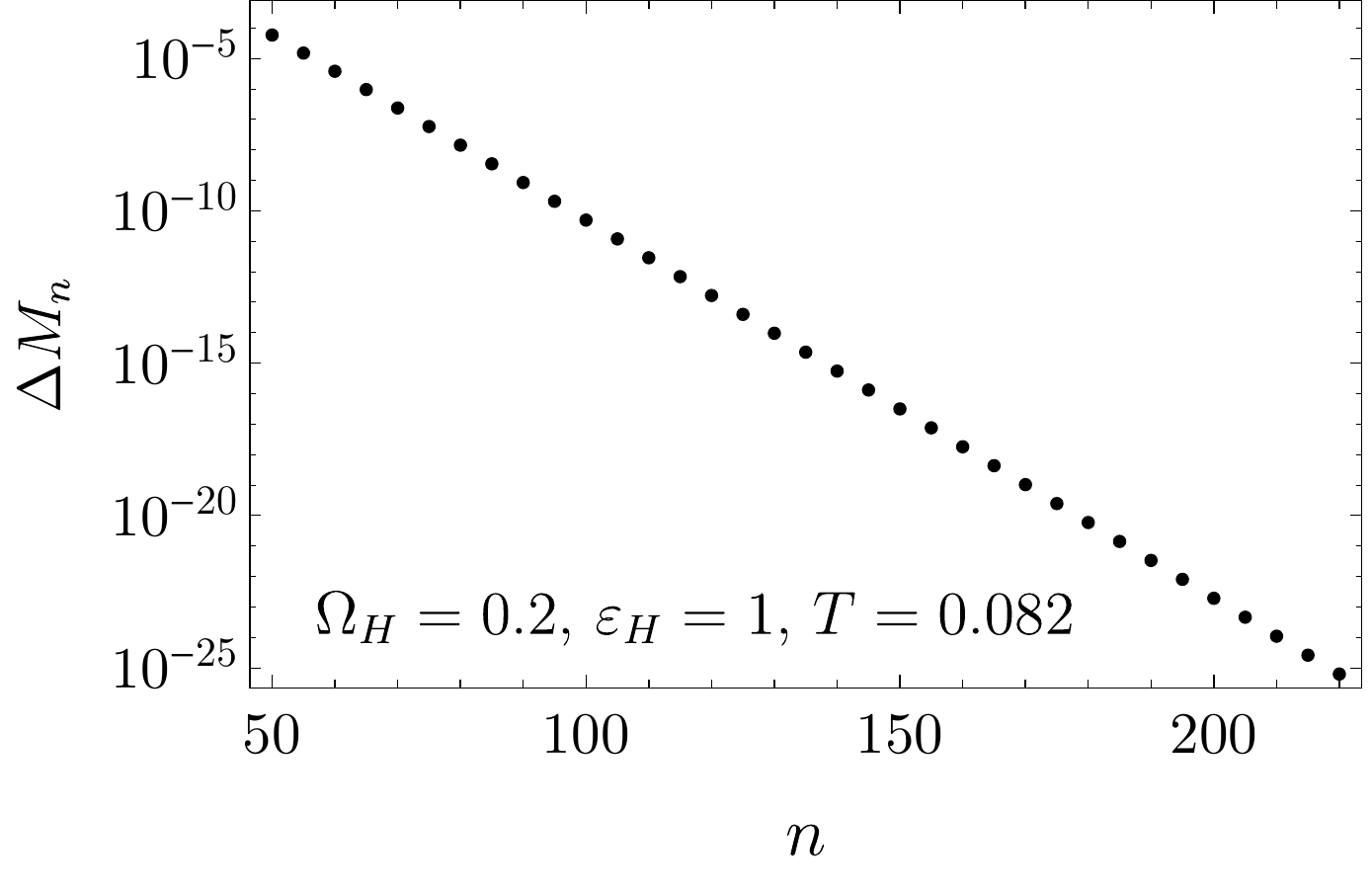}
  \end{minipage}
      \caption{Convergence of fractional error in black hole energy $\Delta M_n=\left|1-M_{n+1}/M_n\right|$ against the grid size $n$. \textit{Left}: Hairy black hole solution in DeTurck gauge. The scale is log-log, and the convergence is a power law. \textit{Right}: Convergence in radial gauge, for the same hairy black hole. The scale now is log-linear, exhibiting an exponential decay.}
        \label{fig:convergence}
\end{figure}

Our ansatz~(\ref{eq:ansatzr}) depends only on the radial coordinate, thus we can employ the radial gauge $\Sigma(r)=r$. The solutions in this gauge exhibit exponential convergence (Fig.~\ref{fig:convergence}, \textit{right}), and it is well suited to study low temperature regime of the hairy solutions. The convergence worsens at the very low temperatures, large angular momenta and large scalar fields, due to steep gradients in the functions. In order to obtain  constant temperature solutions in various thermodynamic ensembles we will also use the DeTurck method \cite{Headrick:2009pv,Figueras:2011va} (for reviews see \cite{Wiseman:2011by,Dias:2015nua}), where we solve the Einstein-DeTurck equation ${G_{\mu\nu}-\nabla_{(\mu}\xi_{\nu)}=0}$. Here ${\xi^{\mu}=g^{\nu\rho}\left[\Gamma^\mu_{\nu\rho}(g)-\Gamma^\mu_{\nu\rho}(\tilde{g})\right]}$, and the metric $\tilde{g}$ is a prescribed suitable reference metric, which remains to be chosen. The reference metric plays an important role, and can significantly alter convergence properties in different parts of the moduli space. Solutions to the DeTurck equation will also be solutions to the Einstein equation (assuming certain conditions)~\cite{Figueras:2011va,Figueras:2016nmo}, and furthermore, the requirement that ${\xi^{\mu}=0}$ will also determine the gauge. The condition can be tracked numerically, and solutions presented in this paper obey the relation $\mathrm{max}\,\xi_\mu\xi^{\mu}=0$ at least to $\mathcal{O}(10^{-10})$ precision.
 
The equations of motion are discretised by pseudospectral collocation on a Chebyshev grid, and the resulting system is solved using a standard iterative Newton-Raphson method. We find that the radial gauge provides exponential convergence with increasing grid size $n$, but in the DeTurck gauge the convergence worsens to a power law (Fig.~\ref{fig:convergence}). Nonetheless, the DeTurck gauge allows us to set the black hole temperature $T$, and we will utilise it to study the canonical ensemble in sections \ref{subsubsec:Isotherms} and \ref{sec:therm}. 
 
\subsection{\label{subsec:set}Numerical setup}
 
In this subsection we will briefly outline the numerical ansatz we used to construct the rotating hairy black hole solutions. First we compactify the radial coordinate $r=\dfrac{y_+}{\sqrt{1-y^2}}$, so that $y=1$ coincides with $r=\infty$, and $y=0$ with $r=r_+=y_+$. The metric \emph{ansatz}~(\ref{eq:ansatzr}) is
\begin{align}
\label{eq:metric1}
\begin{split}
\mathrm{d}s^2&=\frac{1}{1-y^2}\left[- (y_+ y)^2 q_1(y)\,\mathrm{d}t^2+\frac{q_2(y)\,\mathrm{d}y^2}{1-y^2}\right.\\
&\left.+y_+^2 \left\{q_3(y) \left[\mathrm{d}\psi+\frac{1}{2}x\,\mathrm{d}\phi- \Omega(y)\,\mathrm{d}t\right]^2+\frac{1}{4}q_4(y) \left[(1-x^2)\,\mathrm{d}\phi^2+\frac{\mathrm{d}x^2}{1-x^2}\right]\right\}\right],
\end{split}
\end{align}
where $x=\cos{\theta}$ and $\phi$ are the usual polar angles on the $S^2$.

The gauge fields~\eqref{eq:gauge} and the scalar field take the form
\begin{equation}
A_t(r)=y^2q_6(y)-q_5(y) q_7(y)\,,\quad A_\psi(r)=q_7(y)\,,\quad \phi(r)=(1-y^2)q_8(y).
\label{eq:ansatzgaugescalar}
\end{equation}
To change into the frame which is non-rotating at the boundary $y=1$, we factorize $\Omega(y)=(1-y^2)^2q_5(y)$, and at $y=0$ we identify $q_5(0)=\Omega_H$, where $\Omega_H$ is the angular velocity at the black hole horizon. We can choose to impose the boundary condition on $q_5(y)$ either at the horizon or at the conformal boundary, and both of these choices will give a complementary view of the solution space. 

A good reference metric $\tilde{g}$ for the DeTurck method can be obtained from (\ref{eq:metric1}) if we set $\tilde{q}_1=\tilde{q}_2=\tilde{q}_3=\tilde{q}_4=1$, and $\tilde{q}_5=q_5$. Overall this ansatz is incredibly simple, and is allowed by the fact that the hairy black holes can reach arbitrarily low temperatures.

Requiring the correct asymptotics~(\ref{eq:expansion}), in the radial gauge, we find that at the boundary $y=1$ we should specify 
\begin{align}
q_1=q_2=q_3=q_4=1, \\
\label{eq:bc5}
2y_+^2 q_5'-q_5 q_7'^2+4 q_5\left(y_+^2+1\right)+\left(q_6'+2q_6\right) q_7'=0,\\
\label{eq:bc6}
y_+^2 q_8'-2q_6^2q_8=0,\\
q_7=0.
\end{align}
In the DeTurck gauge replace~(\ref{eq:bc5}) by $2 y_+^2 q_5' + 2 q_6 q_7' + q_6' q_7' - q_5 q_7'^2=0$, and~(\ref{eq:bc6}) by $y_+^2 q_8'+q_8\left(y_+^2+1-2 q_6^2\right)=0$. There is some freedom in specifying these conditions, depending on which variables we want to keep fixed as we explore the parameter space. For instance, we can impose the fall-off of the scalar field, however, this condition may not guarantee the unique parametrization of solutions~\cite{Markeviciute:2016ivy}. We will find that it is convenient to fix the angular momentum, which is equivalent to setting the $\Omega(y)$ fall-off at infinity, replacing the condition~(\ref{eq:bc5}) with $q_5=\Omega_\mathrm{inf}$.

At the horizon $y=0$, regularity imposes
\begin{align}
\label{eq:bc1H}
q_1'&=0,\\
q_2'=q_3'=q_4'=q_6'=q_7'=q_8'&=0,\\
\label{eq:bc5H}
q_5&=\Omega_H,\\
q_8&=\varepsilon_H.
\end{align}
 
\noindent Here we choose to specify the scalar field value at the horizon and label it $\varepsilon_H$, and angular velocity at the horizon $\Omega_H$. If instead we specify the angular momentum, we would change~(\ref{eq:bc5H}) to the Neumann condition $q_5'=0$. For the DeTurck gauge, replace~(\ref{eq:bc1H}) by $q_1 - q_2 (1 - q_5^2)=0$. We can also specify the expectation value of the dual operator, by fixing $q_8=\langle \mathcal{O}_\phi\rangle/y_+^2$ at infinity.

We did not find any rotating smooth soliton solutions\footnote{One can globaly rotate the solution~(\ref{eq:solansatz}) using the gauge freedom~(\ref{eq:gaugeres}). However, such solutions don't seem to play any important role in the phase space.}. For completeness, we give the numerical ansatz for the smooth non-rotating soliton, as these solutions play an important role in the phase space of hairy rotating black holes. With the compactified radial coordinate ${r=\dfrac{y}{\sqrt{1-y^2}}}$ it can be written as	
\begin{equation}
\begin{split}
\mathrm{d}s^2&=\frac{1}{1-y^2}\left[- q_1(y)\,\mathrm{d}t^2+\frac{q_2(y)\,\mathrm{d}y^2}{1-y^2} +q_3(y)y^2 \left\{\frac{\mathrm{d}x^2}{1-x^2}+(1-x^2)\,\mathrm{d}\theta^2+x^2\mathrm{d}\psi^2\right\}\right],\\
A(y)&=q_4(y),\\
\phi(y)&=(1-y^2)q_5(y).
\end{split}
\label{eq:solansatz}
\end{equation}
\noindent The boundary conditions at $y=1$ are given by $q_1=1$, $q_2=1$, $q_3=1$, $(q_4^2-1) q_5 - q_5'=0$ and at $y=0$ $q_1'=0$, $q_2=q_3$, $q_3=1$, $q_4'=0$, $q_5'=0$, $q_5=\varepsilon_H$.
 
\subsection{\label{subsubsec:qua}Thermodynamic quantities}

The global charges of the system, $M$, $J$, $Q$ and the chemical potential $\mu_\infty$ can be found from the fall-off of the metric~(\ref{eq:expansion}). The asymptotic Killing vectors $\partial_t$ and $\partial_\psi$ will have associated conserved quantities, which in AdS spacetimes can be computed using the Ashtekar-Das method~\cite{Ashtekar:1999jx}. The mass and the angular momentum in the radial gauge are given by
\begin{align}
M&=\frac{1}{32}y_+^2\left[18+y_+^2\left(18-3q_1''+q_3''\right)\right]_{y=1},\\
J&=\frac{1}{2} y_+^4 q_5|_{y=1},
\label{eq:momentum}
\end{align}
where all functions are evaluated at the $y=1$ boundary, and the background AdS$_5$ has zero mass. The electric $U(1)$ charge is given by the electromagnetic flux integral at infinity
\begin{equation}
Q=\frac{1}{16 \pi G_5}\int_{\mathbb{S}^3_{\infty}}(\star F -F\wedge A)=\frac{1}{4} y_+^2 \left(-q_6'-2q_6+q_5 q_7'\right)|_{y=1},
\label{eq:charge}
\end{equation}
where we note that the Chern--Simons term vanishes due to no magnetic field at the boundary. The chemical potential, which is a thermodynamic conjugate quantity to the electric charge, is given by
\begin{equation}
\mu=\xi^\nu A_\nu|_{H}-\xi^\nu A_\nu|_{\infty}=-q_6|_{y=1},
\end{equation}
where $\xi^\nu\partial_\nu=\partial_t+\Omega_H\partial_\psi$ is the null generator of the horizon. The conjugate potential to the angular momentum is given by
\begin{equation}
\Omega=\Omega_H-\Omega_\infty=q_5|_{y=0},
\end{equation}
\noindent as we work in the frame which is non-rotating at infinity with $\Omega_\infty=0$. The Hawking temperature is found in the usual way by requiring the periodicity of the Euclidean time
\begin{equation}
T=\frac{y_+}{2\pi}\left.\sqrt{\frac{q_1}{q_2}}\right|_{y=0}\, ,
\end{equation}
and the Bekenstein-Hawking entropy of the black hole is given by
\begin{equation}
S=\pi y_+^3\,\sqrt{q_3}\,q_4\,|_{y=0},
\end{equation}
which is proportional to the area of its event horizon.
 
The thermodynamic quantities satisfy the first law
\begin{equation}
\mathrm{d}M=T\mathrm{d}S+3\mu\mathrm{d}Q+2\Omega\mathrm{d}J,
\label{eq:firstlawofTD}
\end{equation}
\noindent and we verified that numerical rotating hairy solutions obey the first law to at least $0.1\%$ accuracy in the DeTurck gauge, and to better than $0.001\%$ in the radial gauge.

\begin{figure}[t]
\centering
  \begin{minipage}[t]{1\textwidth}
    \includegraphics[width=\textwidth]{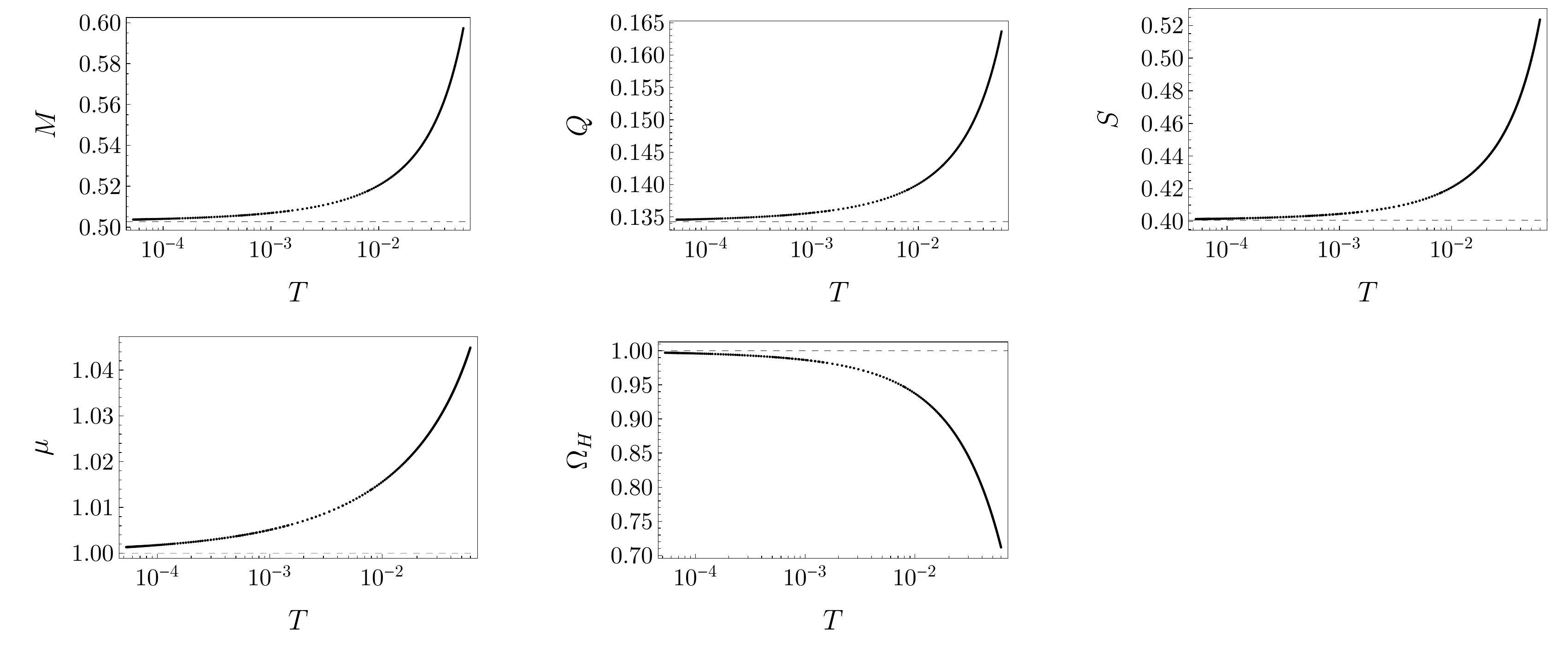}
  \end{minipage}
      \caption{Thermodynamic quantities against the temperature $T$, for black holes with fixed angular momentum $J=0.05$ and horizon scalar field $\varepsilon_H=10^{-4}$ (black data points). The dashed gridlines show the values for the supersymmetric black hole with the same $J$.}
        \label{fig:additional1}
\end{figure}

\section{\label{sec:Phase}Phase diagram of hairy black holes} 
\subsection{Linear instability and the onset plane} 

\begin{figure}[t]
\centering
  \begin{minipage}[t]{0.43\textwidth}
    \includegraphics[width=\textwidth]{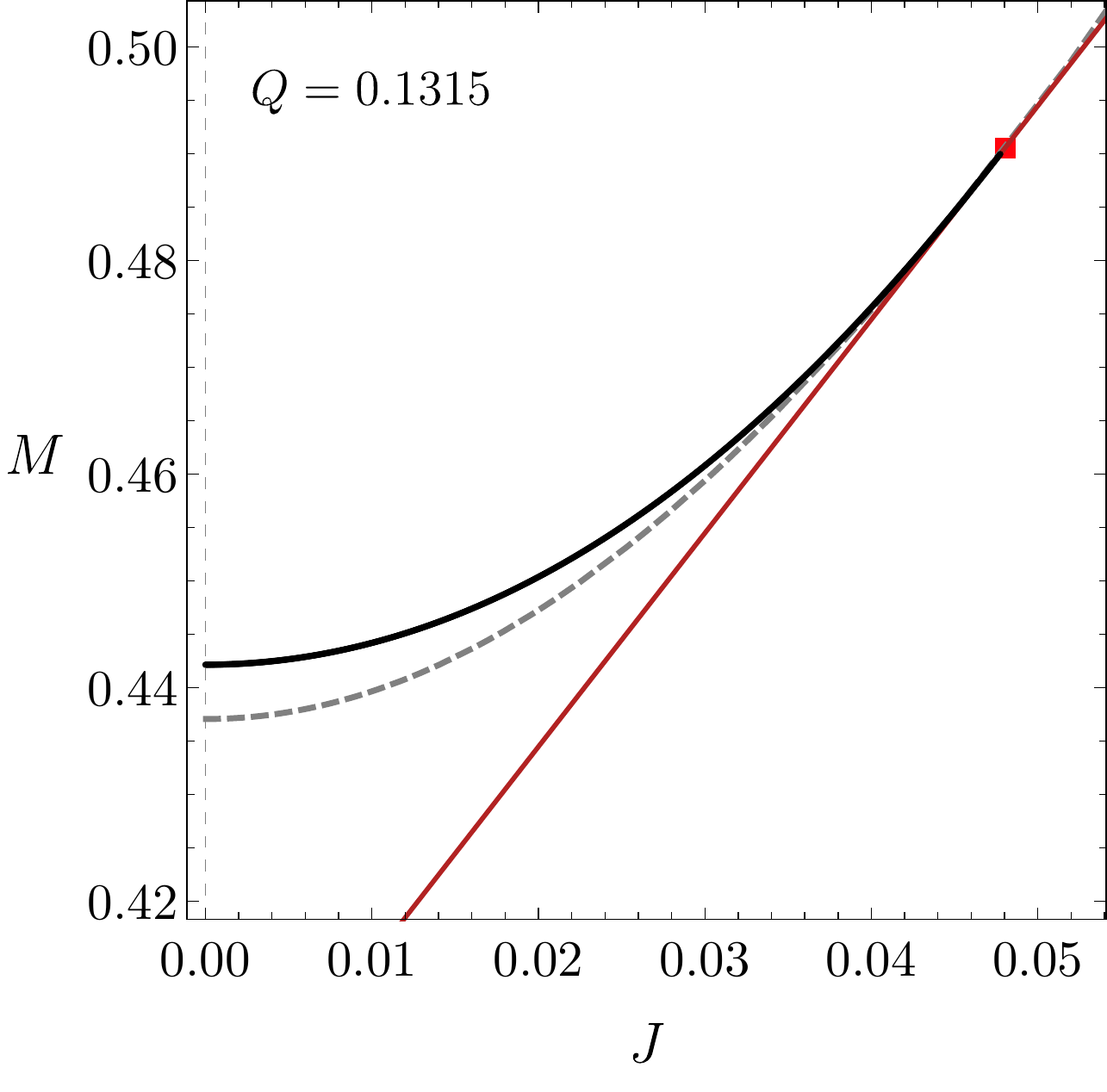}
  \end{minipage}
  \hfill
    \begin{minipage}[t]{0.48\textwidth}
    \includegraphics[width=\textwidth]{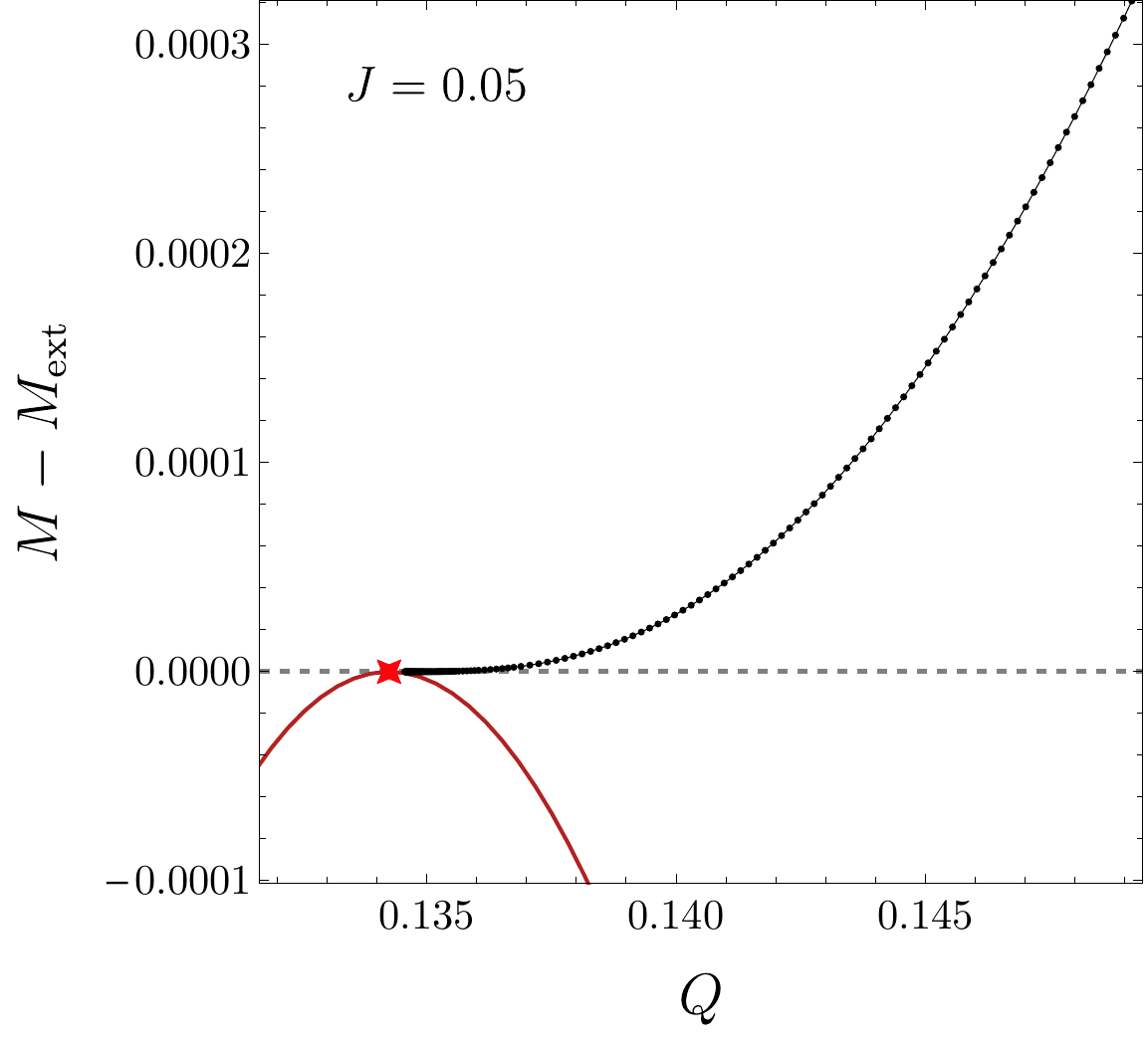}
  \end{minipage}
      \caption{\textit{Left}: The line of solutions showing the onset of the superradiance for the CLP black holes at constant charge $Q=0.1315$ (black points). These solutions extend arbitrarily close to the supersymmetric black holes (red square), with the temperature asymptotically approaching $T=0$. The grey dashed line shows the extremality, and the red bold line shows the BPS bound $M=3Q+2J$. \textit{Right}: The difference $M-M_\mathrm{ext}$ against the charge $Q$, for black holes with fixed $J=0.05$ and $\varepsilon_H=10^{-4}$ (black data points), where $M_\mathrm{ext}$ is mass of the extremal CLP black hole with the same $Q$. Red solid line is the BPS bound, and the red star shows the Gutowski-Reall solution.}
        \label{fig:linear}
\end{figure}

In order to directly compute the onset plane of the superradiant instability\footnote{At large charges the tachyonic instability is the dominant one, however, in this section we are mostly focusing on relatively small black holes.}, we linearise the scalar equation~\eqref{eq:scalar} about the CLP black hole background~(\ref{eq:CLP}). We obtain a second order equation for the infinitesimal perturbation $\delta q_8$ of $q_8$ defined in (\ref{eq:ansatzgaugescalar}), which has three parameters $y_+, j, q$. The lowest order parameter is $q$, and the linearized equation can be solved as an 8th order eigenvalue equation in $q$. Alternatively, we can fix two parameters and regard the third parameter  as an extra variable in the Newton's method. We follow this approach, the details of which can be found in \cite{Dias:2010ma, Dias:2011tj,Dias:2015nua,Markeviciute:2017jcp}, and numerically solve the resulting equation
\begin{equation}
L(y;y_+)\delta q_8(y)=0,
\end{equation}
where $L(y;y_+)$ is a second order differential operator, which is a non--linear function of $y_+$ once we fix $j$ and $q$. This way we can choose $y_+$ to be a parameter which is determined once we fix the overall charge of the black hole $Q$, and approach the $T\rightarrow 0$ limit when $j$ is increased from zero. The boundary conditions can be found by demanding regularity at the horizon and expanding the scalar field equation off the asymptotic infinity. The boundary conditions are
\begin{equation}
\delta q_8'(0)=0,\qquad -2(y_+^2-j^2(1+y_+^2))^2q^2\,\delta q_8(1)+y_+(y_+^4+j^2 q)\,\delta q_8'(1)=0.
\end{equation}
The linear results at constant charge $Q=0.1315$ are presented in~Fig.~(\ref{fig:linear}). We find that for all charges that are feasible to track numerically, the family of black holes at the onset of the instability approaches the supersymmetric black hole, and exists just above the extremality plane of the CLP solutions. 

The plane representing the onset of the scalar field instability can also be generated by solving the non-linear equations~\eqref{eq:eom} by fixing the horizon scalar $q_8(0)=\varepsilon_H$ to be very small, and both methods were found to be in a very good agreement where the comparison is possible\footnote{A direct comparison is somewhat difficult, as we are fixing different conserved charges in each case. In the full non-linear setup it is possible to fix the overall charge $Q$, but this requires the other two free parameters to be $y_+$ and $\varepsilon_H$, which causes $J$ to vary slightly.}. In the radial gauge, we were able to achieve temperatures lower than $T\simeq 10^{-4}$. The results are presented in Fig.~\ref{fig:linear} (\textit{right}), where we plot the mass difference $M-M_\mathrm{ext}$ for fixed $J=0.05$, where $M_\mathrm{ext}$ is the mass of the corresponding extremal black hole in the microcanonical ensemble. As in the linear case, we find hairy black holes arbitrarily close to the Gutowski-Reall black hole. The non-linear equations are much harder to solve near the extremality than the linearised equation. In Fig.~\ref{fig:additional1}, we also show various thermodynamic quantities along the family of a constant $J$ and $\varepsilon_H$, as $T\rightarrow 0$. As expected, the quantities approach those of the corresponding supersymmetric black hole with the fixed $J$, in particular $\Omega_H\rightarrow 1$ and $\mu\rightarrow 1$.
 
\subsection{\label{subsec:constJ}Constant $J$ planes}

\begin{figure}[t]
\centering
  \begin{minipage}[t]{0.45\textwidth}
    \includegraphics[width=\textwidth]{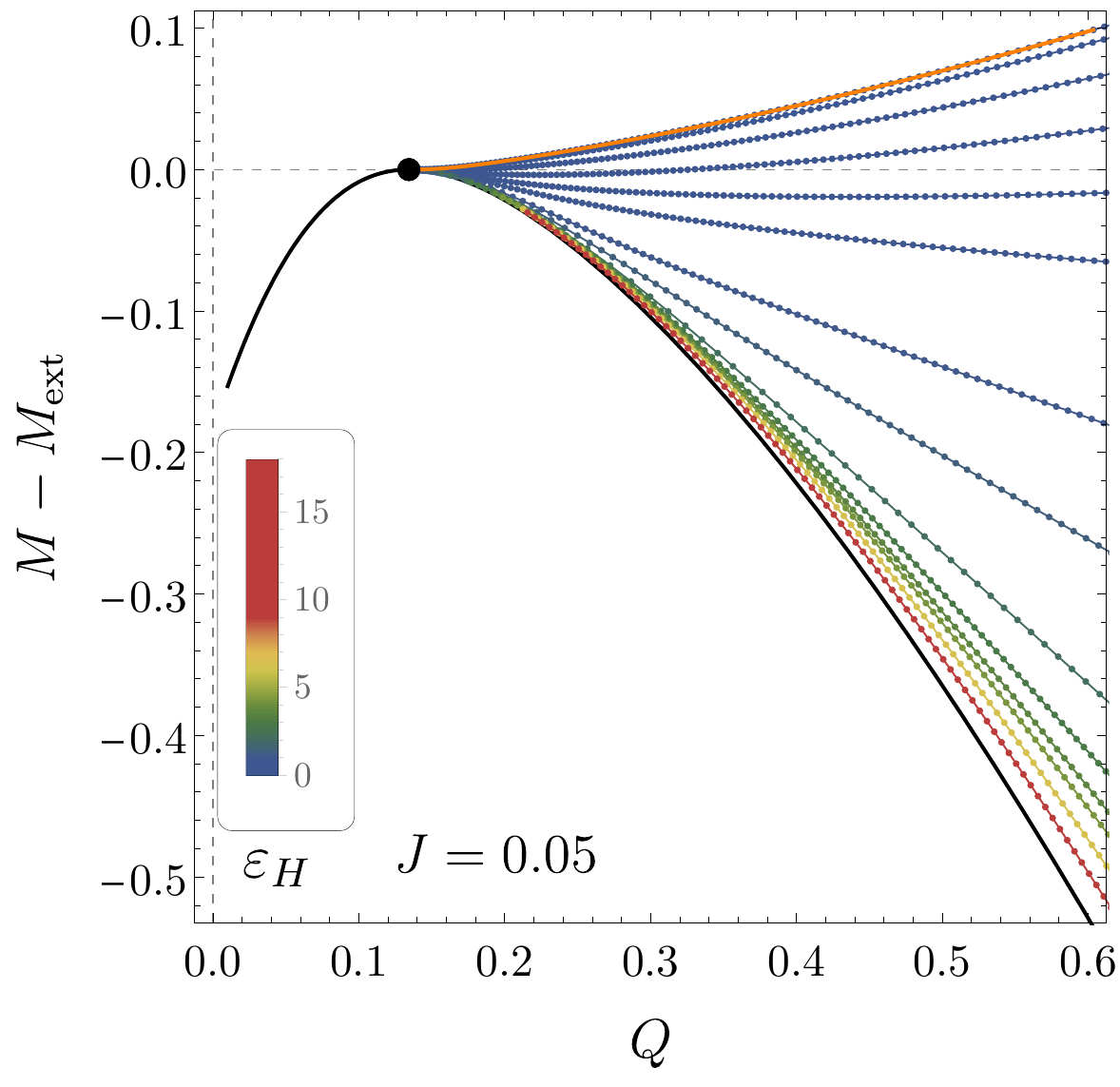}
  \end{minipage}
  \hfill
    \begin{minipage}[t]{0.44\textwidth}
    \includegraphics[width=\textwidth]{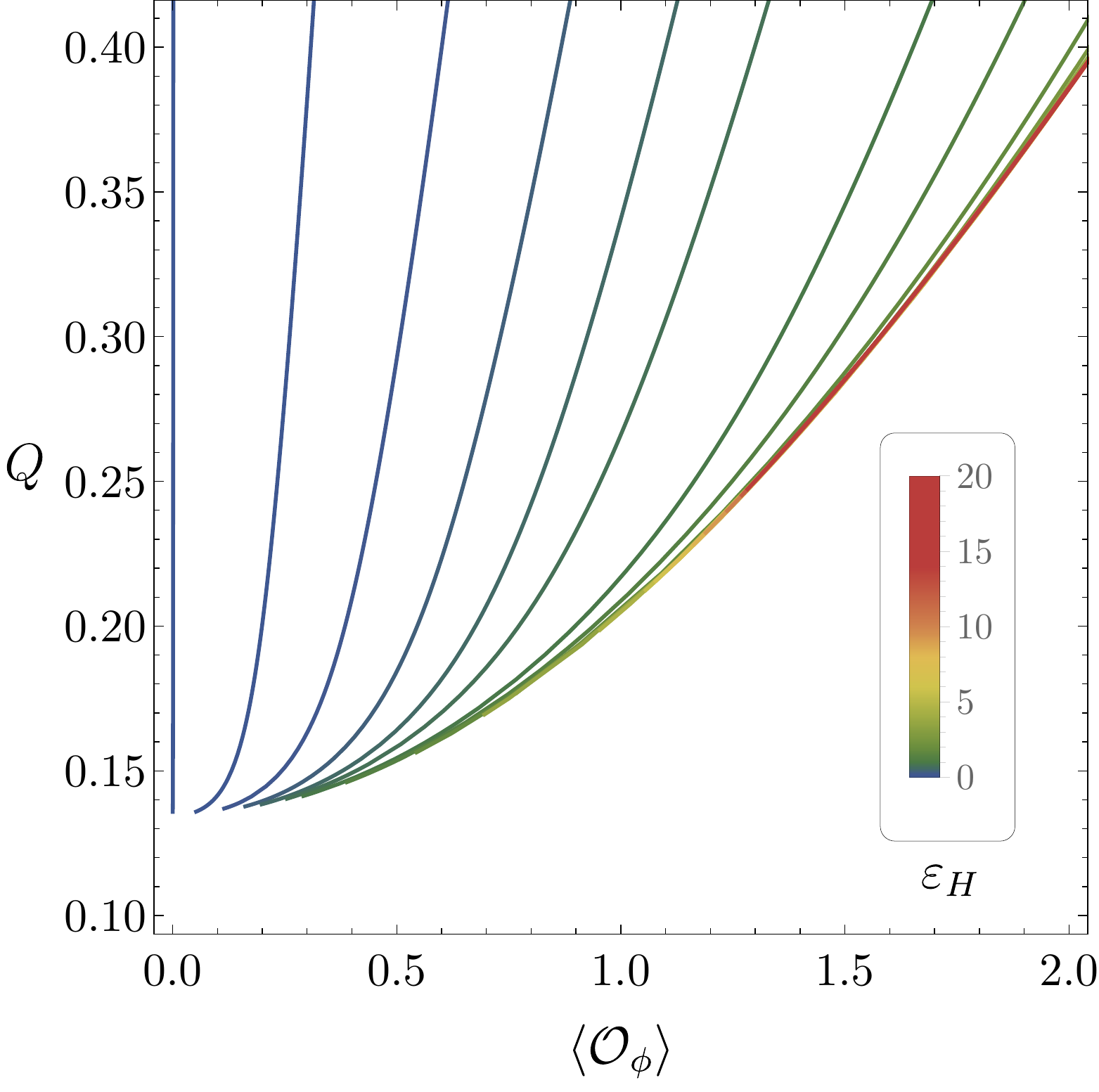}
  \end{minipage}
      \caption{\textit{Left}: The mass difference $M-M_\mathrm{ext}$ against the charge $Q$ for fixed angular momentum $J=0.05$ and different values of the horizon scalar $\varepsilon_H$ (rainbow disks), where $M_\mathrm{ext}$ is the mass of the corresponding extremal CLP black hole with the same $J$ and $Q$. The black disk shows the supersymmetric black hole, the grey solid line shows the BPS bound, the orange solid line shows the instability curve. \textit{Right}: The charge $Q$ against the expectation value of the operator dual to the scalar field, for families of hairy black holes with constant $J=0.05$ and various values of the horizon scalar $\varepsilon_H$.}
        \label{fig:QM}
\end{figure} 

In order to explore the hairy solution space, we need to decide how to fix the free constants at the boundaries. To ensure that our solutions are rotating, we can fix the value of $q_5(y)$ at the horizon, or at infinity, and both of these choices will give us a different way of understanding the hairy solution space. The latter is equivalent to setting the total angular momentum~$J$~(\ref{eq:momentum}). We found in the previous subsection that the hairy black holes at a constant $J$ slice with very small horizon hair $\varepsilon_H$ (just on the merger line), approach the supersymmetric black hole on the BPS bound as we lower $y_+$, and hence the temperature $T$. In such limit $T\rightarrow 0$, and numerically we can achieve temperatures as low as $10^{-4}$ (Fig.~\ref{fig:additional1}).

As we increase $\varepsilon_H$ and decrease $y_+$, the constant $J$ and $\varepsilon_H$ hairy black hole families approach the BPS bound and $T\rightarrow 0$, for all $\varepsilon_H$, with larger horizon scalar curves advancing towards the BPS bound at larger charges (Fig.~\ref{fig:QM}). We were able to reach extremely low temperatures, and the coldest hairy black hole solutions satisfy the supersymmetric bound $M=3Q+2J$ to better than $\mathcal{O}(10^{-6})$ accuracy. In this near-extremal regime the functions acquire large gradients and our numerical method yields slow convergence. We will discuss the approach to the BPS bound in more detail in the next subsection.

\begin{figure}[t]
\centering
    \begin{minipage}[t]{0.45\textwidth}
    \includegraphics[width=\textwidth]{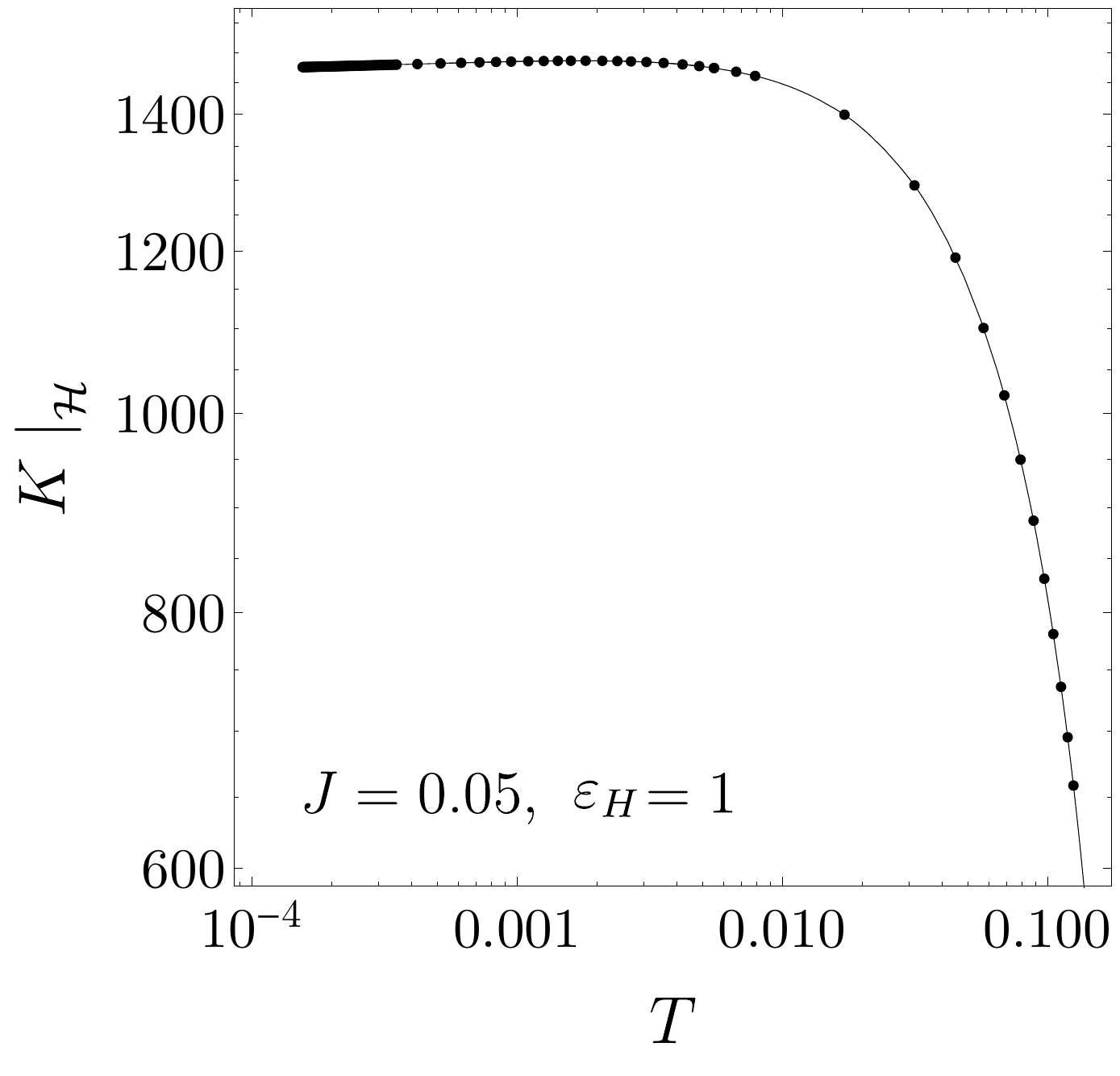}
  \end{minipage}
  \hfill
  \begin{minipage}[t]{0.42\textwidth}
    \includegraphics[width=1\textwidth]{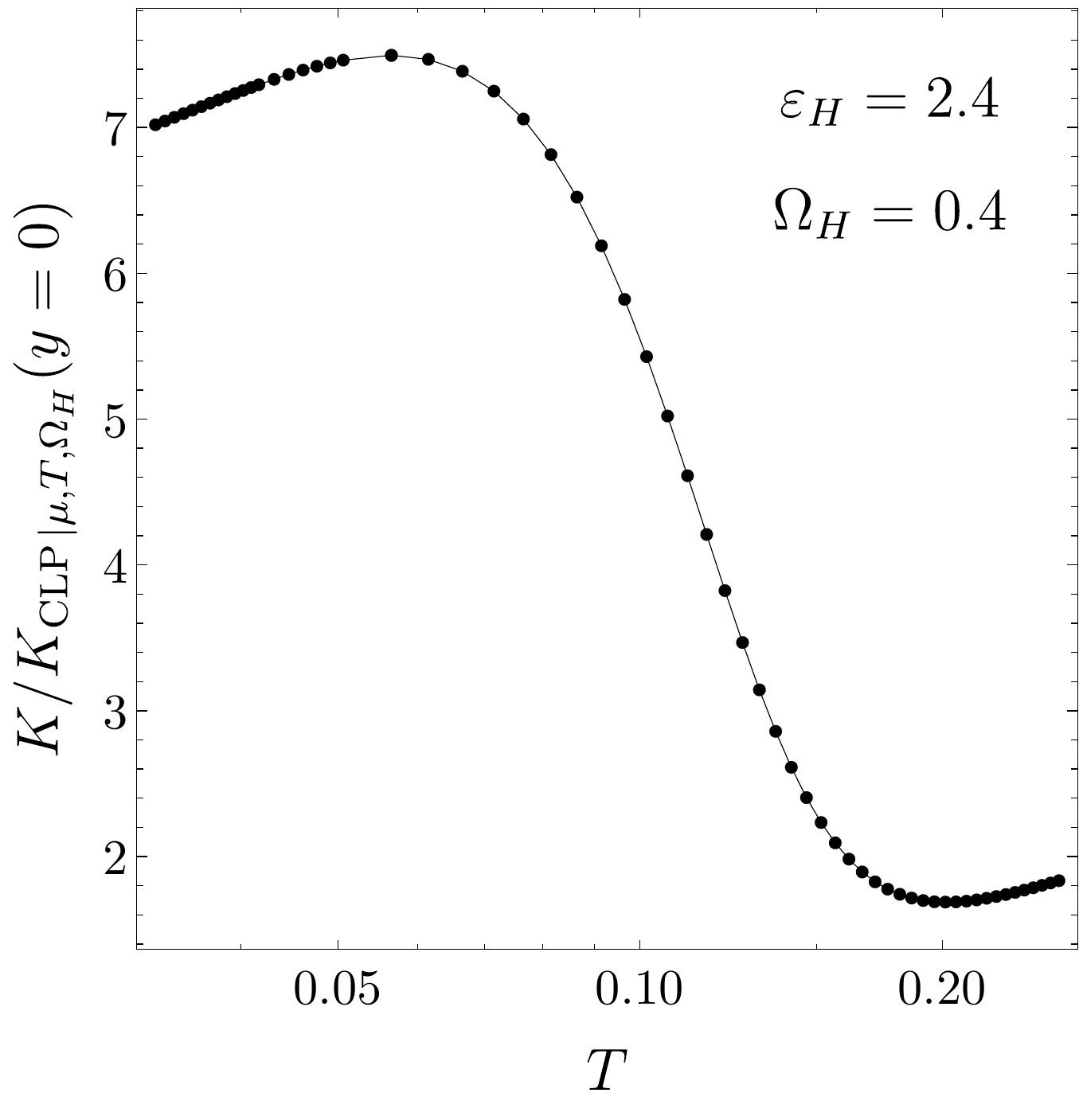}
  \end{minipage}
      \caption{\textit{Left}: The log-log plot of the Kretschmann curvature scalar $K=R_{abcd}R^{abcd}$ against the temperature, for fixed angular momentum $J=0.05$ and horizon scalar field $\varepsilon_H=1$. \textit{Right}: Kretschmann curvature invariant at the horizon $y=0$ for the hairy black hole with constant horizon values $\varepsilon_H$ and $\Omega_H$, scaled by the $K$ value of a corresponding CLP black hole in the grand-canonical ensemble, \textit{i.e.} with the same $\mu$, $T$ and $\Omega_H$. The ratio remains finite as we approach the BPS bound, and the black hole family shrinks to a smooth solitonic solution.}
        \label{fig:curvature}
\end{figure}

The ultra-cold hairy black hole solutions are free of curvature singularities, and in the coordinate frame~\eqref{eq:metric1} the components of the Riemann tensor $R_{abcd}$ and its derivatives are finite everywhere (notably including the horizon $y=0$), which results in the curvature invariants derived from the Riemann tensor being finite as well. We have explicitly checked that curvature invariants $R_{ab}R^{ab}$, $R_{abcd}R^{abcd}$ (presented in Fig.~\ref{fig:curvature} (\textit{left})), $C_{abcd}C^{abcd}$ where $C_{abcd}$ is the Weyl tensor, $T_{ab}T^{ab}$ and $F_{ab}F^{ab}$ are everywhere finite as $T\rightarrow 0$. However, the tidal forces felt by an infalling observer are diverging, and hence there is a parallely propagated curvature singularity. This can be confirmed by computing the Riemann tensor in a freely falling frame~\cite{Horowitz:1997uc}. For a further discussion on the interpretation of the singular nature of the limiting solution, and the compatibility with the supergravity interpretation we refer the reader to~\cite{Markeviciute:2018yal}.

\begin{figure}[t]
\centering
     \begin{minipage}[t]{0.46\textwidth}
    \includegraphics[width=\textwidth]{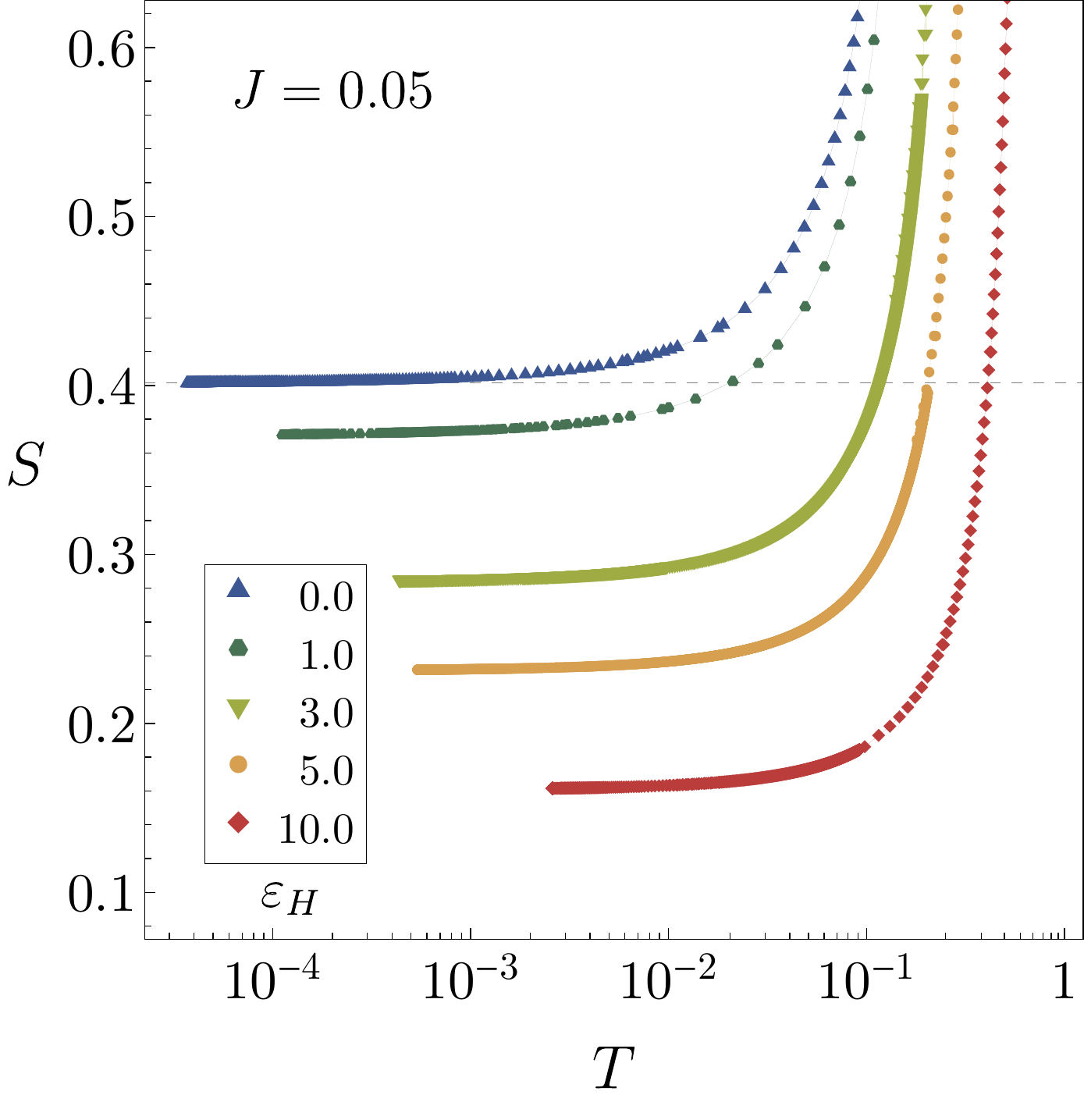}
  \end{minipage}
 	\hfill
    \begin{minipage}[t]{0.47\textwidth}
    \includegraphics[width=\textwidth]{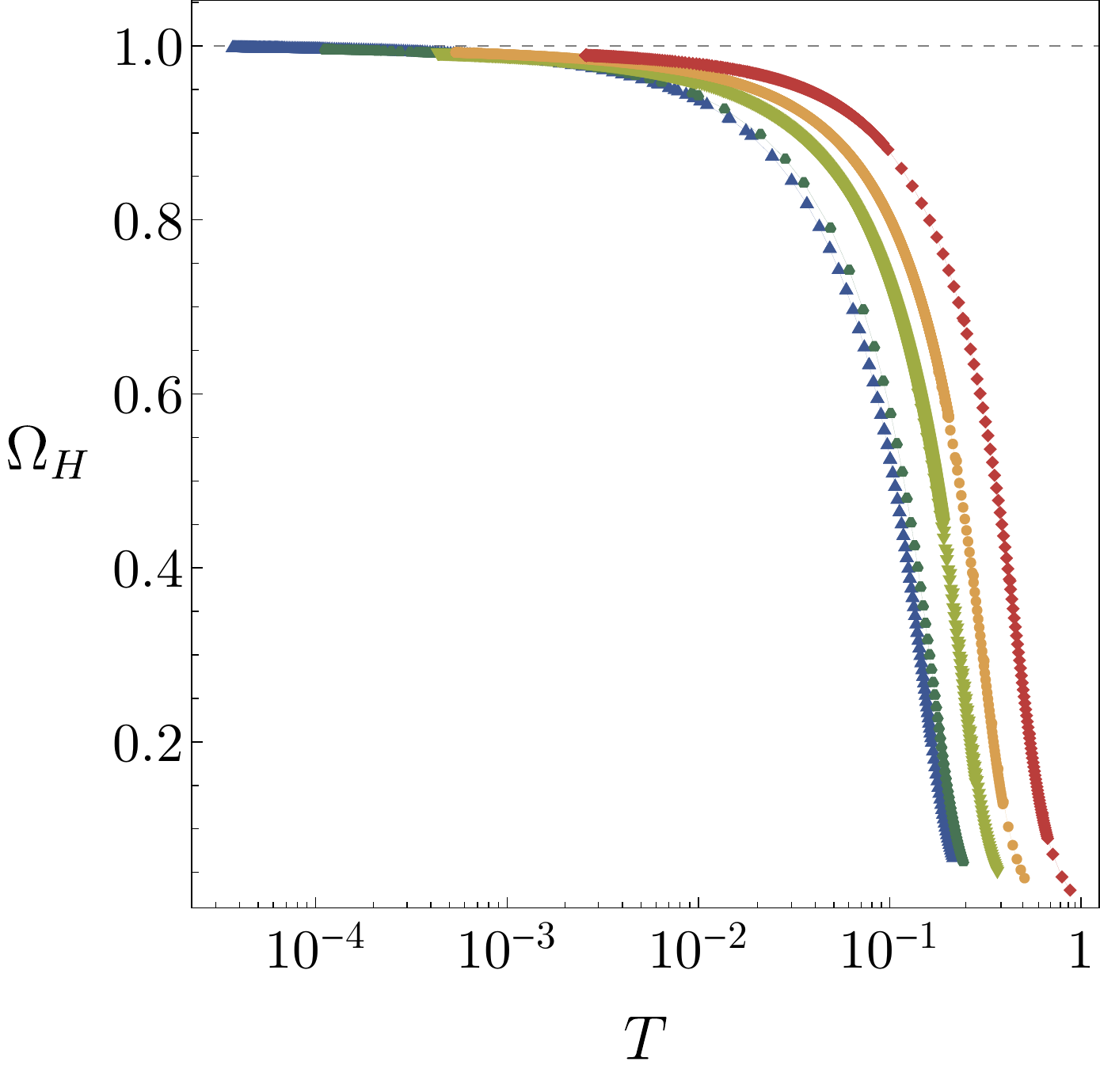}
  \end{minipage}
      \caption{Log-linear plot of the entropy (\textit{left}) and the horizon angular velocity (\textit{right}, the legend is the same as for $S$) of the hairy solutions with fixed angular momentum $J=0.05$ against the temperature, for several values of the horizon scalar $\varepsilon_H$. The horizontal dashed gridline shows the values for the supersymmetric black hole with the same $J$.}
        \label{fig:SOmega}
\end{figure}

Perhaps the most intriguing property of this limit is the fact that these hairy black holes retain finite entropy (Fig.~\ref{fig:SOmega}, \textit{left}) while approaching arbitrarily close the the BPS bound, \textit{i.e.} $\Omega_H\rightarrow 1$ (Fig.~\ref{fig:SOmega}, \textit{right}), $\mu\rightarrow 1$, $T\rightarrow 0$ and $|M-(3Q+2J)|\rightarrow 0$. We conjectured in~\cite{Markeviciute:2018yal}, that the entropy is decaying asymptotically with increasing $\varepsilon_H$, and it is possible that all finite $\varepsilon_H$, $\Omega_H=1$, $J\neq 0$ hairy black holes have non-zero entropy.

\begin{figure}[t]
\centering
  \begin{minipage}[t]{.46\textwidth}
    \includegraphics[width=\textwidth]{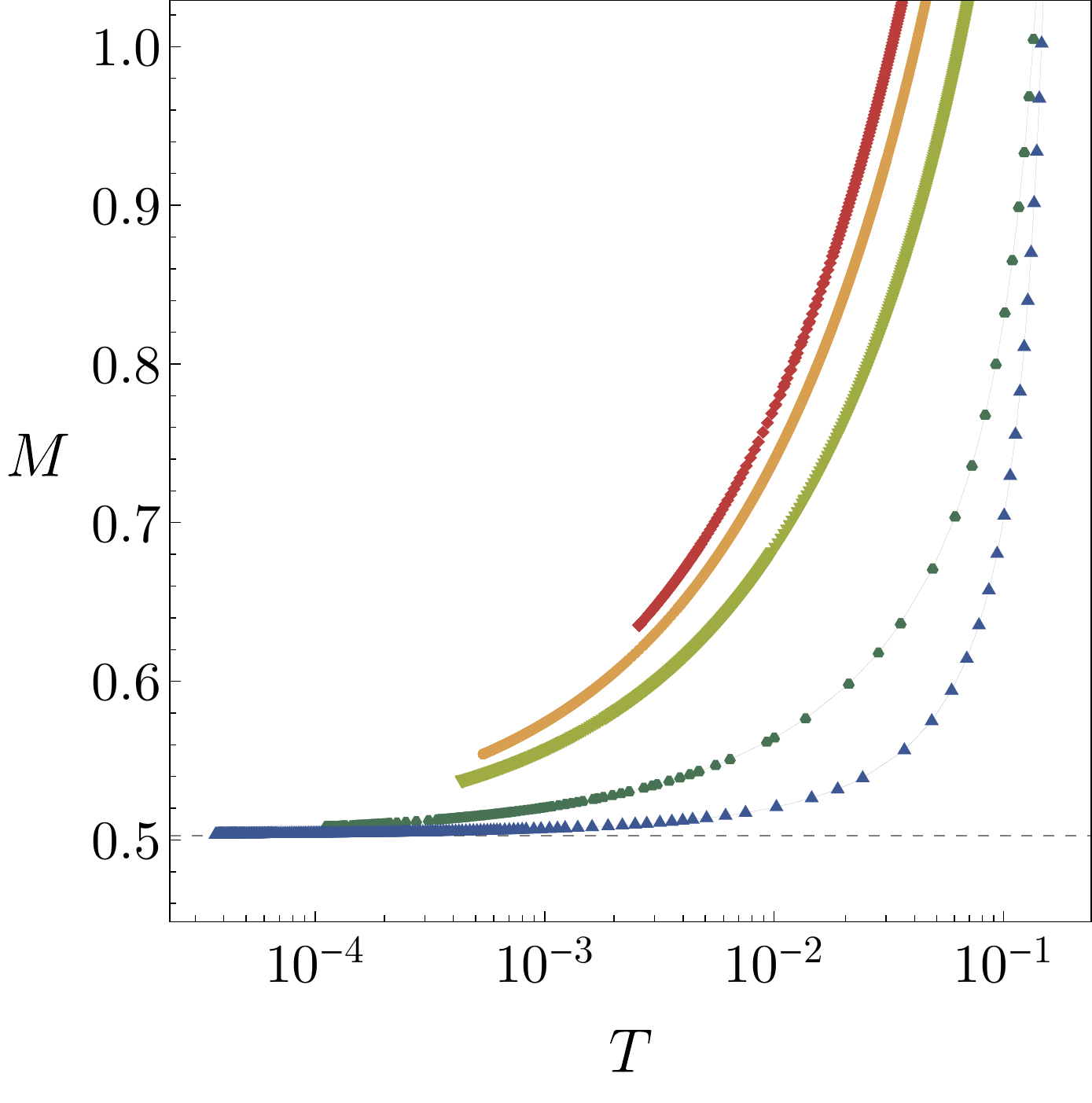}
  \end{minipage}
  \hfill
    \begin{minipage}[t]{.47\textwidth}
    \includegraphics[width=\textwidth]{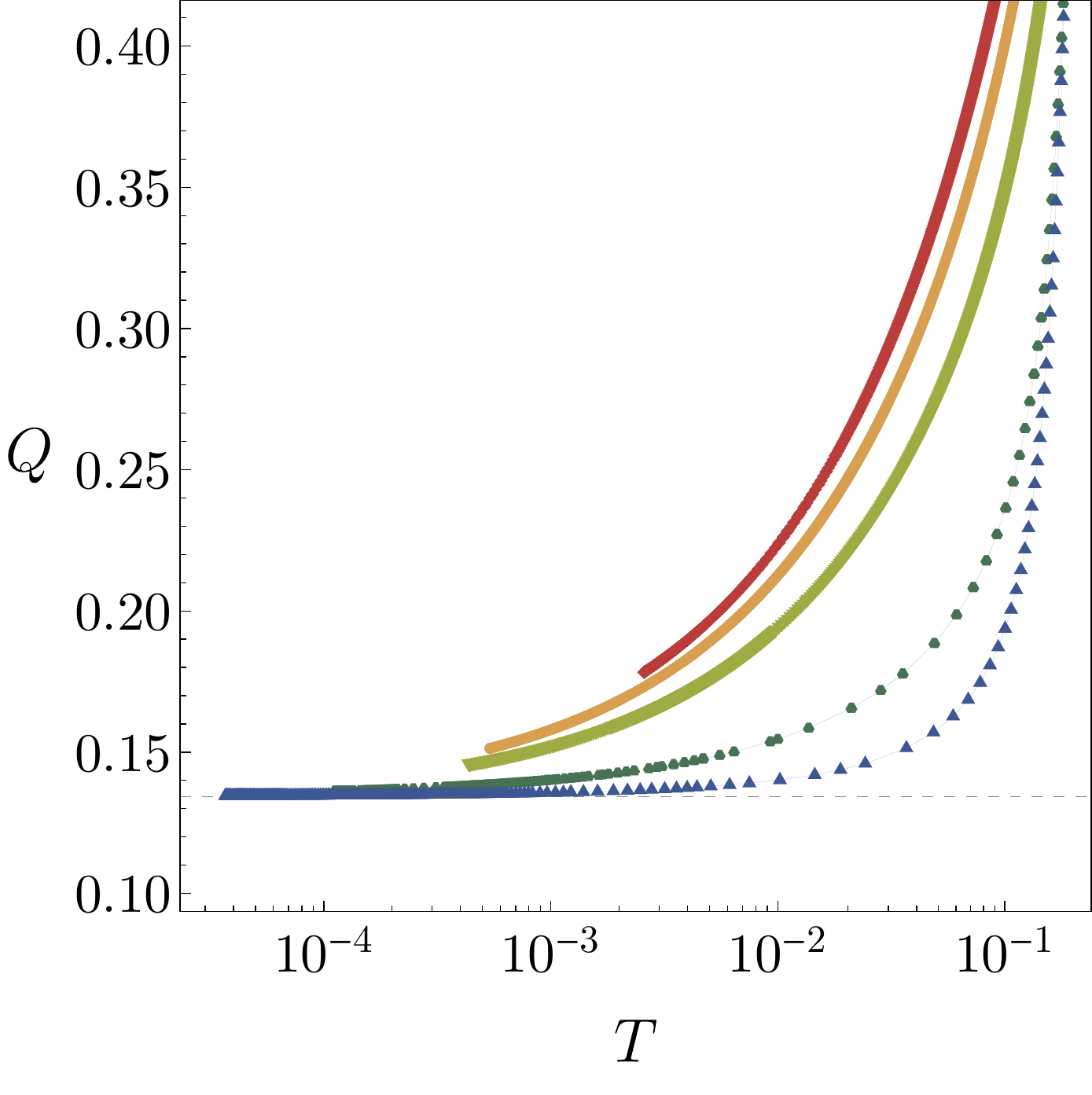}
  \end{minipage}
      \caption{Log-linear plot of the mass (\textit{left}) and the charge (\textit{right}) of the hairy solutions with angular momentum $J=0.05$ versus $T$, for several values of $\varepsilon_H$. The horizontal dashed gridline shows the values for the supersymmetric black hole with the same $J$. The legend is the same as in Fig.~\ref{fig:SOmega}.}
        \label{fig:MT}
\end{figure}

As we increase $\varepsilon_H$, the numerics becomes increasingly difficult, and for larger horizon scalar fields we do not yet see the mass and the charge settle (Fig.~\ref{fig:MT}). Finally, varying $J$ gives a very similar picture. As expected, the larger values of $J$ increase the numerical error.


\subsubsection{\label{subsubsec:Extremal}The extremal limit}

\begin{figure}[t]
\centering
  \begin{minipage}[t]{0.31\textwidth}
    \includegraphics[width=\textwidth]{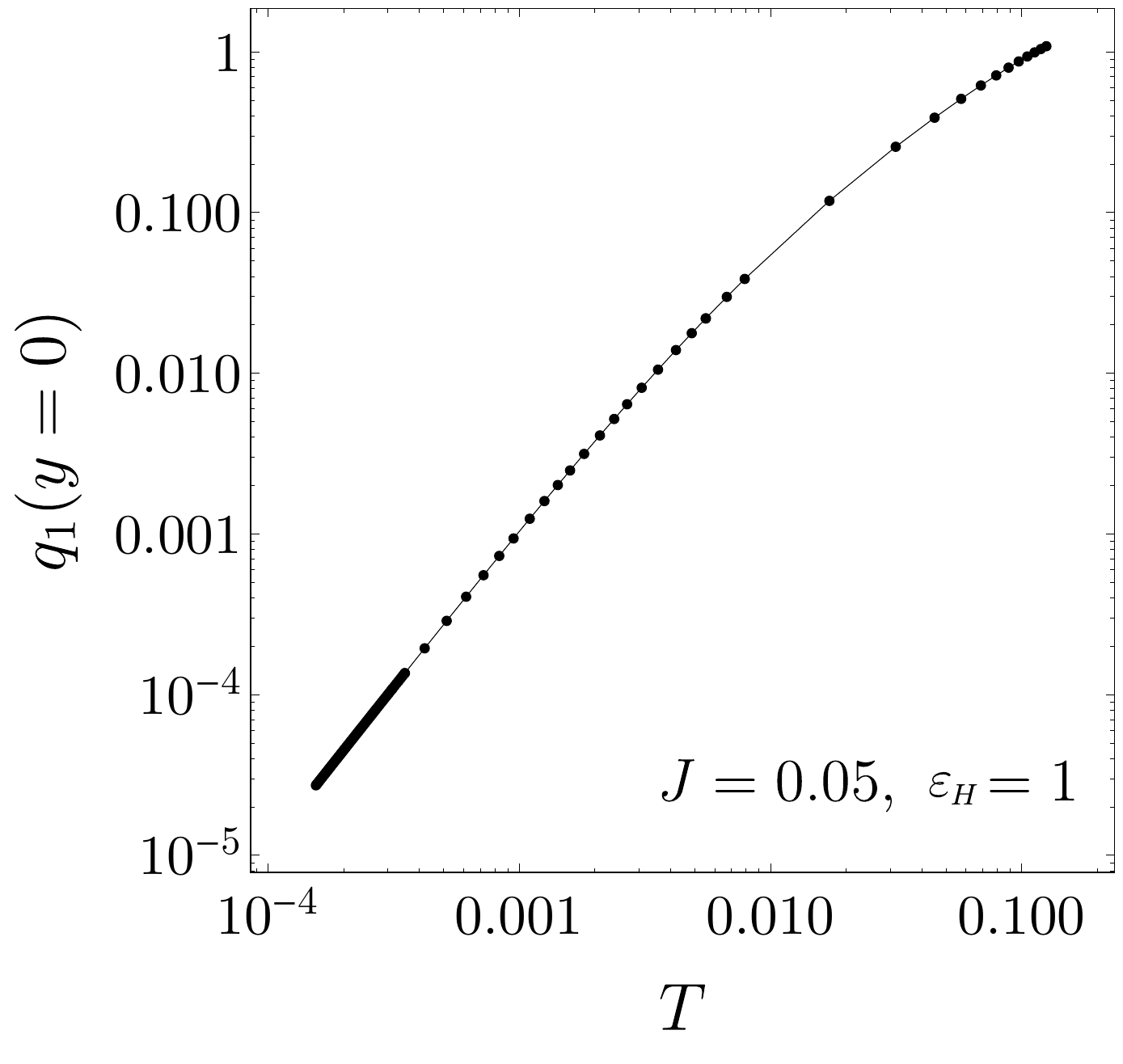}
  \end{minipage}
  \hfill
  \begin{minipage}[t]{0.31\textwidth}
    \includegraphics[width=\textwidth]{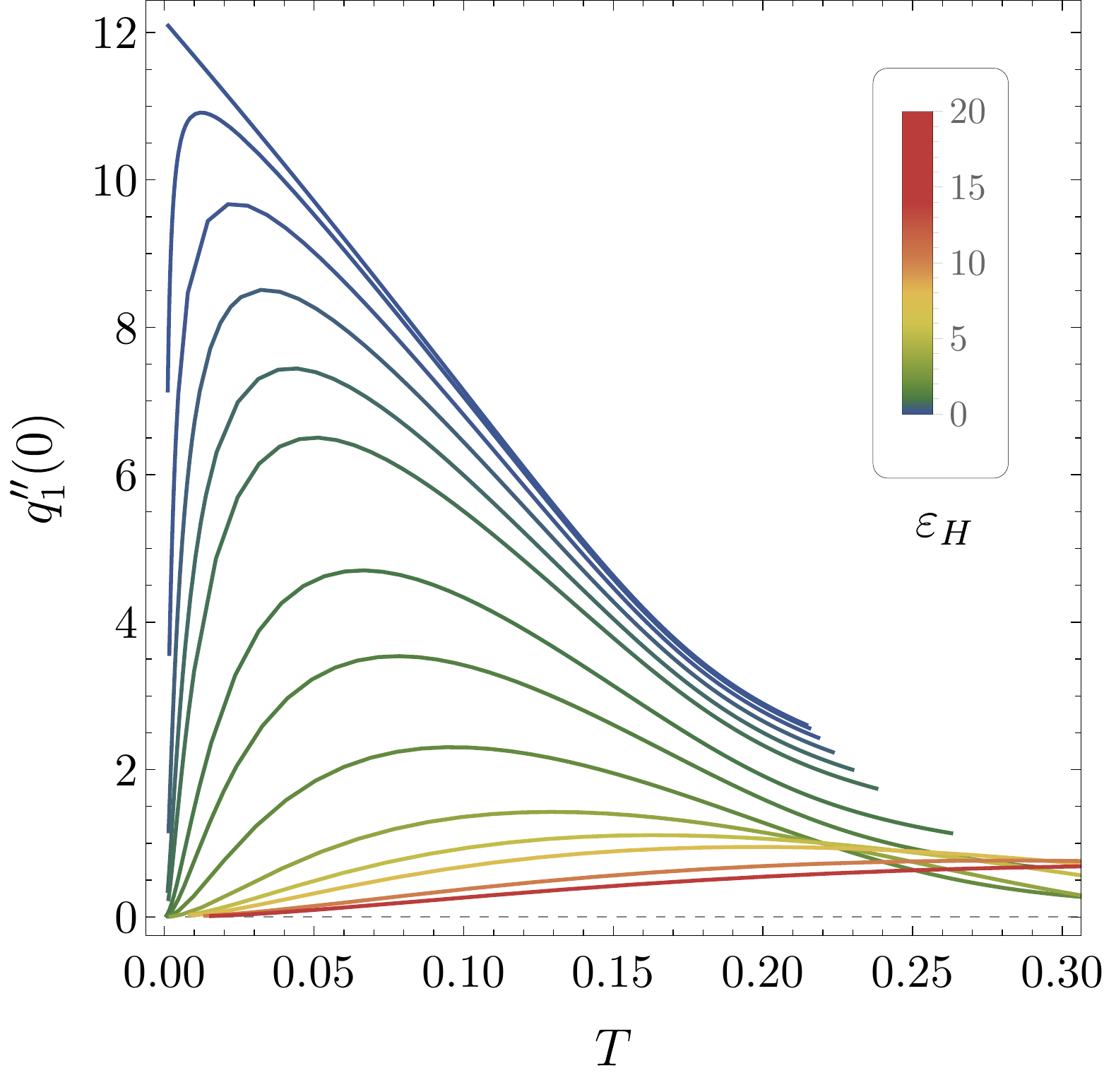}
  \end{minipage}
  \hfill
    \begin{minipage}[t]{0.31\textwidth}
    \includegraphics[width=\textwidth]{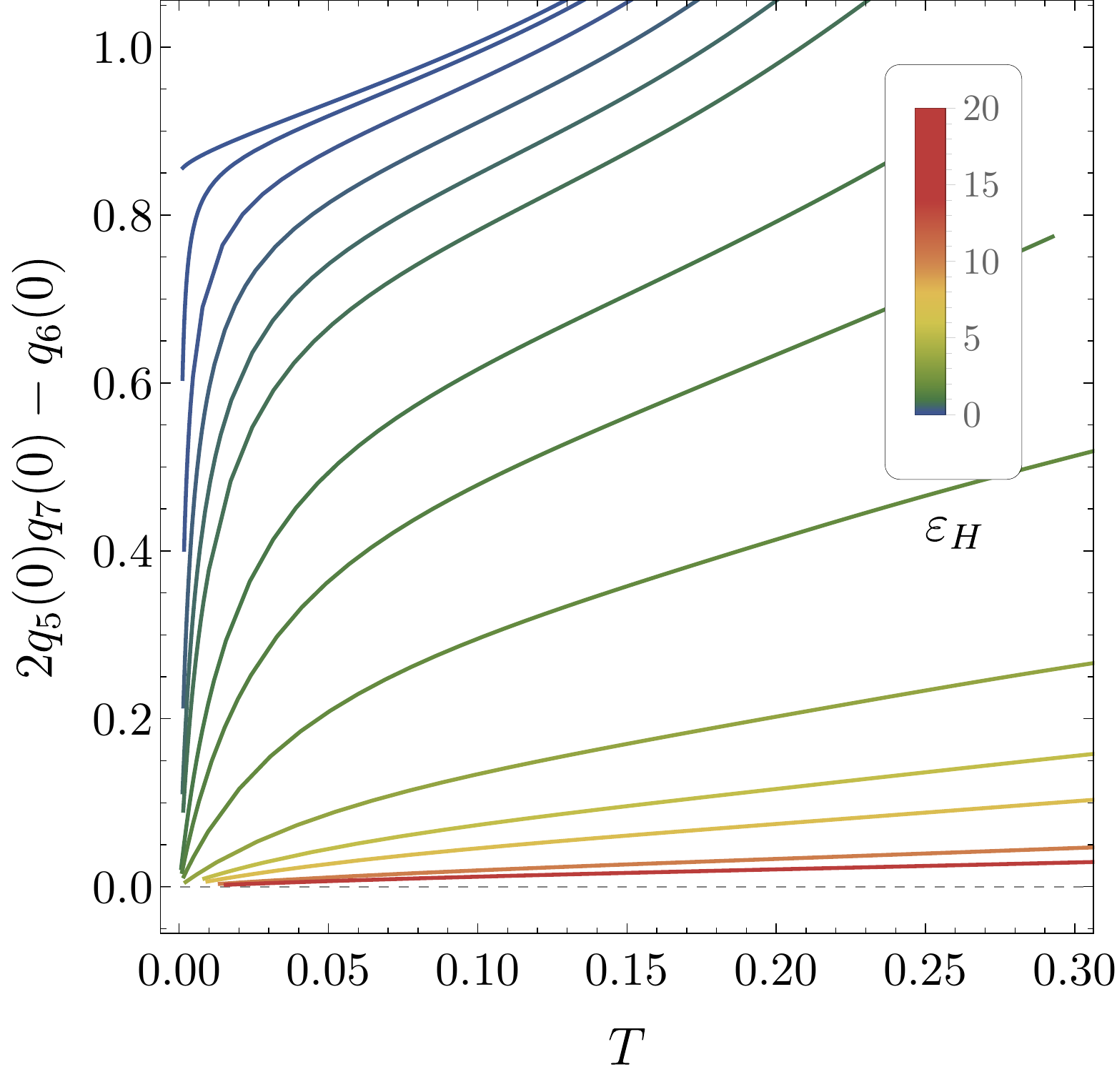}
  \end{minipage}
      \caption{The value of $q_1$ at the horizon $y=0$ against the temperature, for a black hole family with $J=0.05$ and $\varepsilon_H=1$. \textit{Middle}: The value of $q_1''$ against $T$, for several values of $\varepsilon_H$. \textit{Right}: the value of $2q_5 q_7-q_6$ at the horizon against the temperature, for several values of the horizon scalar field.}
        \label{fig:relation}
\end{figure}
 
In this subsection we are considering black hole families with fixed angular momentum $J$ and horizon scalar $\varepsilon_H$, and varying the parameter $y_+$. The temperature for these solutions is given by $T=\dfrac{y_+}{2\pi}\sqrt{\dfrac{q_1(0)}{q_2(0)}}$, where the functions $q_1,q_2$ appear in the ansatz~(\ref{eq:metric1}). As we lower the parameter $y_+$, which tends to some non-zero value, the function $q_2$ tends to a constant at the horizon $y=0$, and $q_1(0)\rightarrow 0$ (Fig.~\ref{fig:relation}, (\textit{left})), and thus $T\rightarrow 0$. This seems to be true for all fixed $\varepsilon_H$ and $J$. 

From the field equations we find the second derivative to be
\begin{equation}
q_1''(0)=\frac{q_3(0) }{y_+^2q_7(0)^2 }\left(q_2(0) q_3(0)-4 y_+^2\right) (2 q_5(0) q_7(0)-q_6(0))^2+\mathcal{O}(T^{1/2}),
\label{eq:relation}
\end{equation}
\noindent which will vanish provided that the condition $2 q_5(0) q_7(0)-q_6(0)=0$ is satisfied (Fig.~\ref{fig:relation}, (\textit{middle})), which does not appear to be the case for small $\varepsilon_H$ (Fig.~\ref{fig:relation}, (\textit{right})). However, when plotting this relation for a range of $\varepsilon_H$, a pattern emerges which suggests that this relation could hold for all $\varepsilon_H$. In fact for $\varepsilon_H=1$, already $q_1''(0)\simeq 0.0025$ at $T=1.2\times10^{-4}$. If the condition is satisfied, then from the field equations it is straightforward to check that $q_1$ derivatives at least up to fifth derivative scale at least as $\mathcal{O}(T)$, regardless of the fixed constants. As $\varepsilon_H$ is increasing, the scaling for $q_1(0)$ tends to $T^2$. If the condition~(\ref{eq:relation}) holds, then as $q_5(0)\rightarrow 1$ we would also have $q_5''(0)=4$.

We also cannot eliminate the case that relation~(\ref{eq:relation}) doesn't hold, in which case $q_1(y)\sim y^2$; for instance, the Gutowski-Reall black hole has $q_1\simeq y^2$, but also $q_2\simeq y^2$. We were unable to find a horizon expansion for the extremal hairy solutions with a finite horizon scalar field, and it seems that in order to satisfy the equations of motion some non-analytic piece must be included. In addition, the near-horizon geometry of the extremal limit of our hairy solutions does not appear to be of AdS$_2\times\mathbb{R}^2$, but could be of Lifshitz~\cite{Kachru:2008yh}, or hyperscaling violating type~\cite{Huijse:2011ef}. For instance, global hyperscaling violating black holes were recently constructed in the Einstein-Maxwell-Dilaton theory~\cite{Pedraza:2018eey}, which retain finite entropy in the extremal limit and have exotic horizon topology.

The attractor mechanism in ungauged supergravity~\cite{Ferrara:1995ih,Strominger:1996kf,Ferrara:1996dd,Ferrara:1996um,Chamseddine:1996pi,Kallosh:1996vy,Ferrara:1997tw} states that the scalar fields at the extremal horizon are usually fixed by the black hole charges, and the Bekenstein-Hawking entropy can be expressed in terms of those charges only. However, the attractor mechanism for BPS rotating black holes in five-dimensional gauged supergravity has not been completely understood, although recently some interesting progress has been made~\cite{Hosseini:2017mds,Hosseini:2018dob}. The entropy of the Gutowski-Reall black hole in terms of its charges is given by~\cite{Kim:2006he}
\begin{equation}
S^2=12\pi^2 Q^2-4\pi^2J.
\end{equation}
We find that this relation is not satisfied by the hairy black hole solutions away from the Gutowski-Reall solution. In particular, for fixed $J$, the entropy should vanish at $Q=\sqrt{3J}/3$, which does not seem to hold (even though such a maximal charge $Q_\mathrm{max}(J)$ might exist, see the discussion in the next subsection). There still may be another relation for the extremal hairy black hole entropy in terms of its charges. A bigger concern may be that the proper distance to the extremal horizon for these solutions appears to remain finite.

\subsubsection{\label{subsubsec:Isotherms}Isotherms}

\begin{figure}[t!]
\centering
  \begin{minipage}[t]{0.45\textwidth}
    \includegraphics[width=\textwidth]{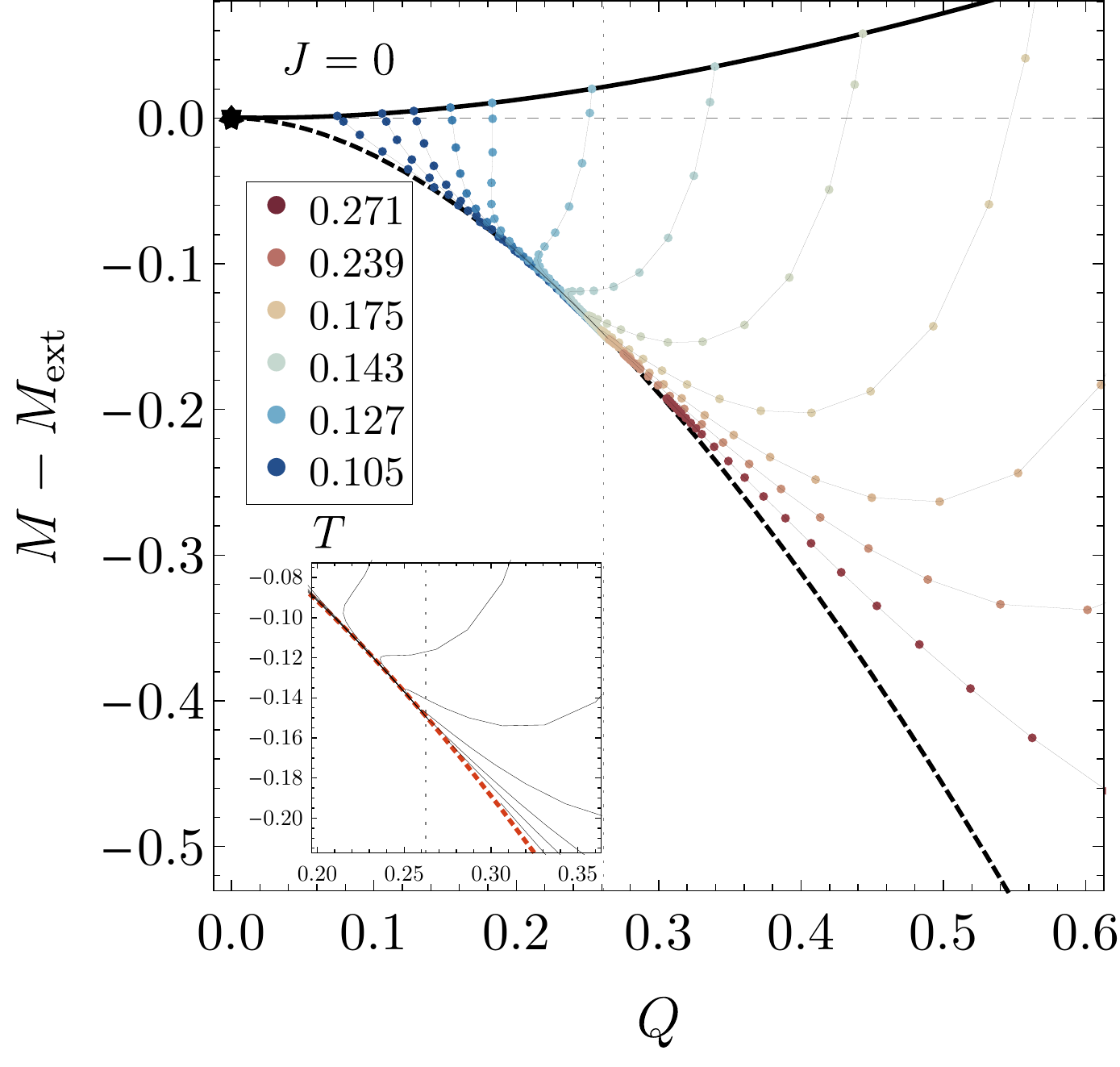}
  \end{minipage}
  \hfill
    \begin{minipage}[t]{0.45\textwidth}
    \includegraphics[width=\textwidth]{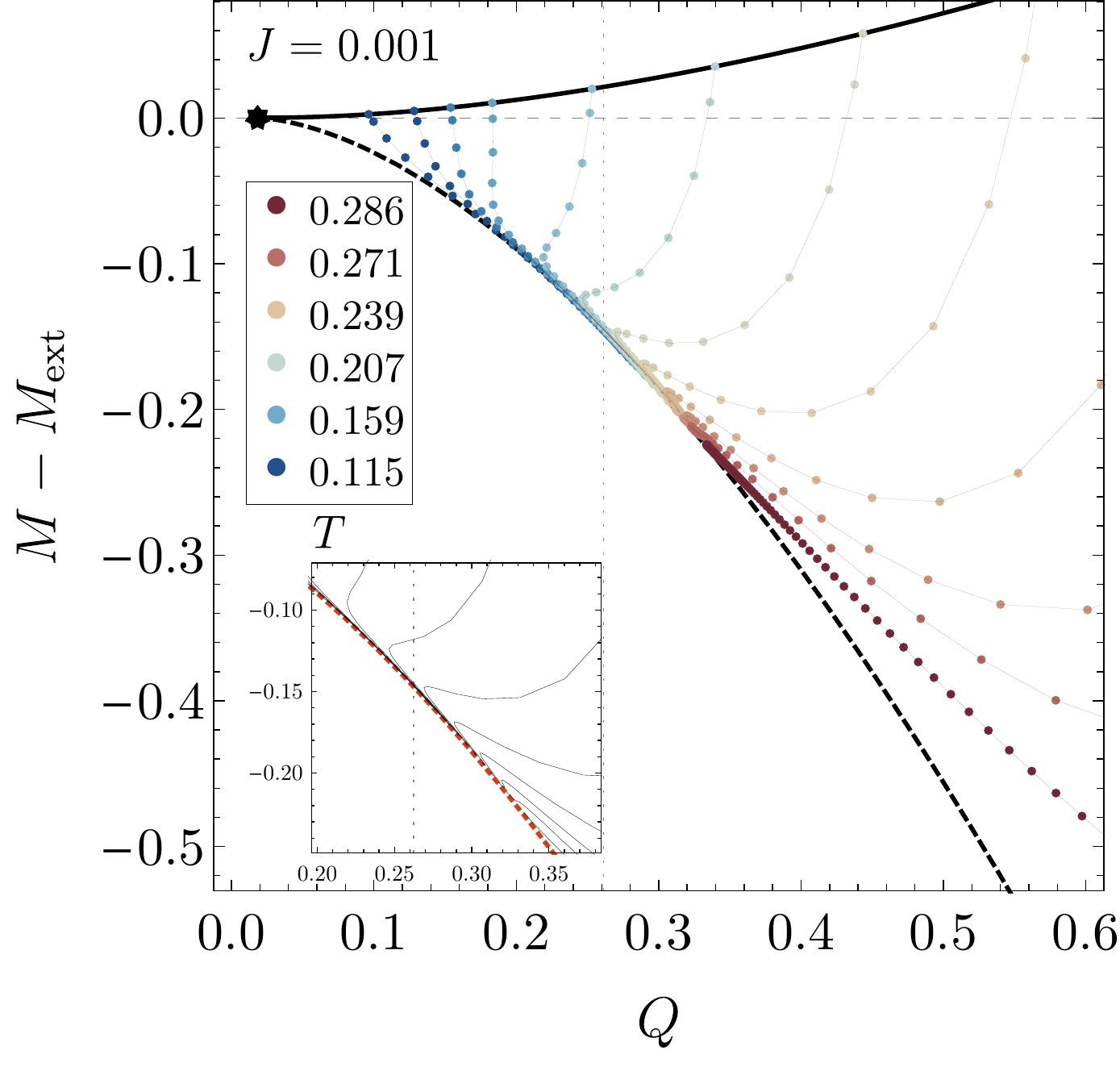}
    \\
  \end{minipage}
      \begin{minipage}[t]{0.45\textwidth}
    \includegraphics[width=\textwidth]{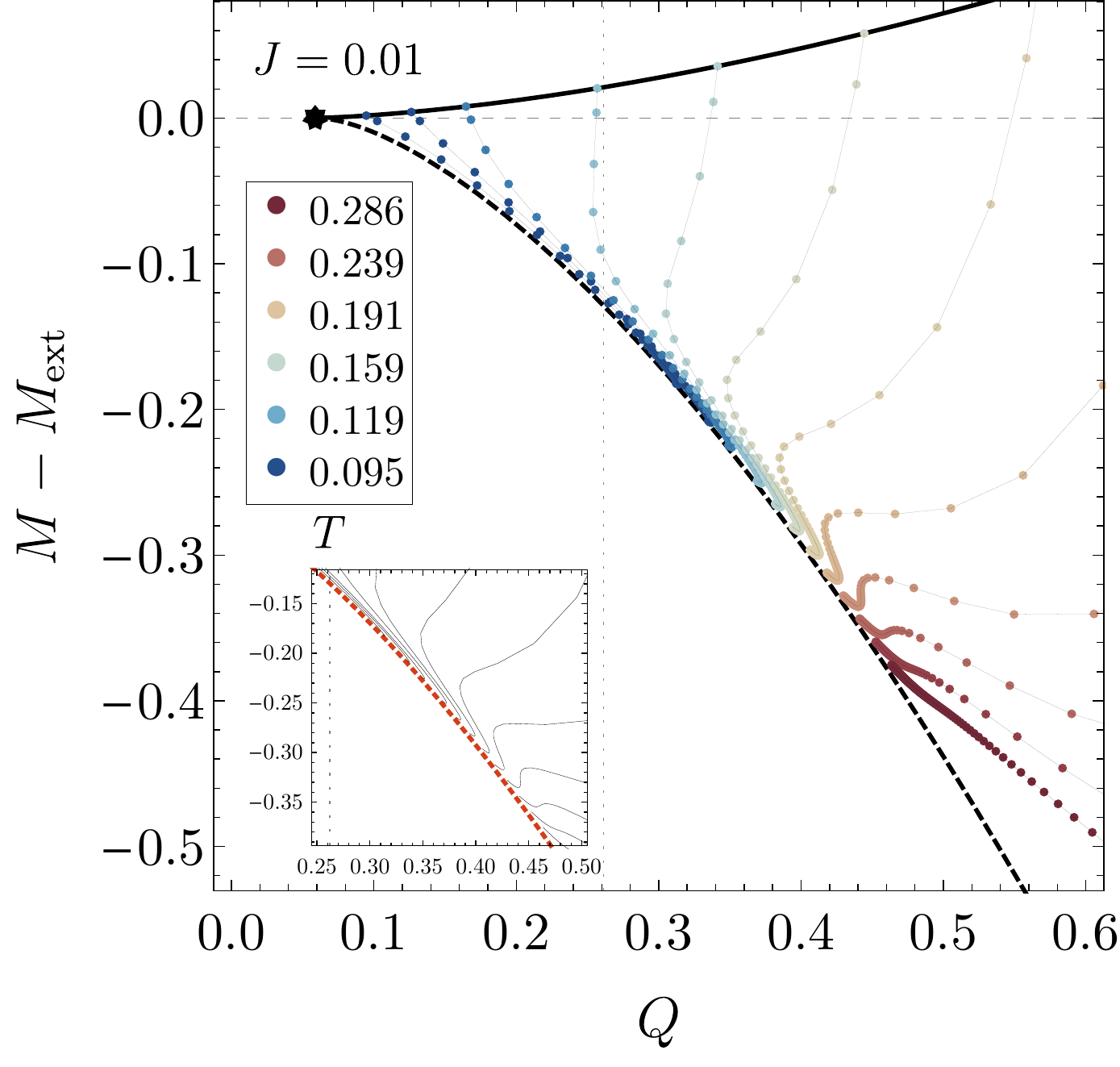}
  \end{minipage}
  \hfill
    \begin{minipage}[t]{0.45\textwidth}
    \includegraphics[width=\textwidth]{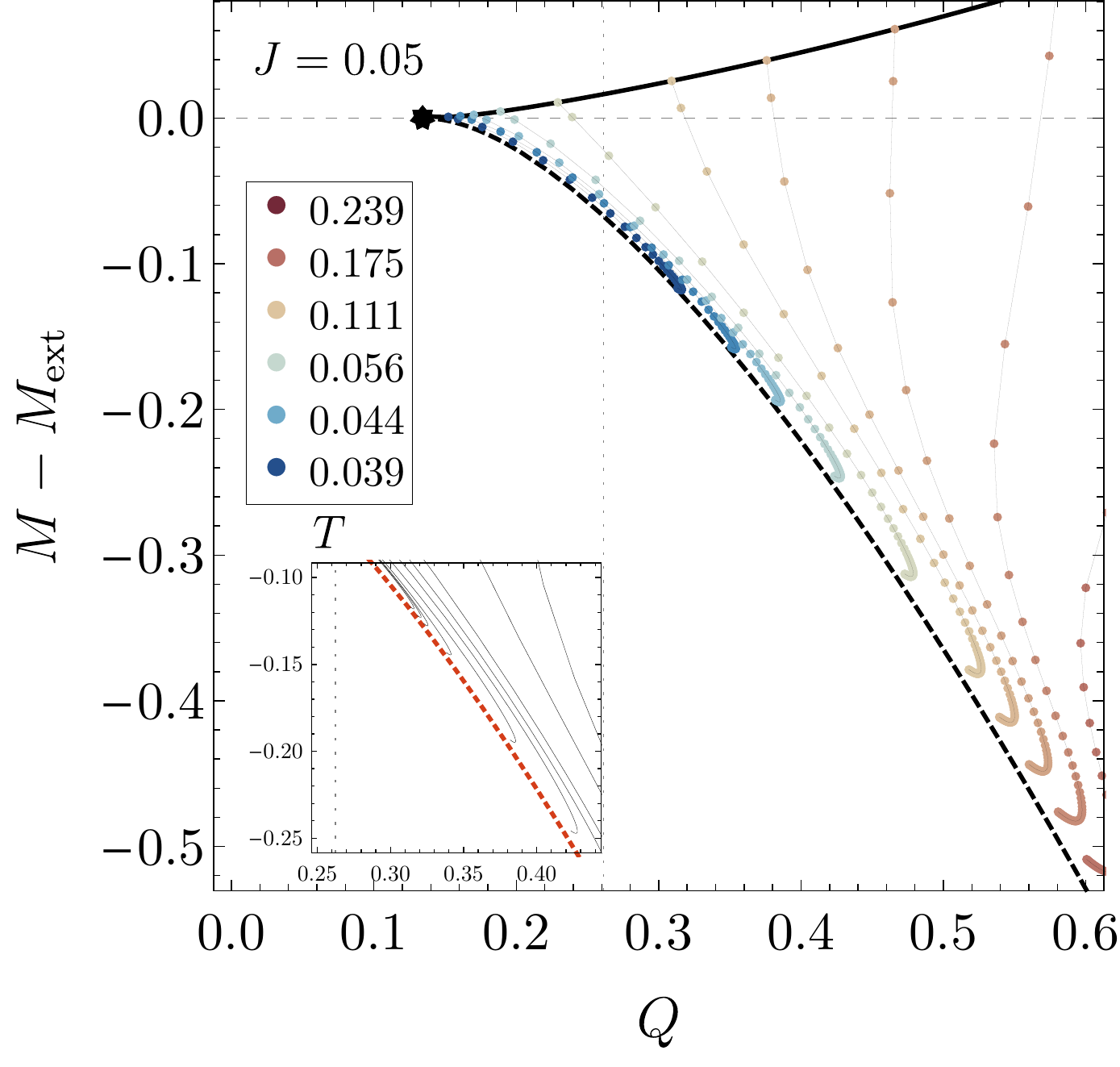}
  \end{minipage}
      \caption{Rotating hairy black holes at fixed temperature, for different fixed angular momenta $J=0,\,0.001,\,0.01,\,0.05$, while varying the horizon scalar field $\varepsilon_H$. We plot the mass difference $M-M_\mathrm{ext}$ against the charge $Q$, where $M_\mathrm{ext}$ is the mass of the corresponding extremal CLP black hole. The solid black line is the scalar instability curve, dashed black line (or red line in the insets) is the BPS bound, black star is the Gutowski-Reall black hole. The vertical dotted line shows the special soliton $Q_\mathrm{max}(J=0)=Q_c$. Insets zoom around interesting isotherm behaviour.}
        \label{fig:Isotherms}
\end{figure}

For $J=0$, zero temperature limit corresponds to the line of the smooth solitons, which terminates at the special soliton $Q_c\simeq 0.2613$ in a spiraling fashion. This is where the $T=0$ and $T=\infty$ limits of the hairy black holes intersect, and all other isotherms are drawn to the vicinity of $Q_c$ (Fig.~\ref{fig:Isotherms}). Therefore if such a point exists for $J>0$, which we will label $Q_\mathrm{max}(J)$, it should be indicated by the behaviour of the isotherms. We conjecture that such a point is the $\varepsilon_H\rightarrow\infty$ and $S\rightarrow 0$ limit of the extremal rotating hairy solutions, and the $T\rightarrow\infty$ limit corresponding to some singular configurations would branch off the special $Q_\mathrm{max}(J)$ solution for large charges. This conjecture is supported by the observed continuity from the $J=0$ case. 

In Fig.~\ref{fig:Isotherms} we present hairy black hole isotherms for a variety of angular momenta\footnote{These results were obtained in the DeTurck gauge, therefore we could only access a limited range of temperature and horizon scalar field values. In these figures, $\varepsilon_H\simeq 40$.}. The system at fixed small angular momentum $J=0.001$ is a small deformation of the non-rotating case. The large $\varepsilon_H$ isotherms extend past the $Q_c$ and curiously exhibit a ``swallowtail'' phase transitions in the canonical ensemble (to be discussed in Section~\ref{sec:therm}). Just near the BPS bound the isotherms turn towards the $Q_c=Q_\mathrm{max}(0)$ solution, and advance up along the BPS bound as $\varepsilon_H\rightarrow\infty$. This gives us an upper bound on $Q_\mathrm{max}(J)$, which for small $J$ is close to the $Q_c$. Unfortunately, the most interesting isotherms have very low temperatures, which we cannot access with the DeTurck method.

Such behaviour is even more apparent for larger values of $J$. For angular momentum $J=0.05$, the isotherm turning point is further from the BPS bound, and the isotherm shape is becoming more pronounced. The canonical phase transition is occurring at larger charges and temperatures, and over a broader range. It could also be the case that, if $J$ is large enough that the charge of the corresponding Gutowski-Reall black hole $Q_\mathrm{GR}(J)$ is larger than $Q_c$, $Q_\mathrm{max}(J)$ would coincide with $Q_\mathrm{GR}(J)$.

We would also like to entertain the idea that $Q_\mathrm{max}(J)=Q_\mathrm{GR}(J)$ for all $J$. This would require the isotherms to have a rather dramatic behaviour near the BPS bound, especially for low angular momenta where we see continuity from the non-rotating case. The isotherms would ``loop'' the supersymmetric black hole, and the $T=0$ limit does not exist, or in fact consists of a single point with $\varepsilon_H=0$. The whole of the BPS plane would be populated by the $T=\infty$ singular solutions, which could in principle be identified by solving the BPS equations directly. We stress, however, that this would only be possible if some large extra scale is involved due to the extremal horizon, and this scenario would be out of reach for our numerical scheme.

\subsection{\label{subsec:constQ}Constant $Q$ planes}
 \begin{figure}[t]
\centering
  \begin{minipage}[t]{.99\textwidth}
    \includegraphics[width=\textwidth]{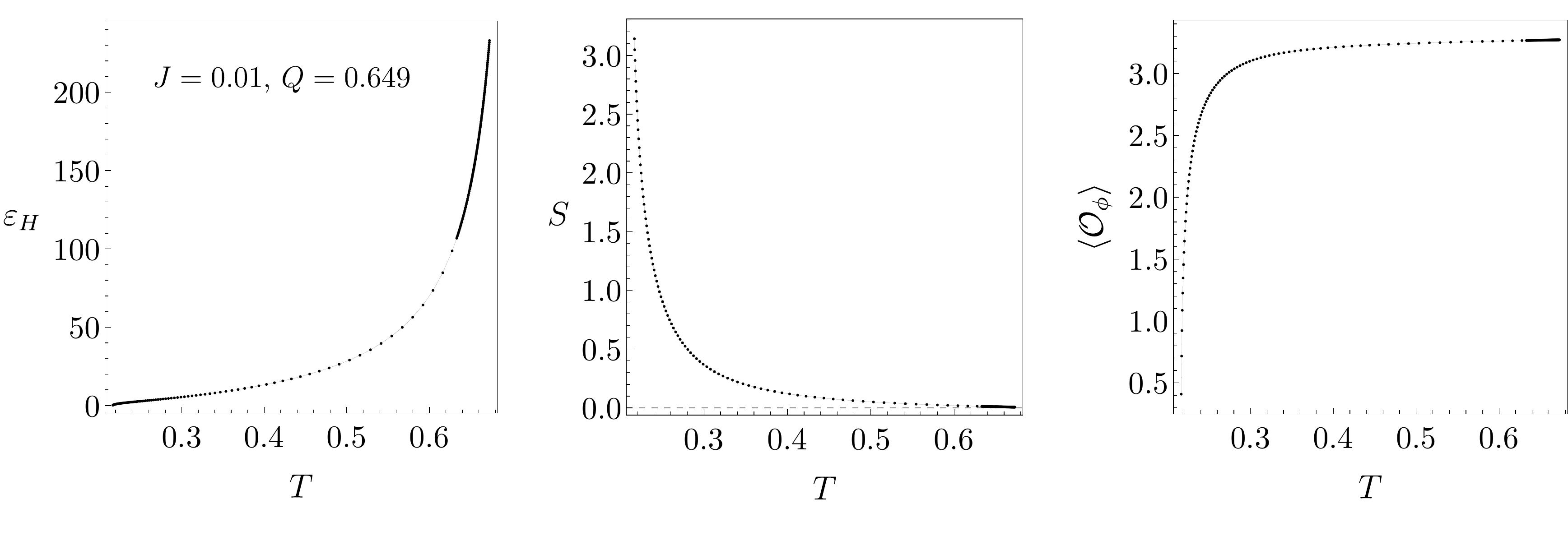}
  \end{minipage}
  \vfill
    \begin{minipage}[t]{.99\textwidth}
    \includegraphics[width=\textwidth]{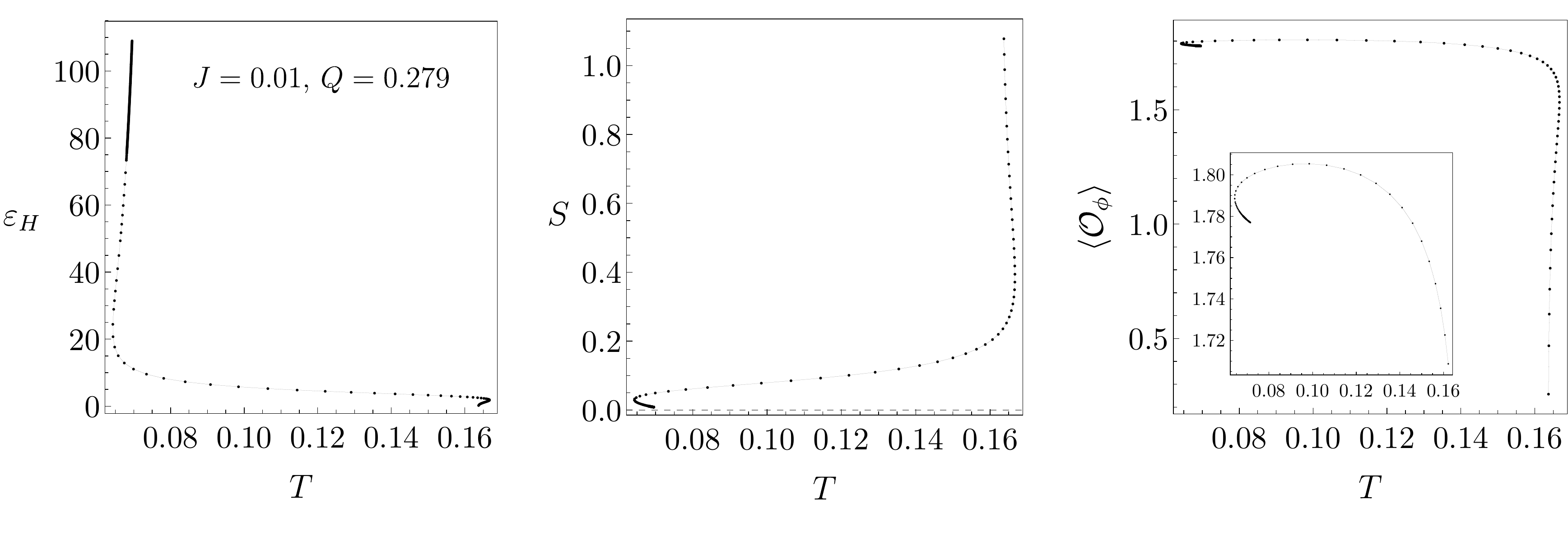}
  \end{minipage}
    \vfill
    \begin{minipage}[t]{.99\textwidth}
    \includegraphics[width=\textwidth]{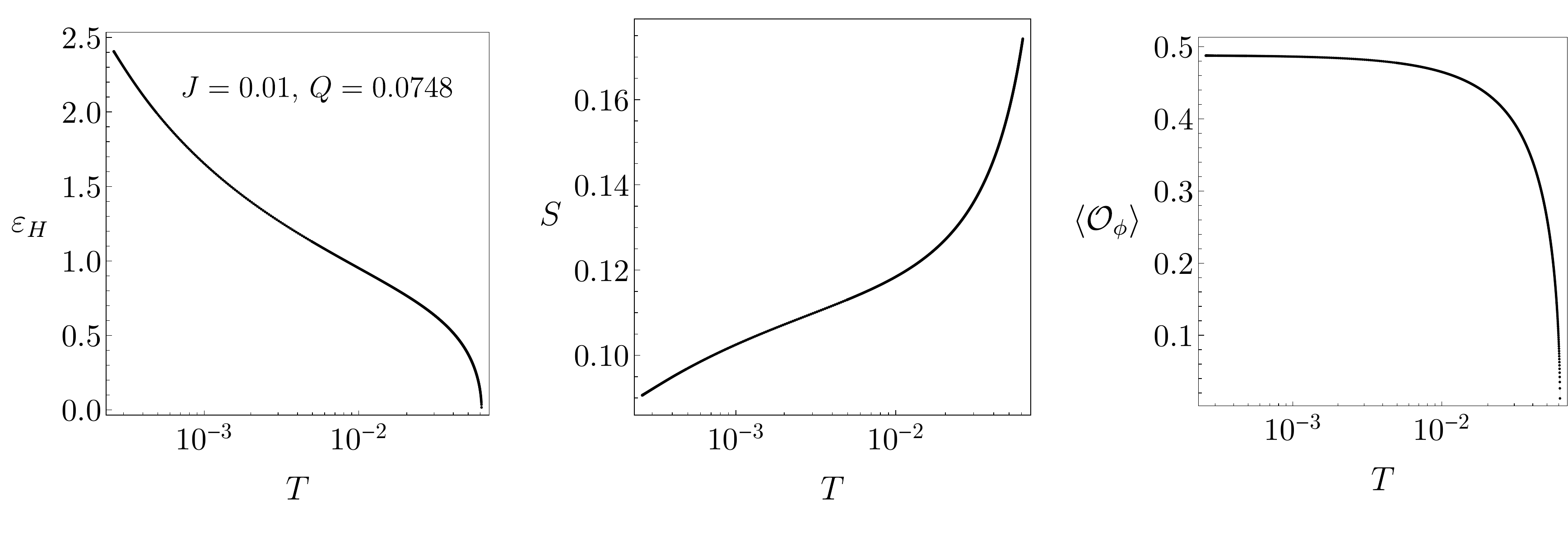}
  \end{minipage}
      \caption{Constant angular momentum $J$ and charge $Q$ solutions, with the parameter $y_+$ being varied. We present the horizon scalar field $\varepsilon_H$ (\textit{left}), the entropy (\textit{middle}), and the expectation value of the dual operator $\langle \mathcal{O}_\phi\rangle$ (\textit{right}), for three different charges $Q=0.0748$ (\textit{bottom row}), $Q=0.279$ (\textit{middle row}), $Q=0.649$ (\textit{top row}).}
        \label{fig:constantQ}
\end{figure}

We cannot rule out the possibility that the constant horizon scalar charge limit is not the most suitable choice to reveal the correct approach to the BPS bound. In the non-rotating case we found that this limit worked incredibly well, and at $T=0$ provided a mechanism to pack an infinite amount of these hairy black hole families onto a small strip of the microcanonical phase space~\cite{Markeviciute:2016ivy}. We can impose~(\ref{eq:charge}) as a boundary condition, yielding a system of three free parameters $\{y_+,Q,q_5(1)\}$, i.e. effectively at the same time we can control $\{T,Q,J\}$. Starting with a hairy seed, we can fix the total angular momentum $J$ and charge $Q$, and lower the parameter $y_+$. If the charge is sufficiently low, the temperature is monotonic in $y_+$ (Fig.~\ref{fig:Isotherms}), however, for larger charge a more complicated picture emerges. At fixed $J$, there is a charge range at which there are at least three hairy black holes with the same temperature\footnote{At least for relatively small values of $J$.} (Fig.~\ref{fig:Isotherms}, Fig.~\ref{fig:Canonical}). 

We present constant $J=0.01$ results in Fig.~\ref{fig:constantQ}, for three different charges $Q$. The supersymmetric black hole has $Q_\mathrm{GR}=0.0588$. The large charge ($Q=0.659$) behaviour is similar to the large charge behaviour of the non-rotating case. As we lower $y_+$, we find that the temperature increases, and the hairy black holes approach the BPS bound. The entropy tends to zero, horizon scalar field diverges and the $\langle\mathcal{O}_\phi\rangle$ is tending to some constant value. We also managed to reach as close as $M-M_\mathrm{BPS}<0.001$ to the BPS bound, with $\mu$ and $\Omega_H$ tending to one. This suggests that we are approaching the supersymmetric bound, where there is a singular solution, which can be interpreted as the $T=\infty$ limit of the hairy black holes. We also observe retrograde condensation, that is the hairy black hole exists for $T>T_c$, which carries over to the planar limit (see section \ref{sec:planar}).

Of course we are most interested in lower charges. For intermediate charges $Q\sim Q_c\simeq 0.26$ we know that there exist at least three black holes with the same temperature. Indeed this is what we find for $J=0.01$, $Q=0.279$. When we decrease $y_+$, the temperature increases initially, then decreases, and finally starts to increase again, sending $\varepsilon_H\rightarrow\infty$ and $S\rightarrow 0$. We also see interesting trend in $\langle\mathcal{O}_\phi\rangle$, perhaps suggesting a loop, as we expect that either $T\rightarrow\infty$, or $T\rightarrow 0$. For these black holes, the closest approach is $M-M_\mathrm{BPS}<0.0006$.

Finally, at low charges, if our conjectures hold, we expect to see $\varepsilon_H$ increasing and tending to a constant value, and $S$ decreasing to a constant non-zero value, while $\langle\mathcal{O}_\phi\rangle$ settles to a non-zero constant. Indeed this is what we observe in Fig.~\ref{fig:constantQ} (\textit{bottom row}). As the low temperature isotherms are very dense near the BPS bound, we do expect to see a sharp growth in $\varepsilon_H$. While we do not observe the values blowing up, we were able to fit both divergent and convergent logarithmic functions to the low temperature data, of the form $a+b\,T^{\alpha} \log{T}$. Numerical convergence at low temperatures gets increasingly worse~\cite{Markeviciute:2018yal}, and it is difficult to extract good quality data for the temperature range where the change in $\langle\mathcal{O}_\phi\rangle$ is very small ($T<10^{-3}$). This is where we would expect the hairy solutions to enter the scaling regime. In addition, at small charges we do not see any sign of $T$ increasing with the parameter $y_+$.

\subsection{\label{subsec:constO}Constant $\Omega_H$ planes}
 
 \begin{figure}[t]
\centering
  \begin{minipage}[t]{0.31\textwidth}
    \includegraphics[width=\textwidth]{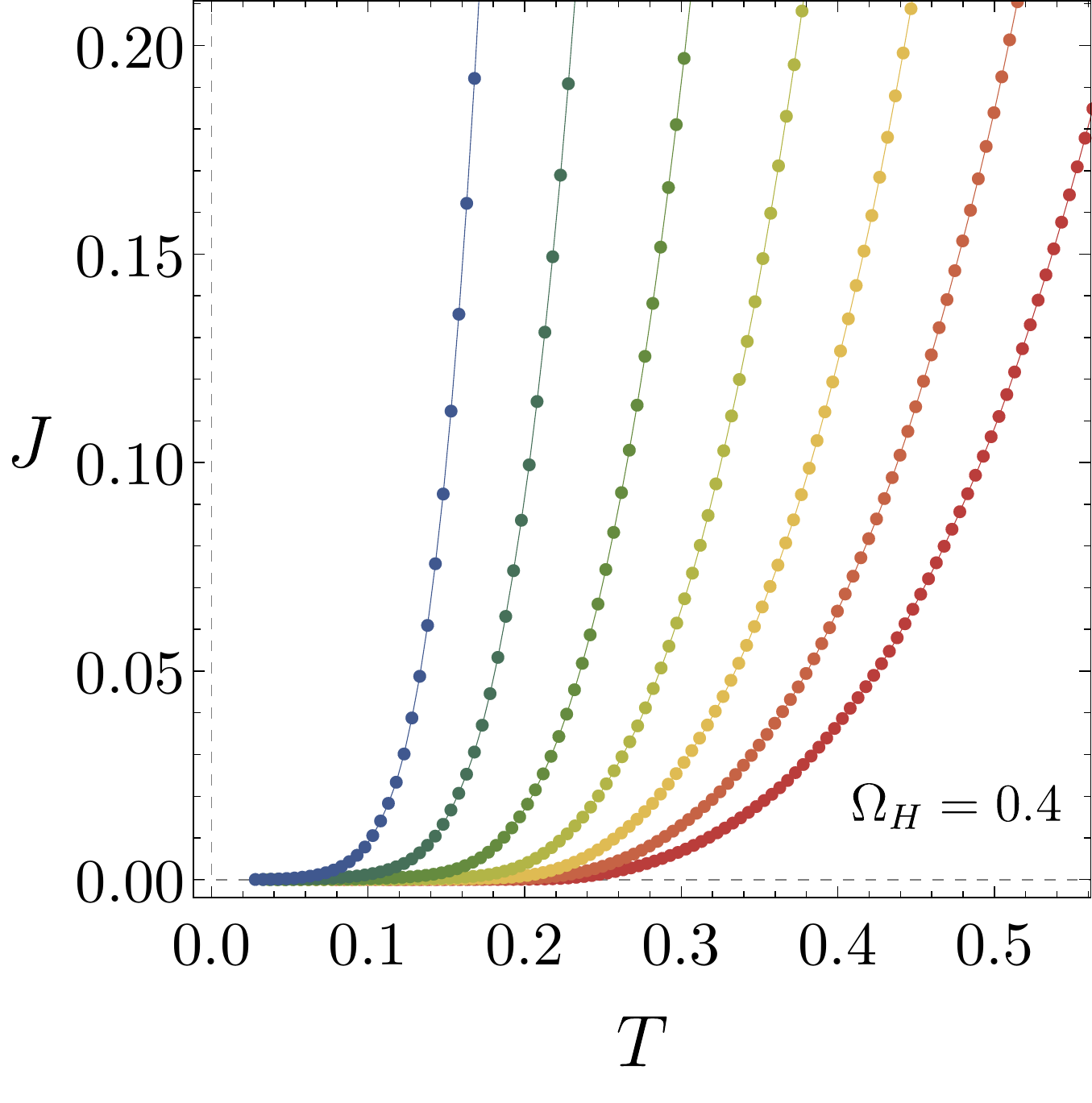}
  \end{minipage}
  \hfill
    \begin{minipage}[t]{0.3\textwidth}
    \includegraphics[width=\textwidth]{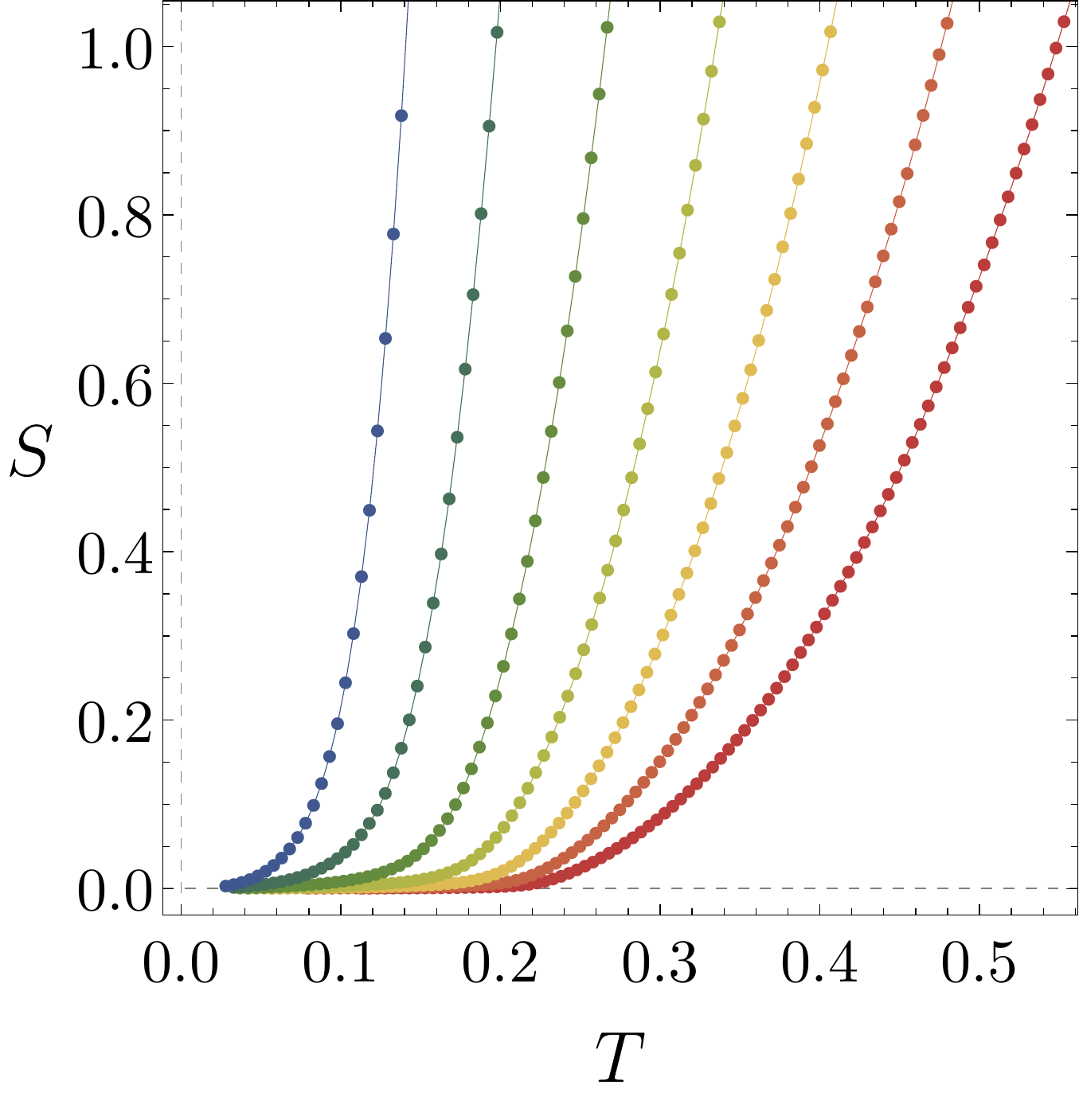}
  \end{minipage}
  \hfill
    \begin{minipage}[t]{0.31\textwidth}
    \includegraphics[width=\textwidth]{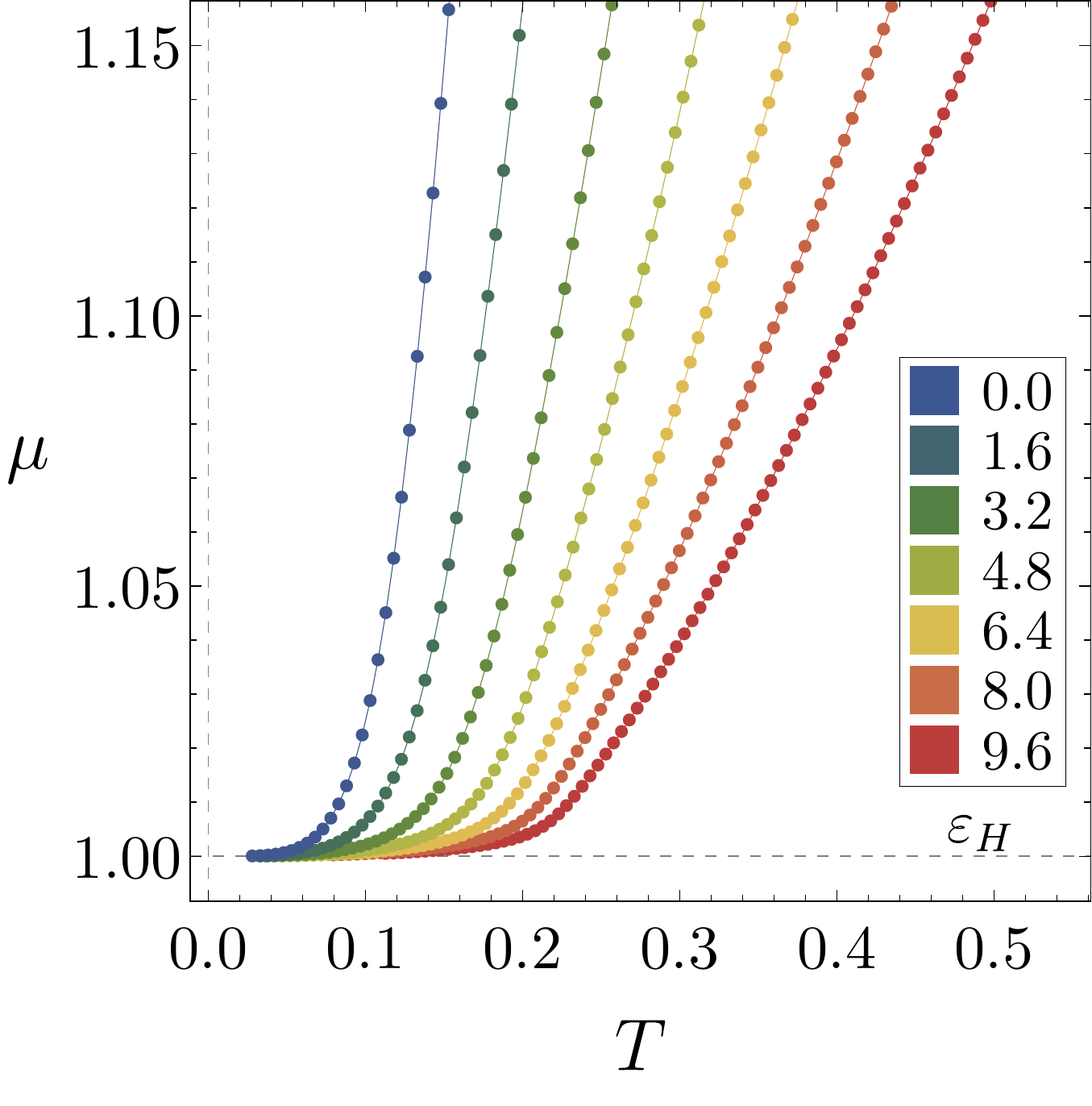}
  \end{minipage}
  \hfill
  \begin{minipage}[t]{0.31\textwidth}
    \includegraphics[width=\textwidth]{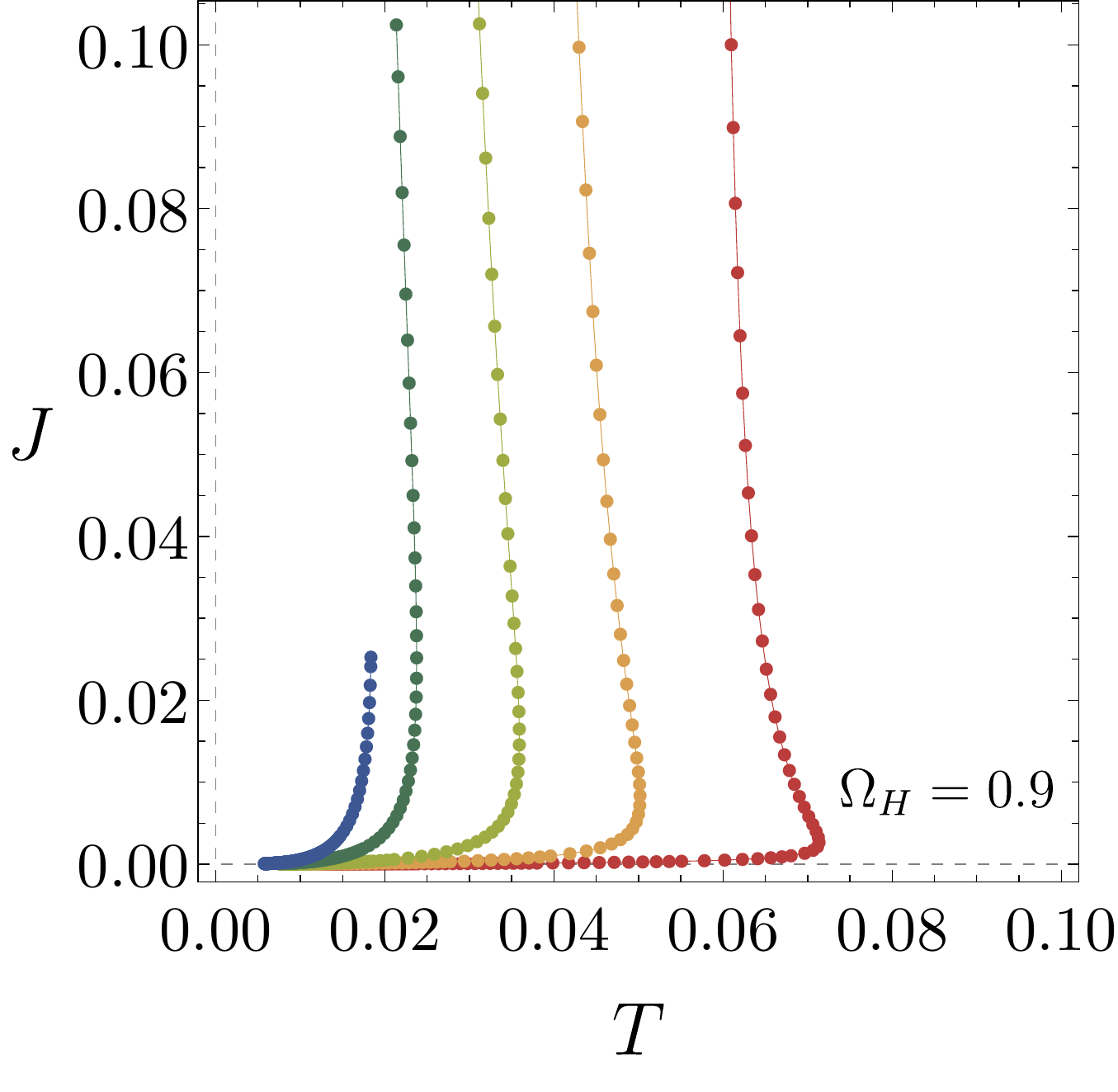}
  \end{minipage}
  \hfill
    \begin{minipage}[t]{0.3\textwidth}
    \includegraphics[width=\textwidth]{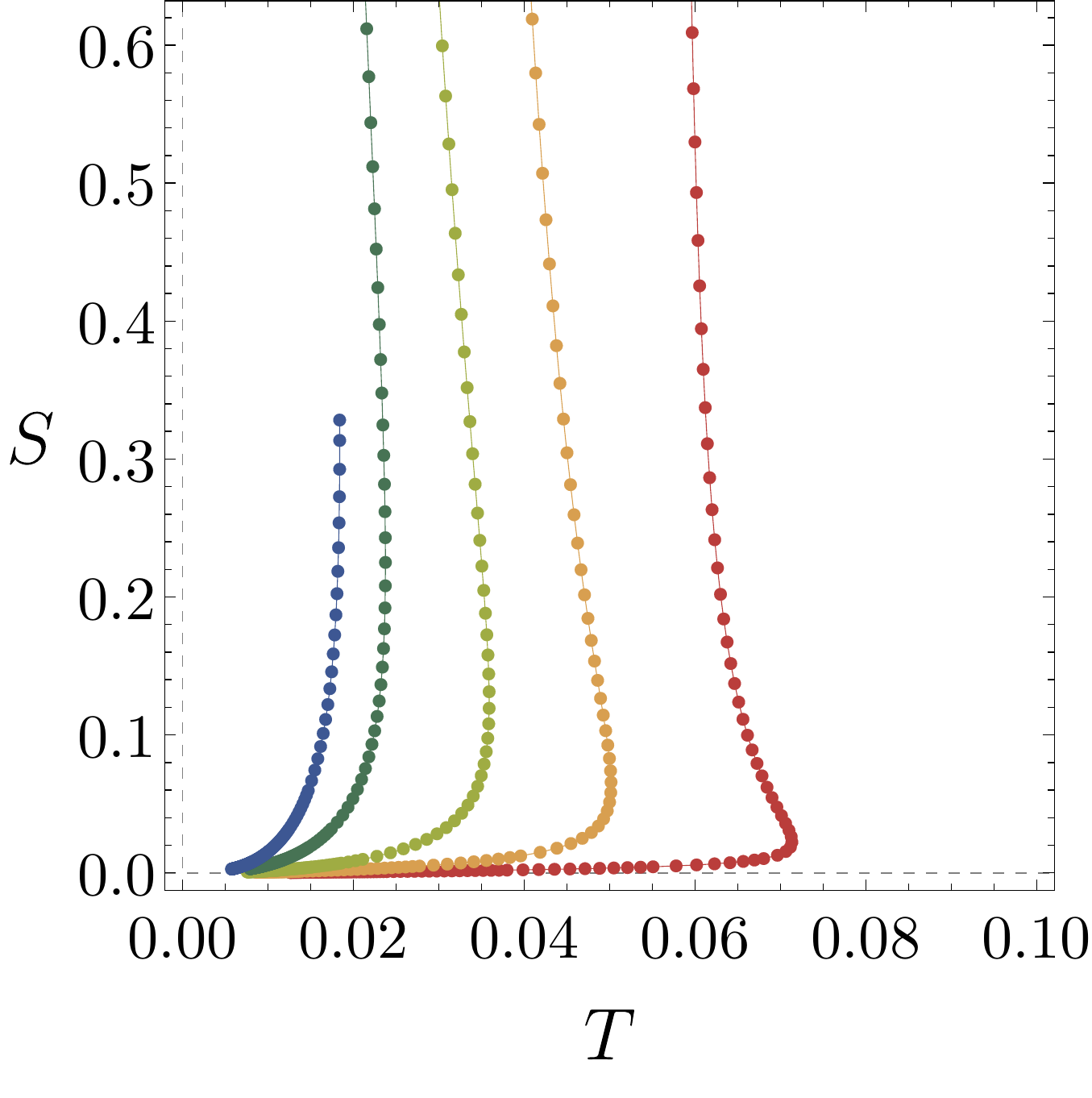}
  \end{minipage}
  \hfill
    \begin{minipage}[t]{0.31\textwidth}
    \includegraphics[width=\textwidth]{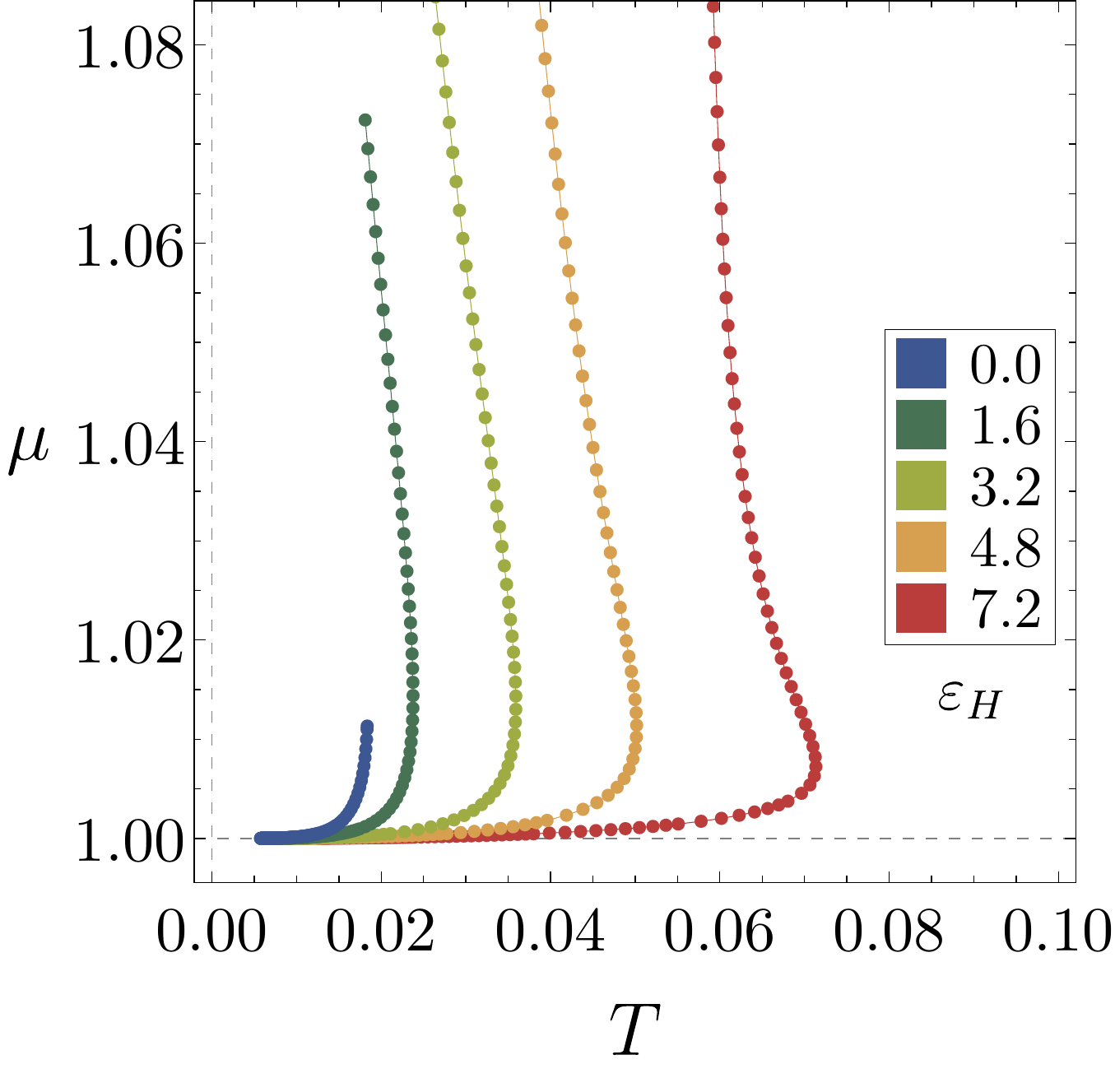}
  \end{minipage}
      \caption{Thermodynamic quantities for the hairy black hole families with a fixed horizon velocity $\Omega_H=0.4$ (\textit{top row}) and $\Omega_H=0.9$ (\textit{bottom row}), and a constant horizon scalar $\varepsilon_H$. As the temperature $T\rightarrow 0$, angular momentum $J\rightarrow 0$ (\textit{left}), entropy $S\rightarrow 0$ (\textit{middle}), and chemical potential $\mu\rightarrow 1$ (\textit{right}). For large horizon velocity, there exists a maximum temperature for the hairy black holes.}
        \label{fig:constOHTD}
\end{figure}

 \begin{figure}[t]
      \centering
      \begin{minipage}[t]{0.45\textwidth}
  	    \includegraphics[width=\textwidth]{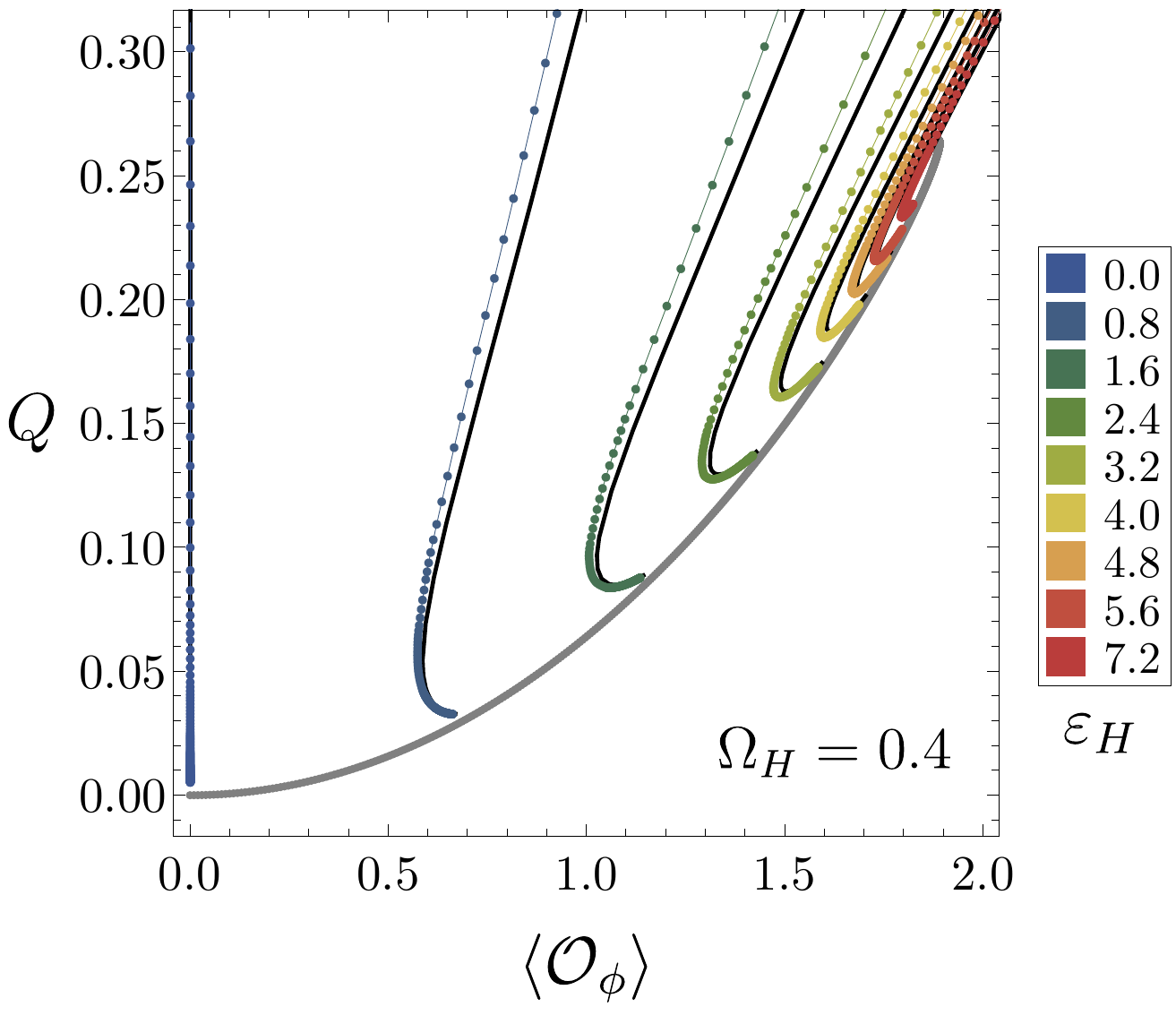}
      \end{minipage}
      \hfill
      \begin{minipage}[t]{0.45\textwidth}
		\includegraphics[width=\textwidth]{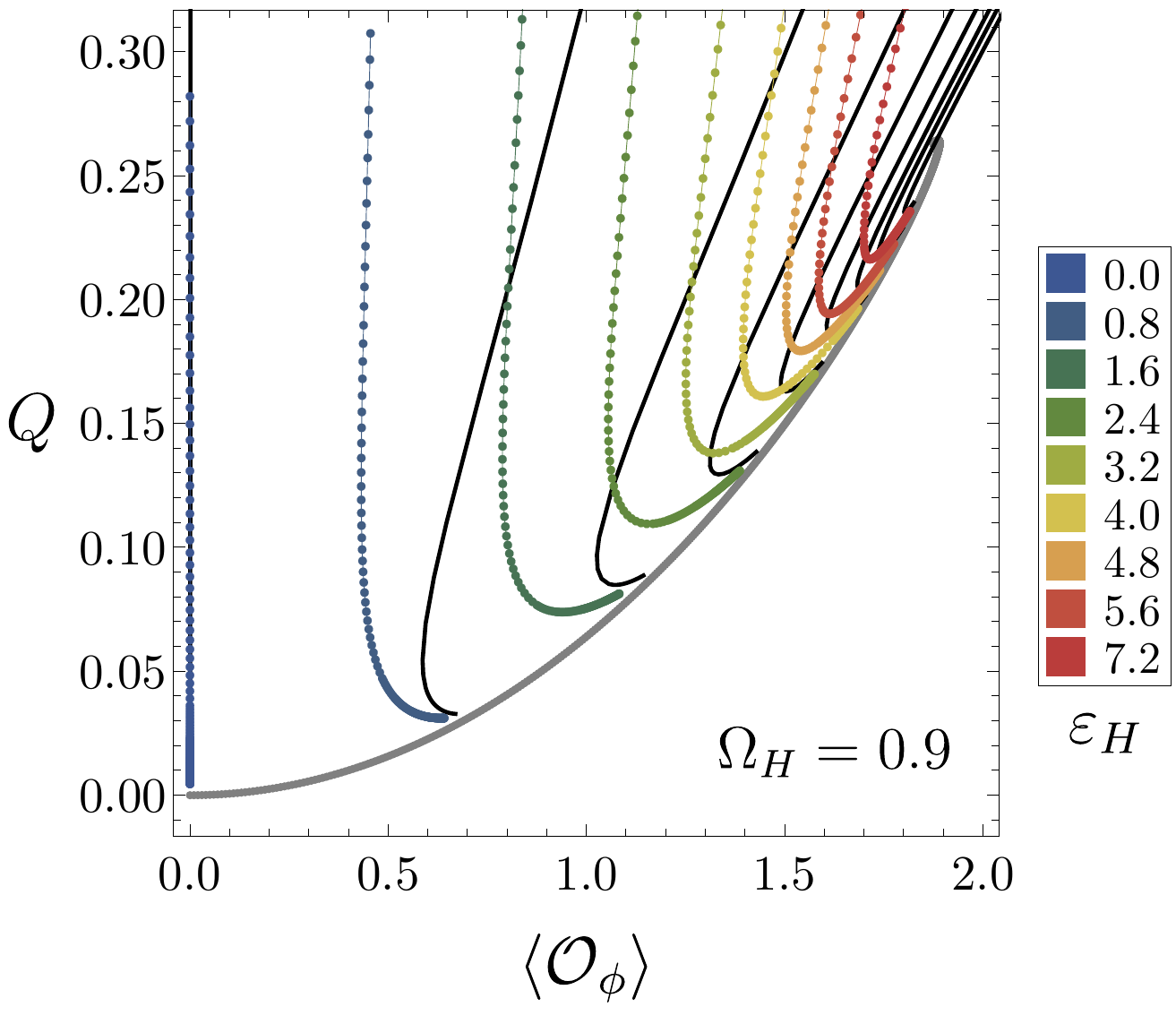}
      \end{minipage}
      \caption{Black hole charge $Q$ against the vacuum expectation value $\langle\mathcal{O}_\phi \rangle$ of the operator dual to the scalar field $\phi$, at fixed horizon scalar charge $\varepsilon_H$, and fixed horizon angular velocity $\Omega_H=0.4$ (\textit{left}), and $\Omega_H=0.9$ (\textit{right}), in the non-rotating frame at infinity. In black are the corresponding values for the non-rotating solutions $\Omega_H=0$, $J=0$. The smooth BPS soliton family is shown in gray. }
       \label{fig:opomega}
 \end{figure}

 \begin{figure}[t]
	 \centering
      \begin{minipage}[t]{0.5\textwidth}
		\includegraphics[width=\textwidth]{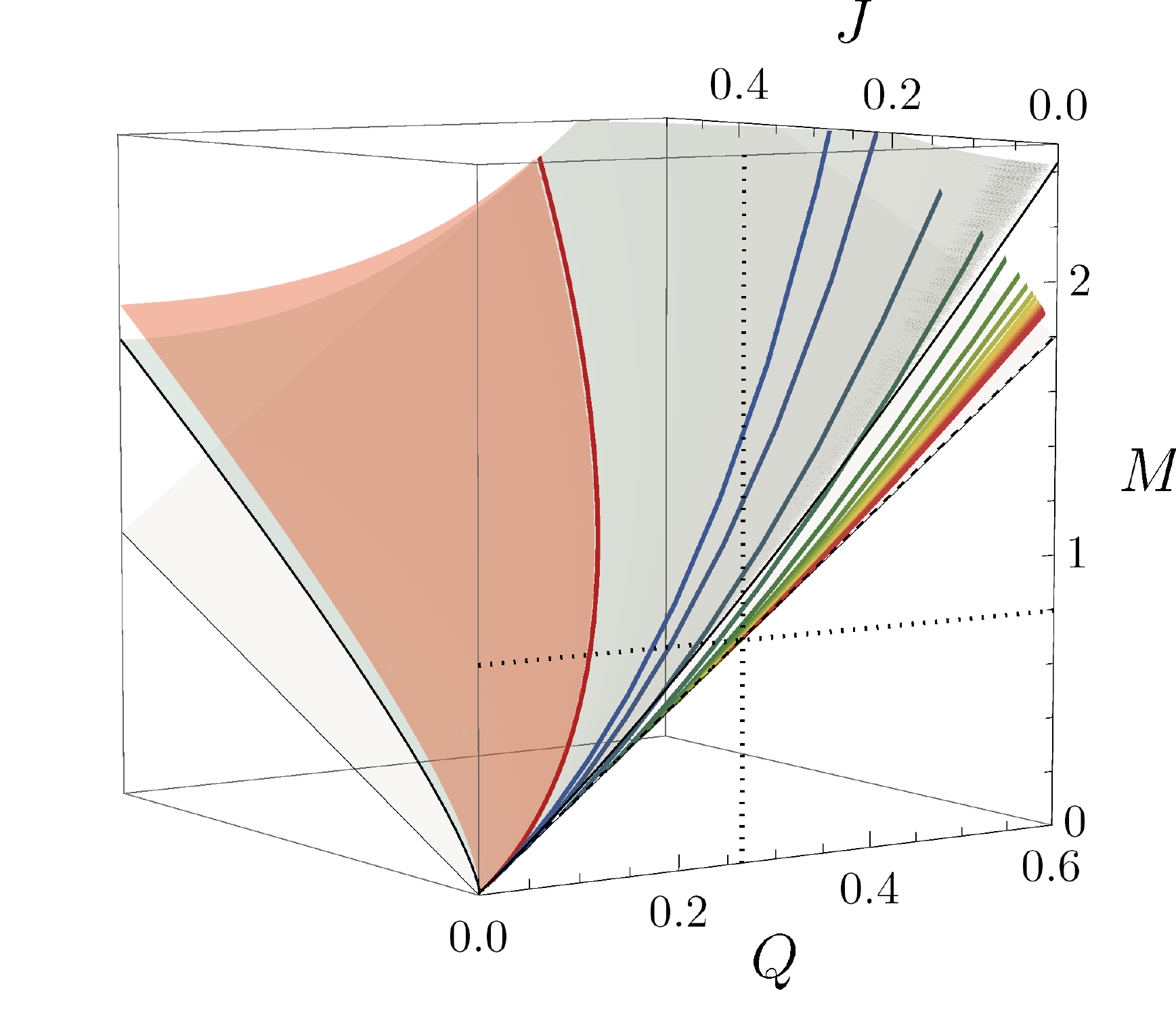}
      \end{minipage}
      \caption{A 3D solution space of the hairy black holes. The CLP black holes exist above the dark gray plane, and intersect the BPS limit $M=3Q+2J$ (the white plane below) at the Gutowski-Reall family (red bold line). The orange plane is the CLP solutions with $\Omega_H=1$, below which the solutions have $\Omega_H>1$. The solid bold lines in rainbow are the hairy solutions with fixed horizon velocity $\Omega_H=0.4$, and fixed horizon scalar value $\varepsilon_H$. The first line in blue is the merger line for the constant $\Omega_H=0.4$, and exists just above the extremal plane. The dotted black gridlines on the front right face show the special soliton with $Q_c\simeq 0.26$.}
        \label{fig:phase3dhair}
\end{figure}

In Subsection~\ref{subsec:constJ} we fix the value of $q_5(y)$ at infinity, $y=1$. Alternatively, we can fix the value at the horizon $q_5(0)=\Omega_H$, where we will find that $0\leq\Omega_H<1$. We again track fixed horizon scalar $\varepsilon_H$ curves, and lower the temperature. For the non-rotating solutions we found that plotting the black hole charge $Q$ against the vev of the boundary operator dual to the scalar field $\langle \mathcal{O}_{\phi}\rangle= y_+^2 q_8(1)$ allowed us to better understand the approach of the hairy solutions to the BPS bound~(Fig.~\ref{fig:opomega} black curves)~\cite{Markeviciute:2016ivy}. They curve towards the smooth soliton curve (in grey), where in the $T\rightarrow 0$ limit they smoothly reduce to the regular soliton solution. The soliton solution has a maximum charge, however there is a regular limiting solution for every such black hole family, allowed by the spiraling behaviour of the smooth soliton branch.

If we turn on the horizon angular velocity (in the non-rotating frame at the infinity), we find a surprisingly similar picture. In~Fig.~\ref{fig:opomega} we present numerical results for $\Omega_H=0.4$ and $\Omega_H=0.9$, and also plot $\Omega_H=0$ for reference. We find that the constant horizon angular velocity and horizon scalar field hairy black hole families in the zero temperature limit approach the same smooth soliton, \textit{i.e.} $\mu\rightarrow 1$, $S\rightarrow 0$ and $J\rightarrow 0$, and close to it, appear to have analogous spiral-like behaviour. As $S\rightarrow 0$, the curvature becomes very large, and we see the Kretschmann invariant $K=R_{\mu\nu\rho\sigma}R^{\mu\nu\rho\sigma}$ blowing up as $T\rightarrow 0$. However, if we compare it with the corresponding CLP black hole in the grand-canonical ensemble, we obtain a finite limit (see Fig.~\ref{fig:curvature}, \textit{right}). This seems to hold for any $\varepsilon_H$ and $\Omega_H$, and again is similar to the non-rotating case \cite{Markeviciute:2016ivy}. In fact, by considering gauge invariant quantities, we find that the limiting solutions are the same non-rotating $J=0$ smooth solitons. These solution curves are shown in the microcanonical diagram in Fig~\ref{fig:phase3dhair}.

In Fig.~\ref{fig:constOHTD} we present some of the thermodynamic quantities (angular momentum $J$, entropy $S$ and chemical potential $\mu$) for such hairy black hole curves. As we increase $\Omega_H$, the hairy black holes exist only up to some maximum temperature $T_\mathrm{max}(\varepsilon_H,\Omega_H)$. This reflects the fact that the limit $\Omega_H\rightarrow 1$ is the extremal limit for hairy black holes. It would be very interesting to know this structure changes as $\Omega_H\rightarrow 1$, however it is difficult to resolve large $\Omega_H$, low $T$ solutions.
 
\section{\label{sec:therm}Thermodynamics} 
\begin{figure}[t]
\centering
  \begin{minipage}[t]{.45\textwidth}
    \includegraphics[width=\textwidth]{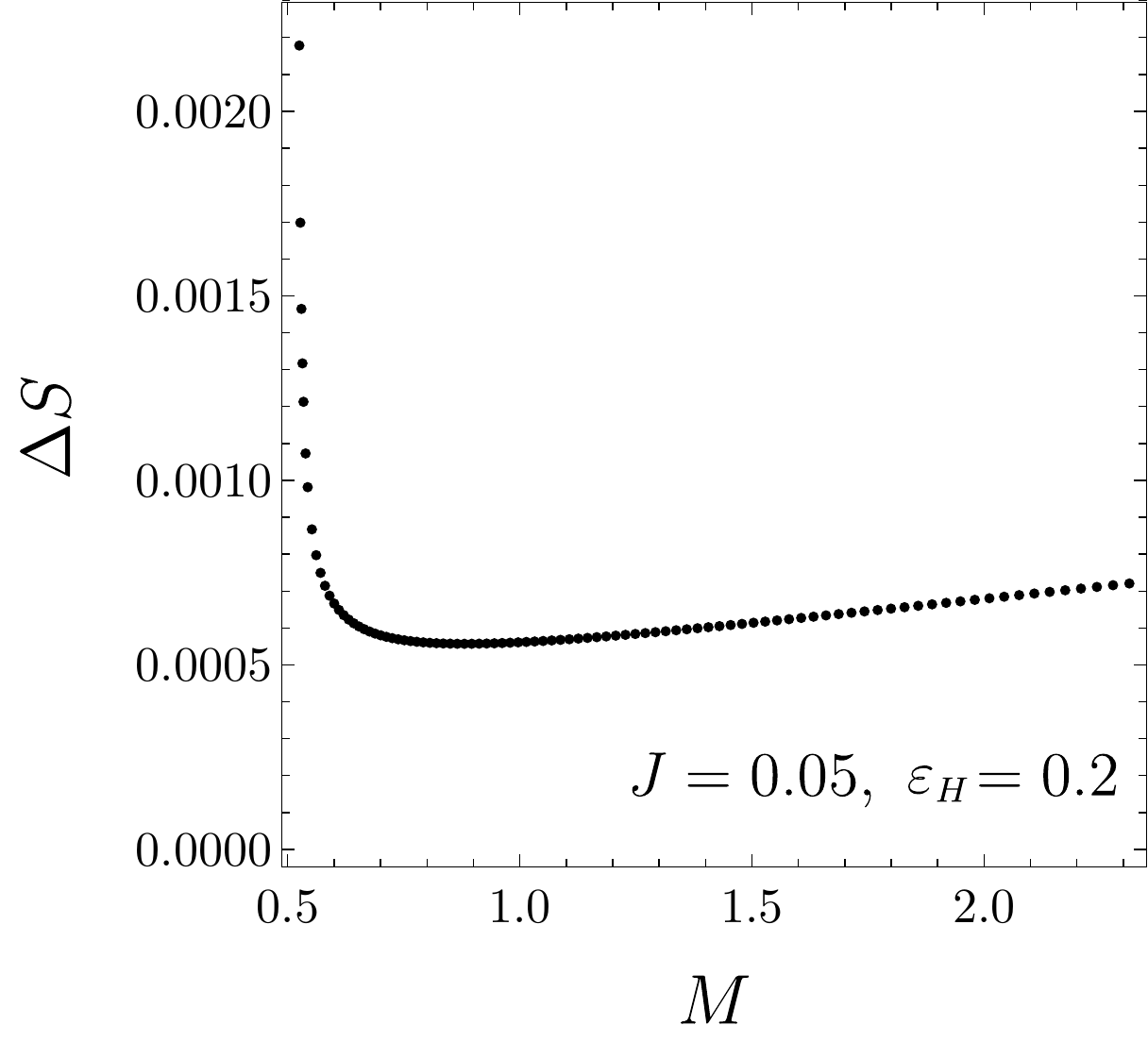}
  \end{minipage}
  \hfill
    \begin{minipage}[t]{.45\textwidth}
    \includegraphics[width=\textwidth]{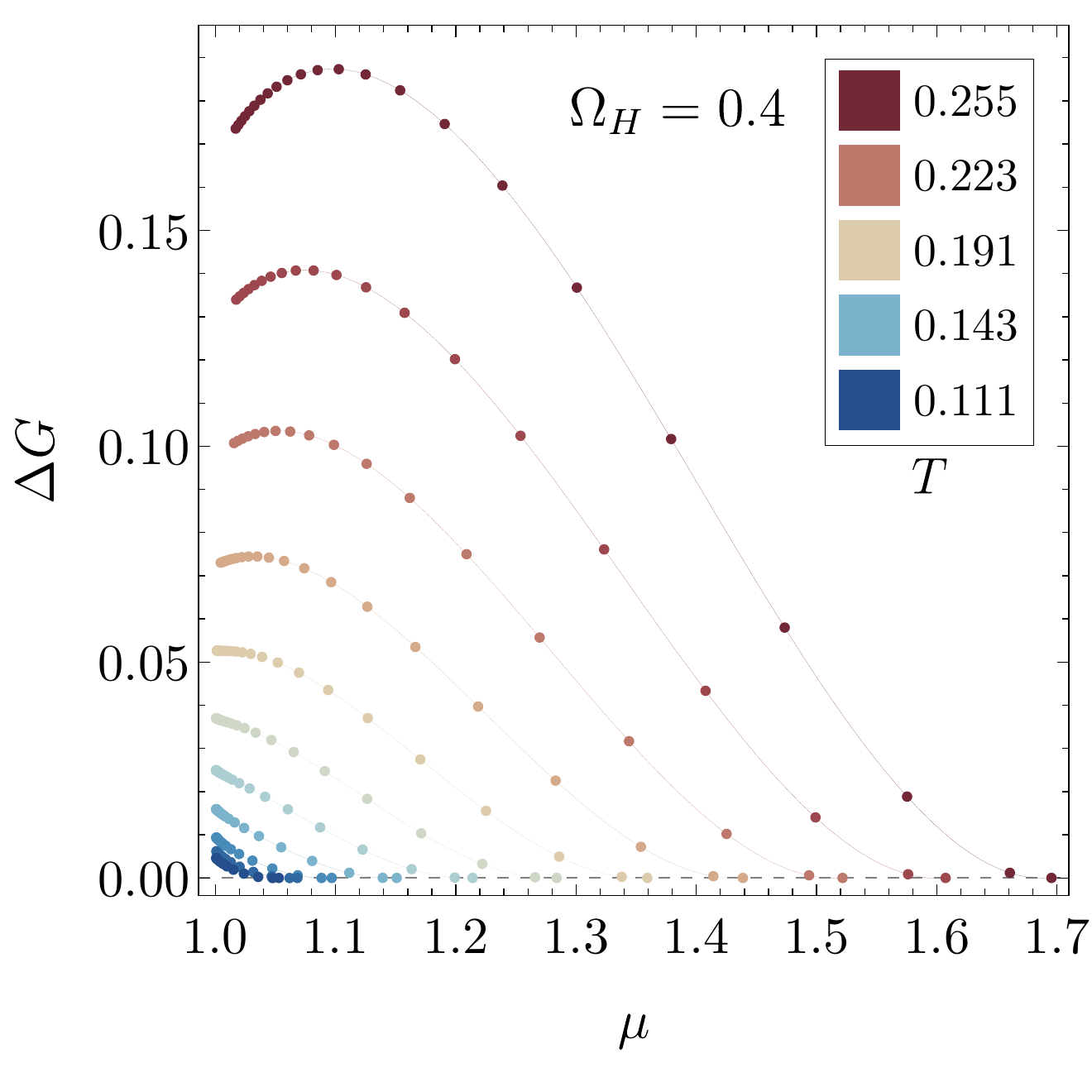}
  \end{minipage}
      \caption{\textit{Left}: The entropy difference $\Delta S=S-S_\mathrm{CLP}$ against the mass $M$, for black holes with fixed $J=0.05$ and $\varepsilon_H=0.2$ (black data points), where $S_\mathrm{CLP}$ is the entropy of the CLP black hole with the same charge $Q$ and angular momentum $J$. \textit{Right}: The difference in Gibbs free energy $\Delta G=G-G_\mathrm{CLP}$, where $G_\mathrm{CLP}$ is the free energy for the corresponding CLP solution with the same temperature $T$, and thermodynamic angular velocity $\Omega_H$. We find that for all hairy solutions have $G<0$, hence at fixed chemical potential $\mu$, ${\Delta G>0}$ indicates that the phase without hair dominates the ensemble.}
        \label{fig:micro}
\end{figure}
 
\begin{figure}[t!]
\centering
  \begin{minipage}[t]{0.47\textwidth}
    \includegraphics[width=\textwidth]{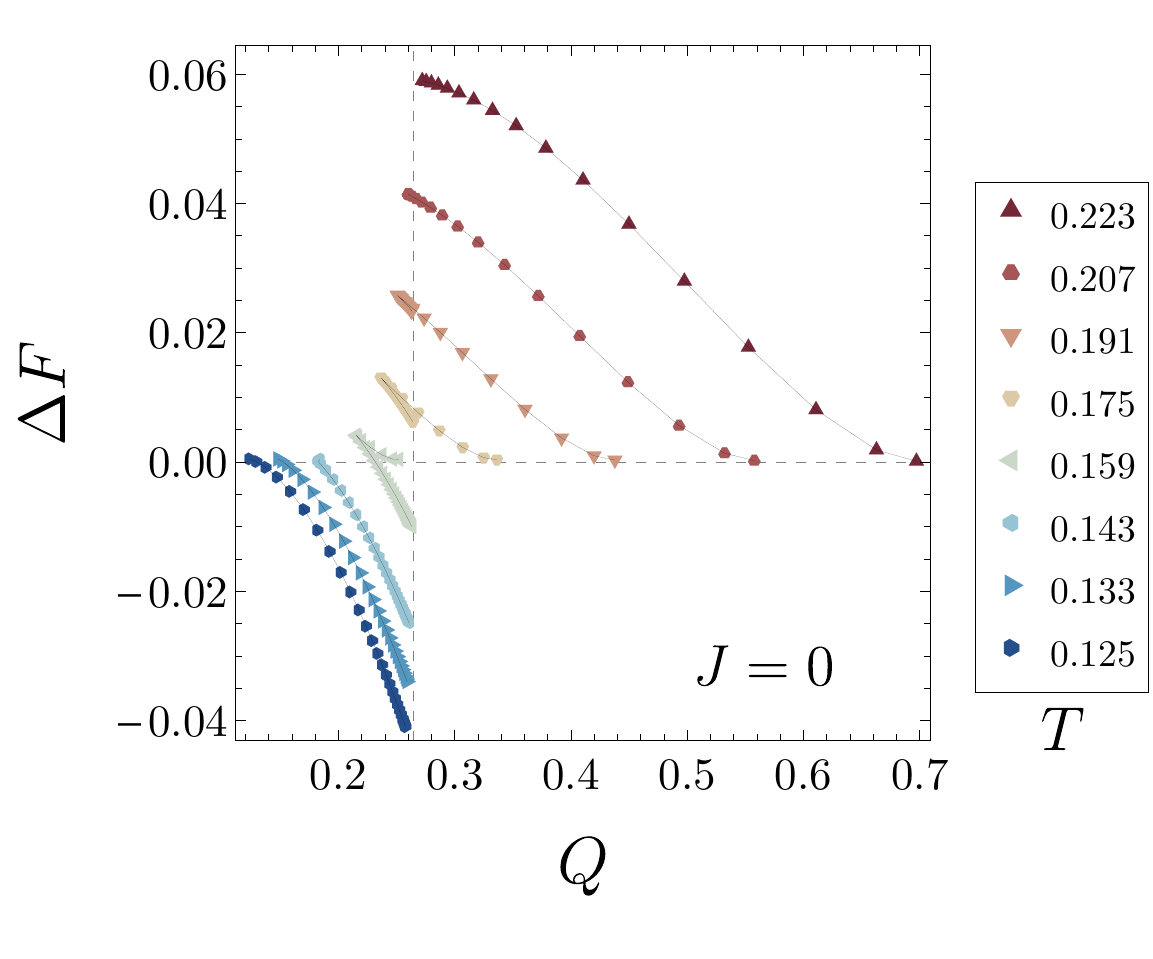}
  \end{minipage}
  \hfill
    \begin{minipage}[t]{0.47\textwidth}
    \includegraphics[width=\textwidth]{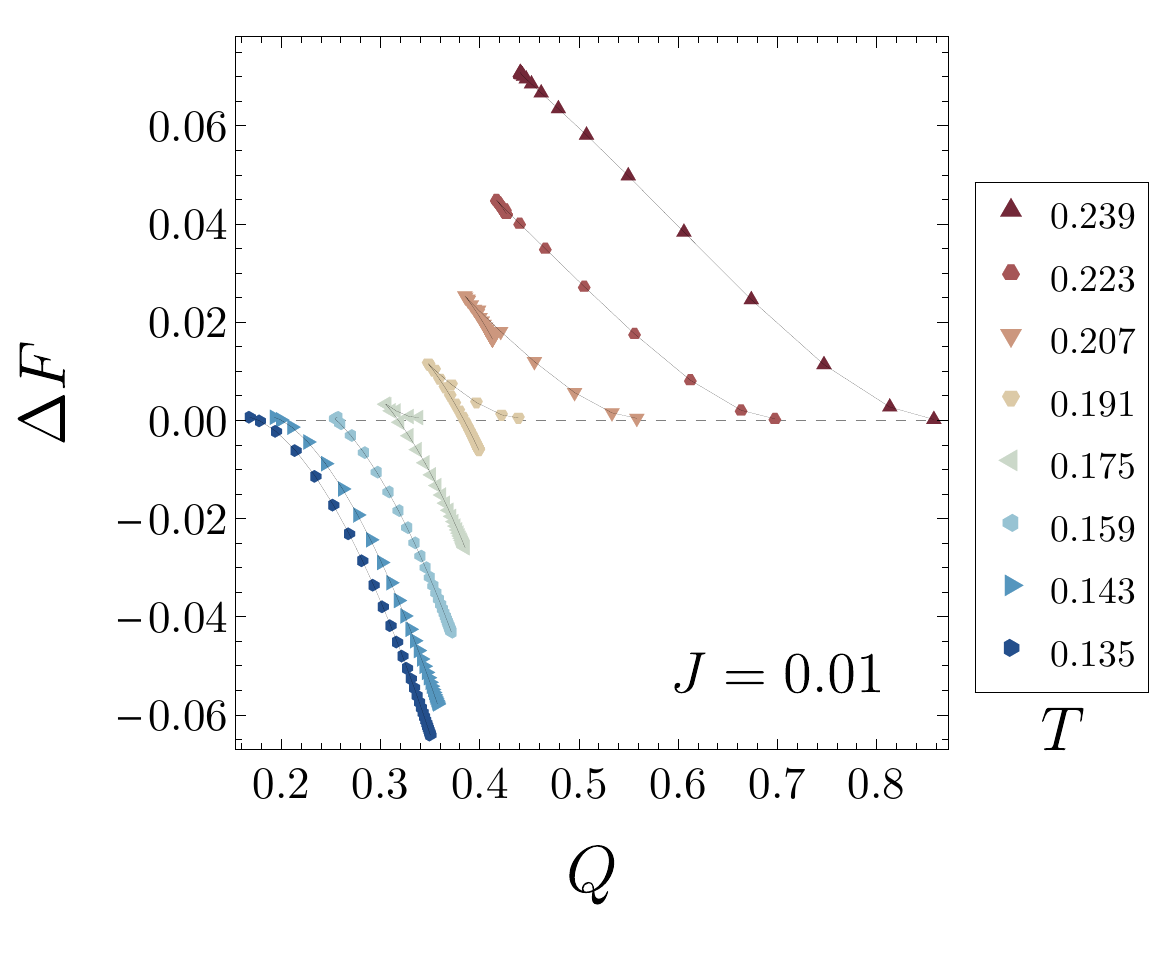}
  \end{minipage}
  \vfill
    \begin{minipage}[t]{0.47\textwidth}
    \includegraphics[width=\textwidth]{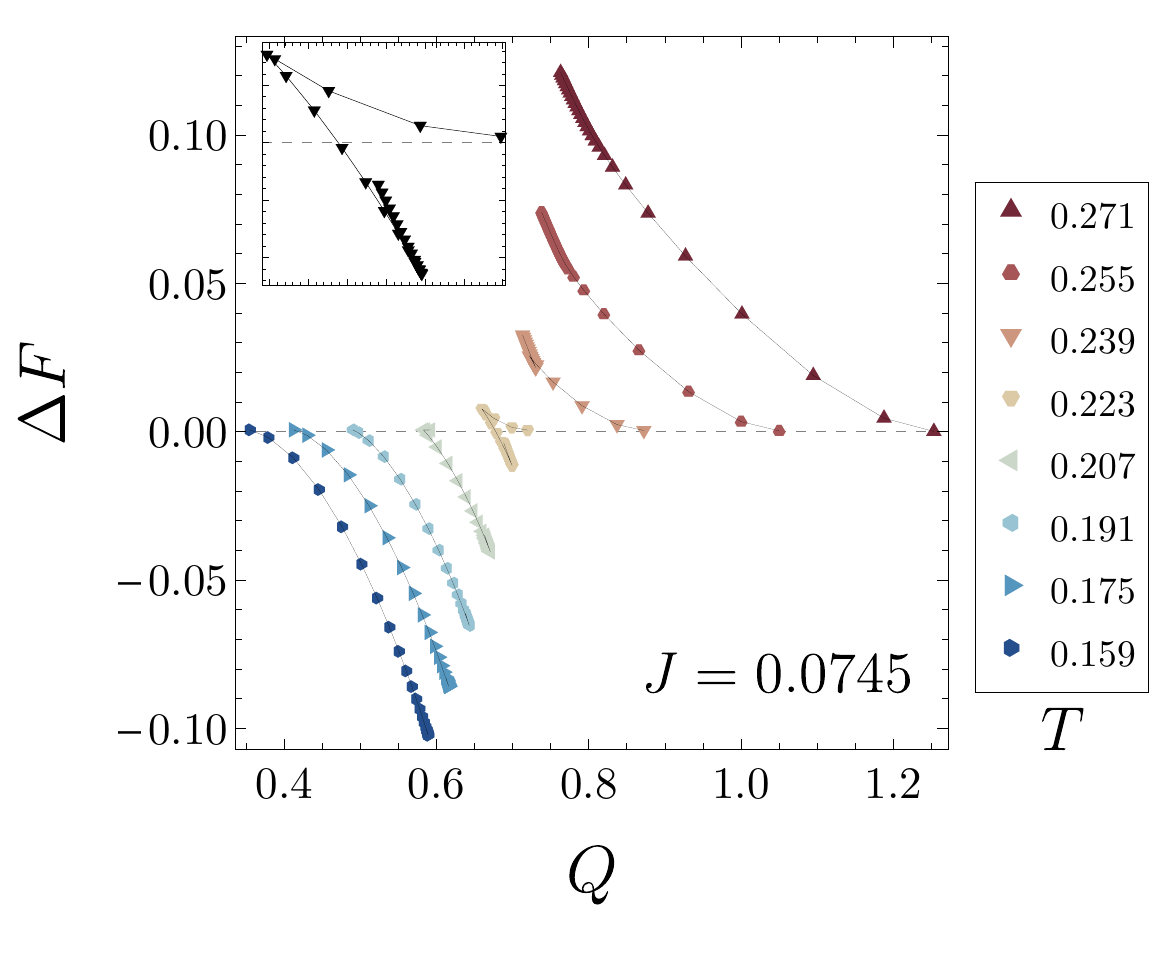}
  \end{minipage}
  \hfill
    \begin{minipage}[t]{0.47\textwidth}
    \includegraphics[width=\textwidth]{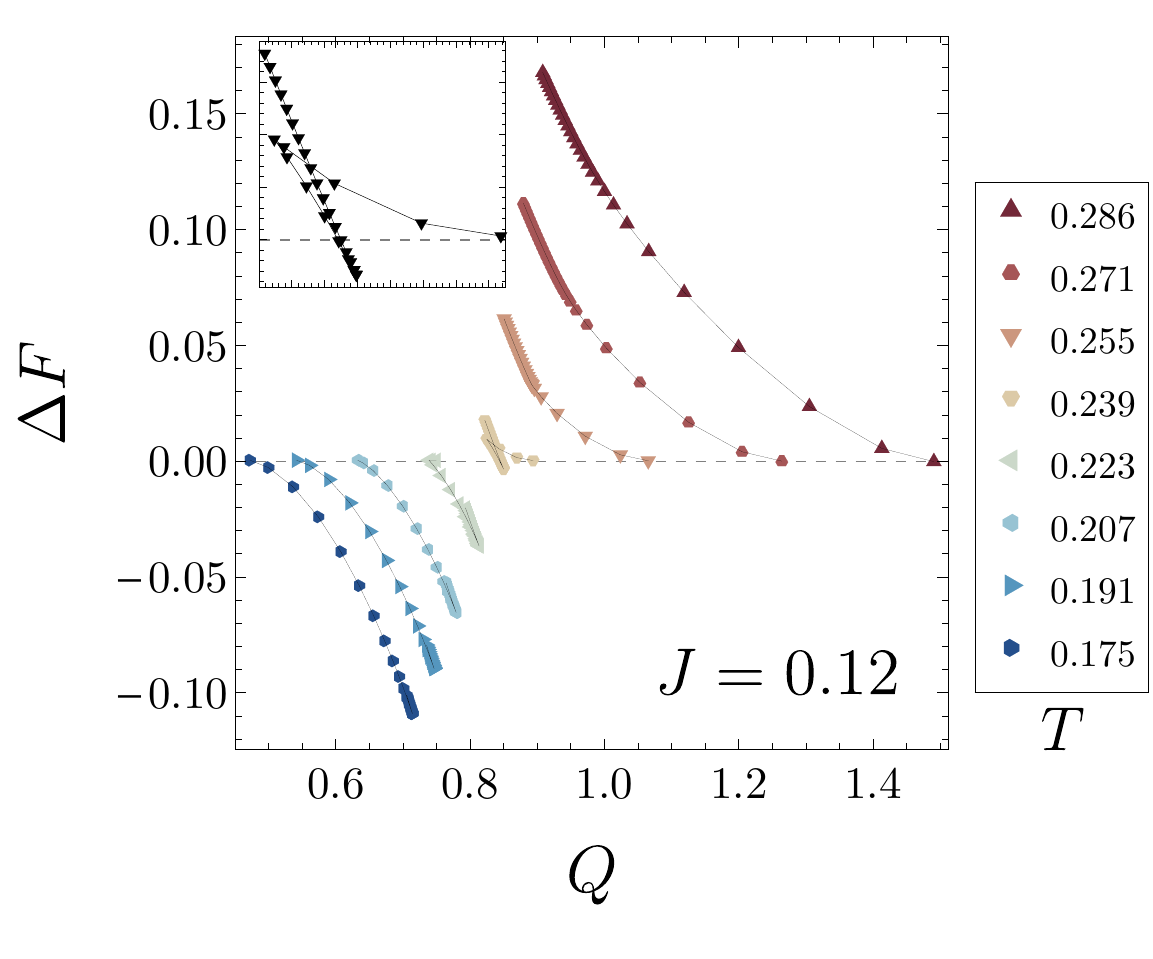}
  \end{minipage}
      \caption{Difference in the Helmholtz free energy $\Delta F=F-F_\mathrm{CLP}$, where $F_\mathrm{CLP}$ is the free energy for the corresponding CLP black hole with the same angular momentum $J$, temperature $T$ and charge $Q$. Here $\varepsilon_H \leq 15$, and the dashed vertical gridline for $J=0$ correspond to the critical charge $Q_c\simeq 0.26$. In general the isotherms demonstrate a complicated behaviour, exhibiting cusp and swallowtail type phase transitions. For all hairy black holes $F>0$ with respect to the thermal AdS.}
        \label{fig:Canonical}
\end{figure}

In this section we analyse the rich and complicated thermodynamic behaviour of the rotating hairy black hole solutions in the microcanonical, canonical and grand-canonical ensembles~\cite{Cvetic:1999ne,Caldarelli:1999xj,Cvetic:2005zi,KUNDURI2005343,Dolan:2014lea}. We find that the thermodynamics is very similar to the non-rotating case~\cite{Markeviciute:2016ivy}. Curiously, in all three ensembles studied, a different phase is the dominant one. In this section we assume that the thermodynamics is dominated by one of the three classes of solutions considered in this paper.
\\\\
 \textit{Microcanonical ensemble}: We find that for all horizon scalar values $\varepsilon_H$ and all $J$ the entropy difference between the hairy black hole and the CLP counterpart at fixed charge, mass and angular momentum $\{Q,M,J\}$, is positive (see Fig.~\ref{fig:micro}, \textit{left}), \textit{i.e.} ${\Delta S=S-S_\mathrm{CLP}>0}$. Hence, the hairy black holes dominate the microcanonical ensemble.
\\\\
 \textit{Canonical ensemble}: In the canonical ensemble the associated thermodynamic potential is the Helmholtz free energy $F=M-TS$, and we fix the temperature, charge and angular momentum $\{T,Q,J\}$. In the non-rotating case~\cite{Markeviciute:2016ivy} we observed complicated phase transitions between the hairy and non-hairy phases in the canonical ensemble, where for large temperatures the hairy black holes have larger free energy compared to the RNAdS black hole, \textit{i.e.} $F>F_\mathrm{RN}$. The opposite behaviour is seen for low $T$, where $F<F_\mathrm{RN}$. However, as $F>0$ for all the black hole solutions, we expect this ensemble to display the Hawking-Page~\cite{Hawking1983} transition to some AdS geometry with the same total charge. 
 
For $J>0$ we observe a similar picture. The results for several values of $J$ are presented in Fig.~\ref{fig:Canonical}. For $J=0$ the isotherms, when the horizon scalar $\varepsilon_H\rightarrow\infty$, cluster around the special soliton $Q_c\simeq 0.26$, and we observe a complicated relation to the corresponding RNAdS black hole phases. When $J>0$, the approach to the extremal limit at fixed $Q$ is non-monotonic in $T$ for a broader range of charges and the isotherms display a new feature near the BPS bound, namely the ``swalowtail'' type phase transition~\cite{Chamblin:1999tk}. The analysis of $\Delta F=F-F_\mathrm{CLP}$ where $F_\mathrm{CLP}$ is the free energy of the corresponding CLP black holes reveals a complicated picture. For any $J>0$ there is a temperature range where at fixed $Q$ and $J$ there are at least three hairy black holes with the same temperature (\textit{e.g.} $J=0.0745$, $T=0.239$ (Fig.~\ref{fig:Canonical}, \textit{bottom left} orange triangles), $J=0.12$, $T=0.239$ (Fig.~\ref{fig:Canonical}, \textit{bottom right} yellow hexagons)). The transition temperature increases with the angular momentum $J$.

Ultimately all the black hole solutions have $F>0$ with respect to the background geometry, thus we expect that these black holes might not be the dominant phase in the ensemble. This analysis also gives us some insight into the complicated behaviour of the isotherms at fixed~$J>0$. The phase transition could suggest the location of the maximal charge $Q_\mathrm{max}(J)$, where in analogy with the $J=0$ case we expect all isotherms to spiral around some special solution separating the $T=0$ and $T=\infty$ limits.

Local thermodynamic stability is ensured by the entropy $S(M,Q,J)$ being concave as a function of the extensive variables $X_i$, \textit{i.e.} the Hessian matrix $\left[\partial^2 S/\partial X_i \partial X_j\right]_{ij}$ being negative definite~\cite{1985tait.book.....C}. In the canonical ensemble the charge and the angular momenta do not vary, thus the sufficient condition for the stability is that the heat capacity $C_{J,Q}=T(\partial S/\partial T)_{J,Q}$ is positive. This is what we find for hairy black holes with sufficiently low charges (Fig.~\ref{fig:constantQ}), however, for general $Q$ the behaviour of $C_{J,Q}$ is complicated. When the charge is sufficiently large, we find $C_{J,Q}<0$, implying that the large $Q$ hairy black holes are locally thermodynamically unstable.
\\\\
 \textit{Grand-canonical ensemble}: The preferred phase in this ensemble minimises the  Gibbs potential given by $G=E-TS-3\mu Q-2\Omega_H J$, where we keep the intensive variables $\{T,\Omega_H,\mu\}$ fixed. We find that for the hairy black holes $G<0$, and the preferred phase is always the non-hairy black hole. The hairy phase always has $\mu>1$, and the CLP black holes have two branches exhibiting the Hawking-Page transition to the pure AdS~\cite{Hawking1983,Caldarelli:1999xj}.

\section{\label{sec:planar}Rotating black branes}
\begin{figure}[t!]
\centering
  \begin{minipage}[t]{0.47\textwidth}
    \includegraphics[width=\textwidth]{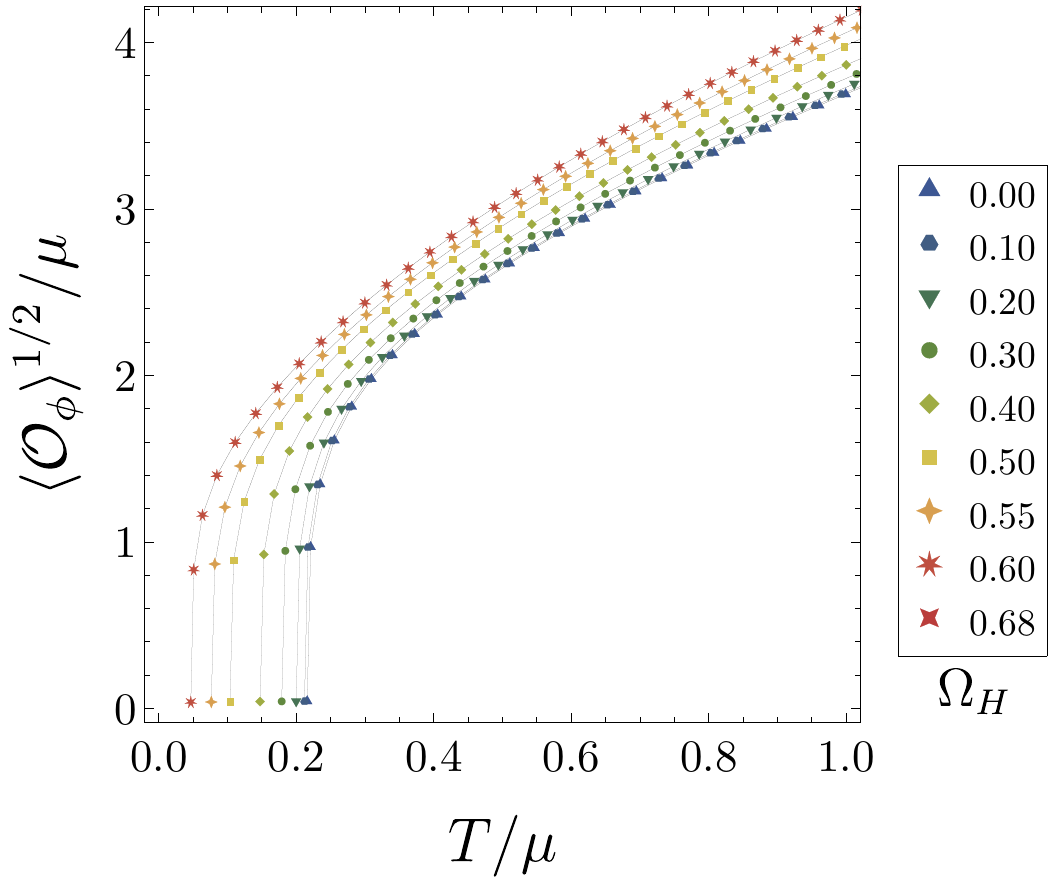}
  \end{minipage}
    \hfill
      \begin{minipage}[t]{0.47\textwidth}
    \includegraphics[width=\textwidth]{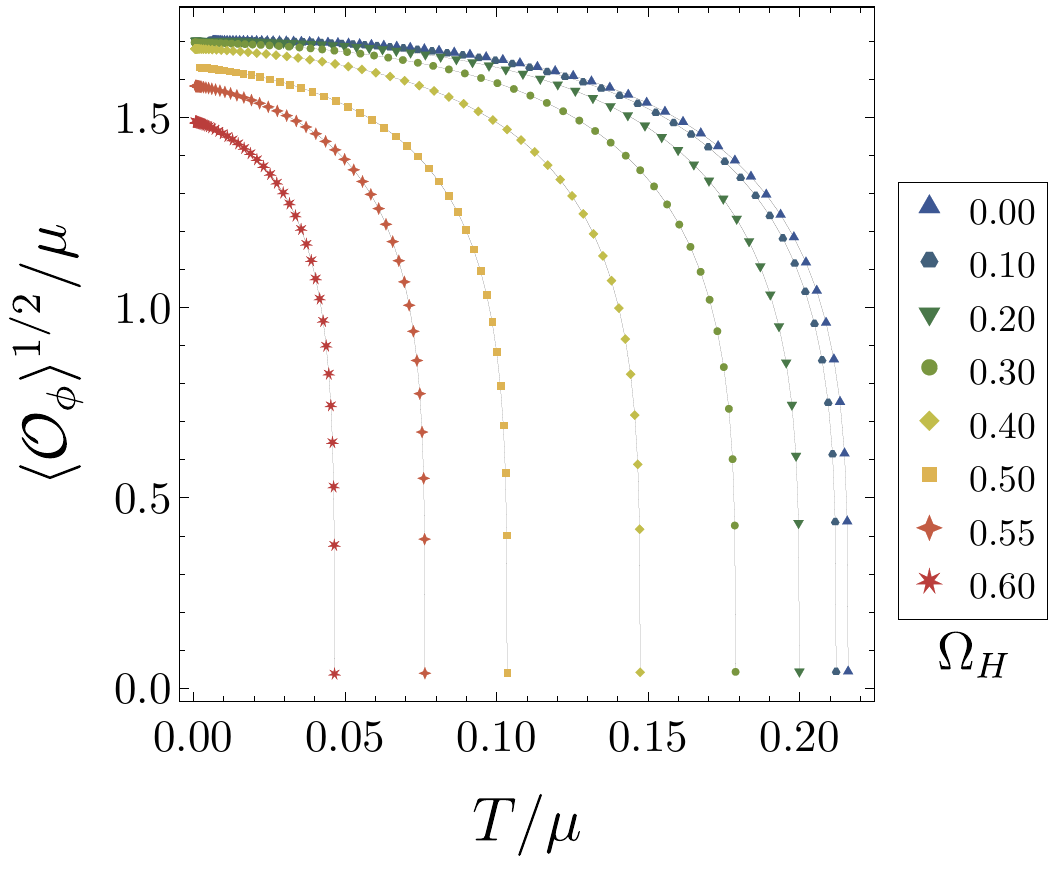}
  \end{minipage}
      \caption{\textit{Left}: Condensate against the temperature, for families of rotating hairy black branes with different values of $\Omega_H$. The critical temperature $T_c$, at which the hairy solutions first appear, decreases with $\Omega_H$. \textit{Right}: Expectation value of the dual operator as a function of temperature for a phenomenological Abelian-Higgs model, as a scale invariant quantity, for various values of the horizon velocity $\Omega_H$. }
        \label{fig:condensate}
\end{figure}

\subsection{\label{sec:planarans}Ansatz}

Even though the focus of this paper is the global solutions, it is also interesting to look at the planar horizon limit due to applications of various hairy black branes to the study of holographic superconductors (\textit{e.g.} \cite{Hartnoll:2008vx,Hartnoll:2008kx,Horowitz:2010gk,Hartnoll:2016apf}, and references therein). \textit{A priori}, one could suspect that our ``rotating'' branes are boosted solutions of the static counterparts which were analysed in~\cite{Aprile:2011uq,Markeviciute:2016ivy}. However, we will find that rotating hairy branes obtained as the scaling limit of~(\ref{eq:ansatzr}) cannot be boosts of the non-rotating hairy branes. In the gauge where $A_t=0$ on the horizon, a Lorentz boost will keep both $A_t=0$ and $A_\psi=0$ on the horizon. However, we find that $A_\psi$ does not vanish on the horizon, regardless of the residual transformations~(\ref{eq:exact}), and that the current in the dual description is non-zero. Similar solutions with a non-trivial source were studied in \cite{Sonner:2010yx} and in a consistent type IIB truncation \cite{Arean:2010wu}.

The planar limit of the ansatz~(\ref{eq:ansatzgaugescalar}) with a flat $\mathbb{R}^3$ horizon can be obtained via the following scaling
\begin{equation}
t\rightarrow\alpha\,t,\qquad y_+\rightarrow\frac{y_+}{\alpha},\qquad\psi\rightarrow\alpha\,x_1,\qquad\phi\rightarrow\alpha\,x_2,\qquad x\rightarrow\alpha\,x_3\,
\end{equation}
\noindent and taking the limit $\alpha\rightarrow 0$. We can set $y_+=1$, leaving us a two-parameter family of branes. This limit can also be obtained by boosting, for instance, a planar Schwarzschild-AdS$_5$ along the $x_1$ direction, and promoting angular velocity to be a function of the radial coordinate. This analysis gives us the following ansatz for the hairy rotating black branes with a planar horizon
\begin{equation}
\label{eq:metricplanar}
\mathrm{d}s^2=\frac{1}{1-y^2}\left[- y^2 q_1(y)\,\mathrm{d}t^2+\frac{q_2(y)\,\mathrm{d}y^2}{1-y^2}+ \left\{q_3(y) \left[\mathrm{d}x_1- \tilde{\Omega}(y)\,\mathrm{d}t\right]^2+\frac{1}{4}q_4(y)\left[\mathrm{d}x_2^2+\mathrm{d}x_3^2\right]\right\}\right],
\end{equation}
where $\tilde{\Omega}(y)=(1-y^2)^2q_5(y)$, and the gauge and scalar fields are given by
\begin{equation}
\tilde{A}_t(y)=y^2q_6(y)-q_5(y) q_7(y)\,,\quad \tilde{A}_{x_1}(y)=q_7(y)\,,\quad \tilde{\phi}(y)=(1-y^2)q_8(y).
\label{eq:ansatzgplanar}
\end{equation}
\noindent The boundary conditions at the horizon $y=0$ are given by $q_i'=0$ for all $i\neq 5$, and $q_5=\Omega_H$. At infinity we have $q_i=1$ for $i=1,2,3,4$ and 
\begin{equation}
-q_5'-\frac{1}{2}q_7' \left(-q_5 q_7'+q_6'+2q_6\right)=0,\qquad  q_8'-2 q_6^2 q_8=0,\qquad q_7=0.
\end{equation}
\noindent We will work in the radial gauge where $q_4=1$, and will find that $q_3>1$, therefore breaking the anisotropy of the homogeneous brane. The rotational symmetry in $x_2-x_3$ plane is preserved, and there is also a translational invariance along the direction in which the momentum is carried. The expectation value of the current in the dual field theory is proportional to $\tilde{A}_{x_1}(1)$, which is non-vanishing, and is not sourced. 

The CLP solution~(\ref{eq:CLP}) admits the scaling limit
\begin{equation}
t\rightarrow\alpha\,t,\quad r\rightarrow \frac{r}{\alpha},\quad\psi\rightarrow\alpha\,x_1,\quad\phi\rightarrow\alpha\,x_2,\quad x\rightarrow\alpha\,x_3,\quad q\rightarrow\frac{q}{\alpha^3},\quad r_+\rightarrow\frac{r_+}{\alpha},\quad j\rightarrow j,
\end{equation}
which yields a two-parameter family of planar black holes describing the normal phase. They are related to the planar RNAdS holes by the boost 
\begin{align}
\begin{split}
t&=\frac{1}{\sqrt{1-j^2}}\,\tilde{t}+\frac{j}{\sqrt{1-j^2}}\,\tilde{x}_1\,,\\
x_1&=\frac{j}{\sqrt{1-j^2}}\,\tilde{t}+\frac{1}{\sqrt{1-j^2}}\,\tilde{x}_1.
\end{split}
\end{align}
\noindent Their thermodynamic quantities are given as
\begin{align}
\hat{M}&=\frac{(3+j^2)((1-j^2)q^2+r_+^6)}{4r_+^2(1-j^2)}, \quad T=\frac{\sqrt{1-j^2}(2r_+^6-(1-j^2)q^2)}{2\pi r_+^5}, \quad\Omega_H=j, \\
\hat{J}&=j\frac{(1-j^2)q^2+r_+^6}{2r_+^2(1-j^2)}, \quad \mu=q\frac{1-j^2}{r_+^2}, \quad\hat{Q}=\frac{q}{2},
\end{align}
\noindent which satisfy the first law~(\ref{eq:firstlawofTD}).

\begin{figure}[t!]
\centering
    \begin{minipage}[t]{0.32\textwidth}
    \includegraphics[width=\textwidth]{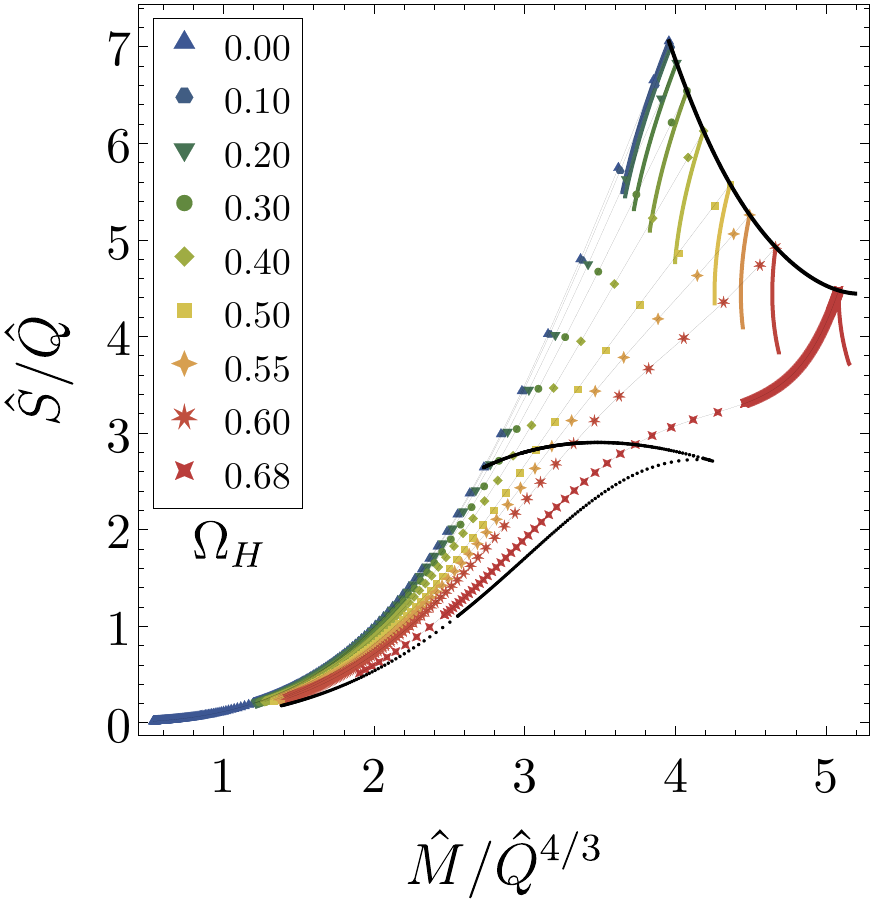}
  \end{minipage}
  \hfill
  \begin{minipage}[t]{0.32\textwidth}
    \includegraphics[width=\textwidth]{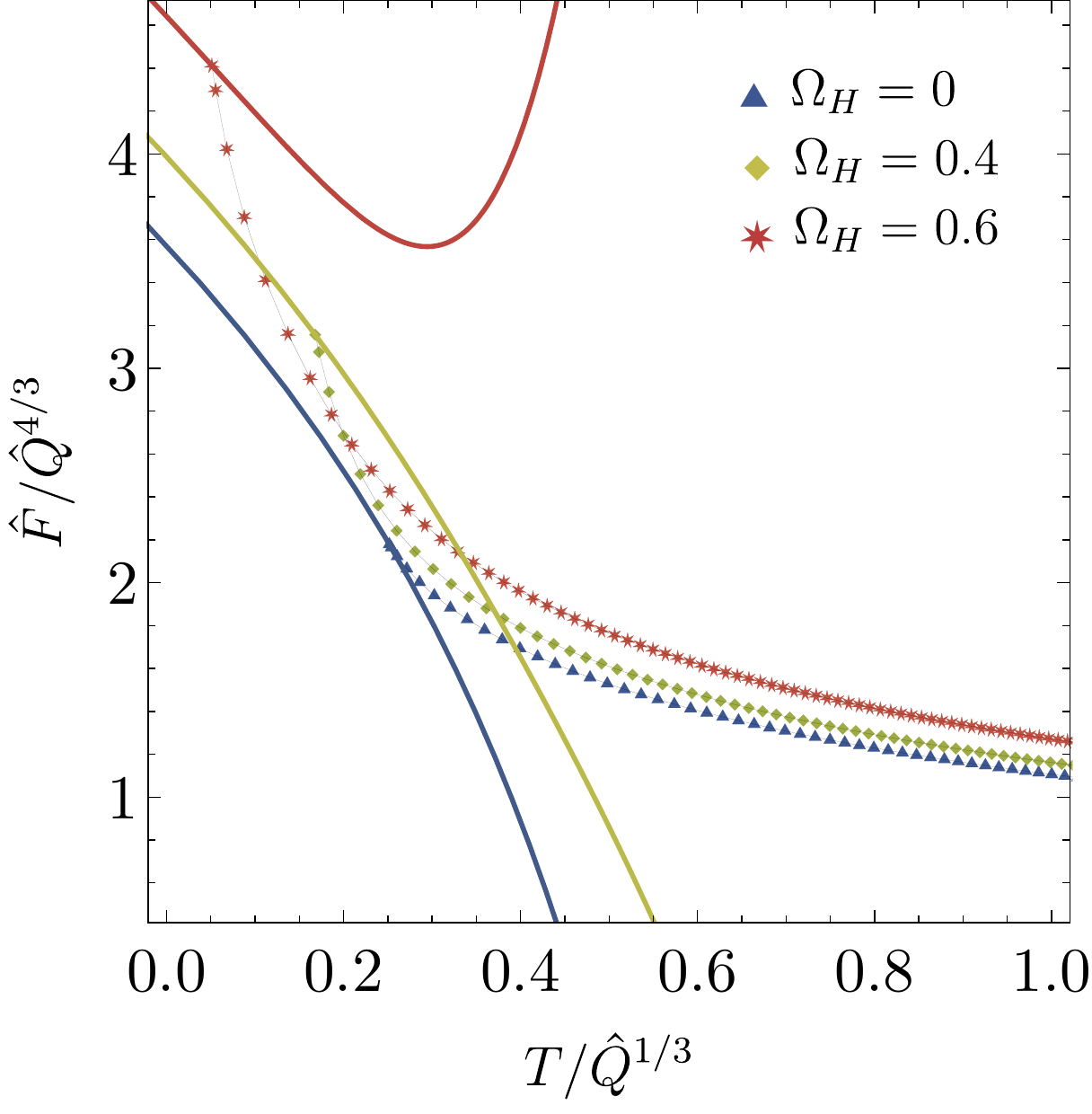}
  \end{minipage}
  \hfill
    \begin{minipage}[t]{0.32\textwidth}
    \includegraphics[width=\textwidth]{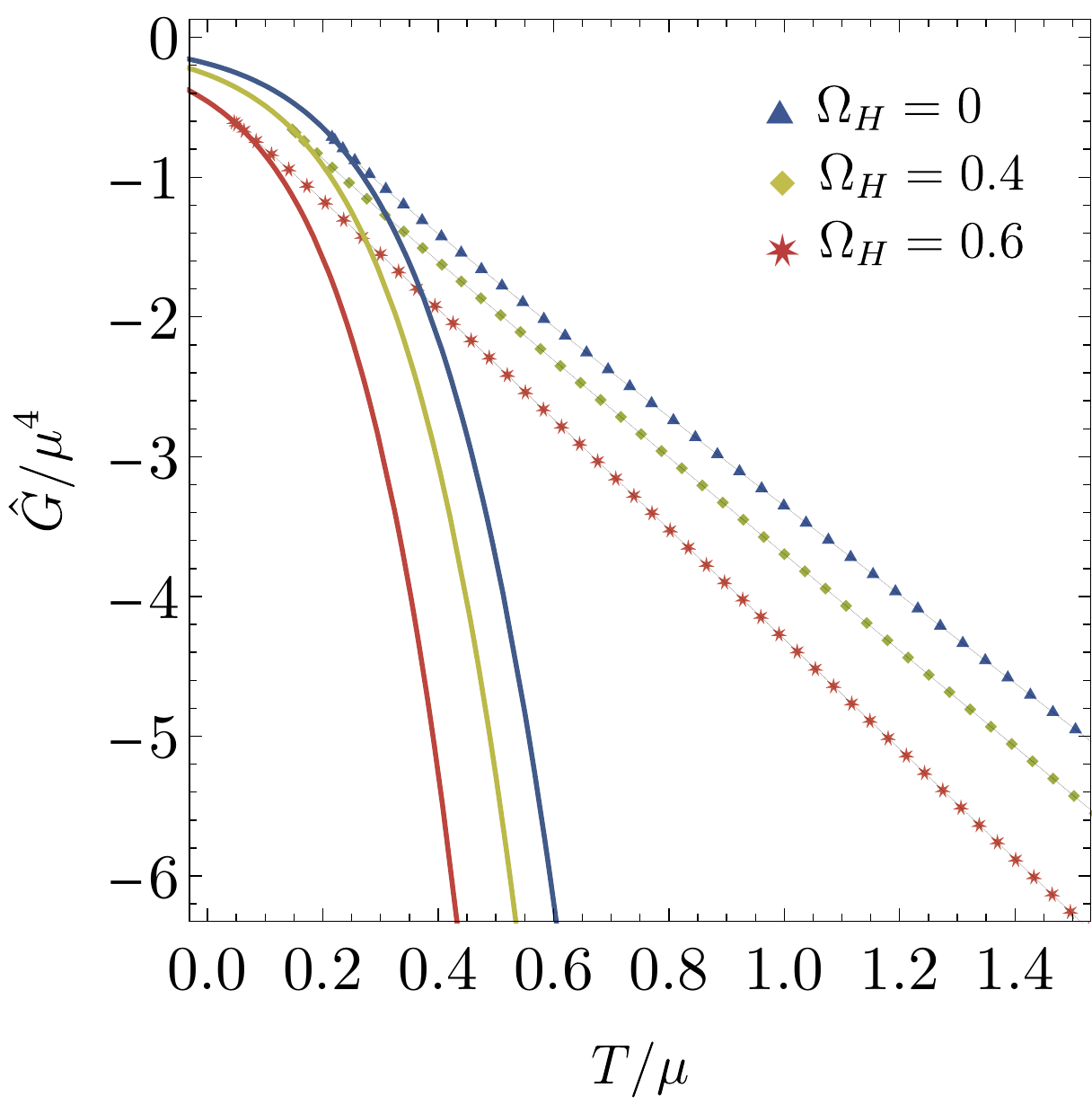}
  \end{minipage}
  \caption{\textit{Left}: Rotating hairy brane entropy density as a function of the mass density in the microcanonical ensemble, for different values of $\Omega_H$ (data points in rainbow). Solid black line shows the onset of the instability, and solid rainbow lines are the corresponding solutions without hair. The black data points show two hairy black hole families, one of constant $\varepsilon_H=3.5$, and one of constant $\Omega_H=0.72$. \textit{Middle}: Hairy black brane free energy density as a function of temperature in canonical ensemble, for different values of $\Omega_H$ (see the inset legend). The solid lines show the corresponding solutions without hair. \textit{Right}: The free energy density in the grand-canonical ensemble, against the temperature.}
        \label{fig:braneFG}
\end{figure}
\subsection{\label{sec:thermplanar}Solution space and thermodynamics}

To generate the two parameter family of hairy branes, we will vary the black hole horizon velocity $\Omega_H=q_5(0)$, and the horizon scalar field $\varepsilon_H$. The non-rotating solutions exhibit the retrograde condensation, i.e. exist for $T>T_c$, which remains unchanged if we turn on the linear momentum (Fig.~\ref{fig:condensate}, \textit{left}). Such phenomena is commonly observed in branes with supergravity potentials, and is usually associated with a subdominant phase~\cite{Buchel:2009ge,Aprile:2011uq,Donos:2011ut,Gentle:2012rg,Cai:2013aca,Banks:2015aca}.

In~\cite{Markeviciute:2016ivy} it was shown that in the large temperature\footnote{Equivalently, the large horizon scalar field limit.} limit the hairy branes connect to the singular soliton branch through the planar limit of the singular, large charge soliton solution
\begin{equation}
\label{eq:exact}
\mathrm{d}s^2=-r^2\mathrm{d}t^2+\frac{r^2\mathrm{d}r^2}{1+r^4}+r^2\mathrm{d}\mathbf{x}^2,\quad\phi(r)=\frac{2}{r^2},\quad A(r)=0.
\end{equation} 
We find that the rotating solutions in the large temperature limit also approach the same solution\footnote{This can be confirmed, for instance, by computing the scalars $g_{tt}$, $g_{\psi\psi}$ in the DeTurck gauge, which fixes the conformal class.}. Another interesting feature of the solution space is that the maximal value of $\Omega_H$ depends on the scalar field $\varepsilon_H$; the merger curve extends to $\Omega_H\simeq 0.699$, while for instance hairy branes with fixed $\varepsilon_H=3.5$ can reach up to $\omega_H\simeq 0.720$. These results are illustrated in Fig.~\ref{fig:braneFG} (\textit{left}).

Rotating hairy branes are the large charge limit of the global hairy black holes, which dominate only the microcanonical ensemble. In the moduli space where the CLP and hairy branes coexist, the hairy solutions have the higher entropy density at the fixed charge density $\hat{Q}$ and therefore dominate the microcanonical ensemble (Fig.~\ref{fig:braneFG}, \textit{left}). 

The canonical ensemble exhibits first order phase transition between the hairy and non-hairy phases, when we vary $\Omega_H$ (Fig~\ref{fig:braneFG}, \textit{middle}). For any $\Omega_H>0$, there exists a temperature range for which the hairy phase is dominant for some $T_c(\Omega_H)\leq T \leq T_\mathrm{max}(\Omega_H)$. For sufficiently large $\Omega_H$, the hairy phase is always dominant (Fig.~\ref{fig:braneFG}, \textit{middle}). These results are somewhat surprising, as the large charge $Q$ and fixed $J$ black holes in global AdS have larger free energy $F$ than the hairy solutions. However, we observe that with the increasing $J$ the transition temperature increases (Fig.~\ref{fig:Canonical}), therefore it is plausible that such a phase transition is also seen in the planar limit. However, these branches have negative specific heat and are locally thermodynamically unstable.

Finally, the hairy branes never dominate the grand-canonical ensemble (Fig.~\ref{fig:braneFG}, \textit{right}) just as in the global case.

\subsection{\label{sec:abelian}The Abelian-Higgs model}

The retrograde condensation typically indicates that a condensed phase has larger free energy and therefore represents an unstable branch. The thermodynamic behaviour is largely governed by the truncation. The scalar potentials from supergravity truncations are complicated and can induce rich dynamics. In this subsection we contrast the supergravity model (\ref{eq:action}) to the minimally coupled charged scalar field (also see \cite{Aprile:2011uq}); in what follows, the Chern-Simons term will not be important.

A holographic dual to the $d=5$ planar Abelian-Higgs model with minimally coupled charged scalar field $V(|\phi|)=m_\phi^2 \phi \phi^\dagger$ was studied by \cite{Horowitz:2008bn,Herzog:2010vz}.  As before, we take the mass saturating the five-dimensional BF bound $m_\phi^2=-4$ with $e=2$, and consider the stationary ansatz (\ref{eq:metricplanar}). We find that the condensate $\langle\mathcal{O}_\phi\rangle$ exists for $T<T_c$ (Fig.~\ref{fig:condensate}, \textit{right}), and saturates at low $T$. Furthermore, the hairy branes dominate over the normal phase in all three thermodynamic ensembles. Once we turn on the ``rotation'', the superconducting behaviour is preserved and the transition temperature $T_c$ decreases with increasing $\Omega_H$, which was also observed in holographic rotating superfluids~\cite{Sonner:2009fk} and holographic superfluids with supercurrents~\cite{Sonner:2010yx}. We also find a critical value of $\Omega_H$ above which there are no hairy solutions, \textit{i.e.} there exists a critical magnetic field which destroys the superconductivity.

The ansatz (\ref{eq:metricplanar}) provides a simple way to realise a spontaneously generated current in the holographic dual description, and merits further investigation. While we do not find the stable condensed phase in this particular truncation, such phases have been found in different consistent truncations of the five-dimensional $\mathcal{N}=8$ gauged supergravity (\textit{e.g.} \cite{Aprile:2011uq,Gentle:2012rg}). It would be interesting to study the spontaneous currents in these models.
 
\section{\label{sec:discussion}Discussion and future directions}

In this paper we have analysed rotating, charged, hairy black holes in five-dimensional AdS, in both global and planar spacetimes. These black holes can be embedded in the $\mathcal{N}=8$ $SO(6)$ gauged supergravity model with the three diagonal $U(1)$ charges set equal. We have analysed the three dimensional solution space by considering various limits, and we have shown that these black holes extend the CLP solutions to the BPS bound for all charges. These new hairy states above the supersymmetry contribute to the state space of $\mathcal{N}=4$ SYM at energies $\mathcal{O}(N^2)$ at large $N$, and demonstrate the existence of new phases at all charges. We have also examined the non-trivial approach to the BPS bound in detail. Facilitated by the radial gauge we were able to reach extremely low temperatures, however, we were not able to fully unravel the nature of the extremal hairy solutions.

In~\cite{Markeviciute:2018yal} we conjectured the existence of a one parameter extension of the supersymmetric Gutowski-Reall black hole. The constant horizon scalar $\varepsilon_H$ hairy black hole families retain a finite entropy in the zero temperature limit, as well as possess regular curvature scalars. In the present manuscript we provide further evidence to support this conjecture, namely the indication of the zero temperature limit by the isotherms, and regularity of the fixed low charge limit.

It is important to note, however, that the extremal limit is challenging to resolve numerically, and the slow, possibly logarithmic, approach renders some of our results inconclusive. It is not clear to what extent the rotating hairy black holes continuously connect to the $J=0$ case. It would be very interesting if a small amount of rotation allows a nucleation of a small rotating black hole at the centre of the scalar field cloud. Numerical results support the idea of some continuity, albeit that we cannot rule out a completely different mechanism governing the near BPS regime. Numerically we found it difficult to locate the maximal charge for the $T=0$ family $Q_\mathrm{max}(J)$, and we do not know if it is connected to $Q_c=Q_\mathrm{max}(0)$. We also cannot eliminate the possibility of the extremal hairy solutions being singular. However, this would require dramatic near extremal behaviour, and such a picture is hard to reconcile with our numerical results. Lastly, we have not analysed large $J$ hairy black holes in detail as these solutions also have large charge, and it is conceivable that some such solutions could become singular.

Numerics at low temperatures become increasingly challenging, and we can only infer the expected qualities of the limiting solution. It would be preferable to study the extremal limit analytically, by directly using the supersymmetry. However, our numerical results indicate non-analytic behaviour which would complicate the near horizon expansion of the BPS solutions, and the near horizon geometry does not appear to be of the usual AdS$_2\times$S$^2$ type. This could explain how the no-hair theorem of~\cite{FernandezGracia:2009em} is evaded for such extremal hairy black holes. We can also look for singular (large charge) BPS solutions with $J>0$ directly, and find the lowest allowed charge. Work in these directions is in progress.

Perhaps it is possible to choose a better ansatz and to simplify the equations enough that some analytic treatment is possible. We found that the Abelian-Higgs scalar potential $V(\phi)=-4 \phi\phi^\dagger$ displays some similarities, and in particular a similar zero temperature, finite entropy limit appears to exist. Varying the Chern-Simons coupling slightly, or setting it to zero also appears to maintain the aforementioned limit\footnote{Only for the supergravity coupling the marginal mode extends to the BPS bound, and the lowest mass rotaing hairy black holes never saturate it.}. The phase space is reminiscent of the $e=e_c$ case in \cite{Dias:2011tj}. Evidence of five-dimensional finite entropy rotating, hairy extremal black holes has been also seen in rotating black holes with a complex doublet scalar field~\cite{Dias:2011at}, and black holes in Einstein-Maxwell-Scalar theory~\cite{Brihaye:2011fj}. It might be worthwhile to study the limiting solutions in these models further. Finally, we have not analysed $Q<0$ solutions in detail, however they exhibit distinctly different behaviour. The fixed $\varepsilon_H$, $J$ solutions when lowering the temperature tend to some singular $S=0$ configurations and have $\Omega_H<1$, and furthermore due to the highly singular behaviour it is not clear if $T=0$ limit exists.

An immediate way to extend this work is to consider non-equally rotating black holes, as the general supersymmetric black hole has two angular momenta~\cite{Gutowski:2004yv,Chong:2005da}. It would be intriguing if the onset of the superradiance still coincides with the BPS bound. A more complicated task is to analyse black holes with independent diagonal $SO(6)$ charges, including when some of the charges vanish. This would necessarily involve consistent truncations with more scalar fields, such as in \cite{Aprile:2011uq}.

The scalar field $\phi$ considered in this paper is dual to the lightest chiral SYM operator with dimension $\Delta=2=e$, and the small black holes lie at the edge of the superradiant instability. We expect these hairy phases to be thermodynamically relevant at small charges. It is not known whether the hairy black holes found in this paper are indeed the thermodynamically dominant saddle points above the supersymmetry from the full ten-dimensional point of view. It is very difficult to tackle this problem in full generality, especially in the non-linear, near-BPS regime.  On the other hand, the hairy black holes constructed in this paper can be oxidized to ten dimensions, which could reveal further insights into the approach to the BPS bound. For instance, singular solutions from the five-dimensional perspective can have regular higher dimensional embedding.

We expect that different scalars are relevant across the moduli space (\textit{e.g.} see \cite{Aprile:2010ge,Aprile:2011uq}, and references therein), depending on which instability dominates. We find that large charge hairy black holes have negative specific heat and are locally thermodynamically unstable. The large charge hairy SUSY black holes are singular, and it would be intriguing if there exist scalars which produce finite entropy hairy rotating BPS black holes for $Q>Q_\mathrm{max}(J)$. 

The planar limit solutions display retrograde condensation, \textit{i.e.} exist for $T>T_c$ ; in \cite{Aprile:2011uq} it was found that in the ensemble with three equal charges a scalar with a larger mass and larger charge condenses for $T<T_c$, and has lower free energy. It would be very interesting to construct large charge global hairy holes in this truncation, which was first studied by \cite{Gubser:2009qm}. The scalar has $\Delta=3=e$, therefore we also expect rich phase diagram at lower charges. Within the $\mathcal{N}=8$ supergravity truncation we anticipate the operators with low dimensions and higher charge to condense first\footnote{This might not be generic for non-minimally coupled scalars.} (\textit{e.g.} see \cite{Gubser:2009qm}), and thus we consider operators dual to light modes of the $\mathcal{N}=8$ supergravity; however, there is clearly some competition between different influencing factors.

Finally, it would be instructive to construct rotating, charged, hairy global black hole phase space in consistent truncations of AdS$_4\times$S$^7$ and AdS$_7\times$S$^4$. In particular, equally rotating black holes in odd dimensions preserve cohomogeneity-one, so the problem is readily tractable. It would be fascinating if there exist new families of hairy supersymmetric black holes which account for missing BPS states in the stringy entropy calculations.

\acknowledgments
We would like to thank J. E. Santos for suggesting this project, and for many useful discussions. We are grateful to C. V. R. Board, H. S. Reall and J. E. Santos for reading an earlier version of the manuscript. JM was supported by an STFC studentship.

\appendix 

\section{\label{sec:comparison}The non-interacting thermodynamic model}

\begin{figure}[t]
\centering
  \begin{minipage}[t]{1.0\textwidth}
    \includegraphics[width=\textwidth]{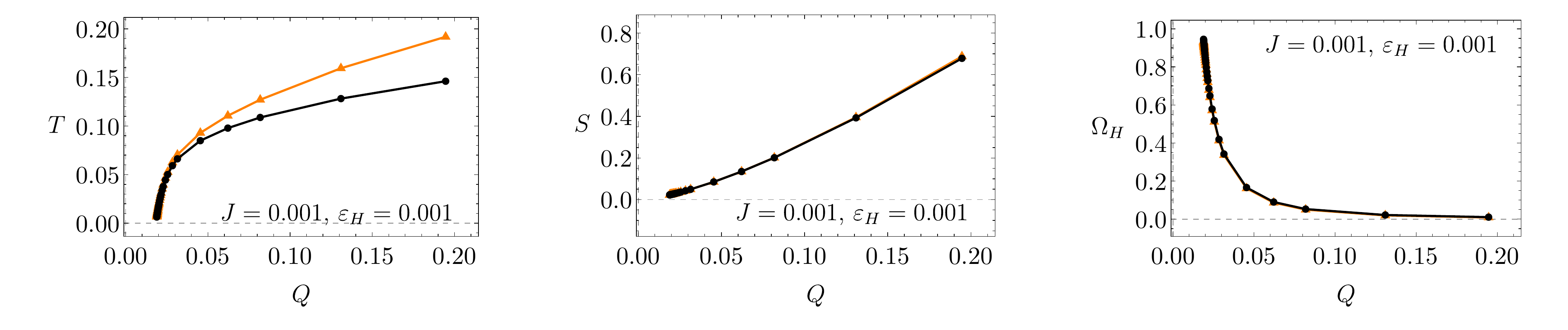}
  \end{minipage}
    \begin{minipage}[t]{1.0\textwidth}
    \includegraphics[width=\textwidth]{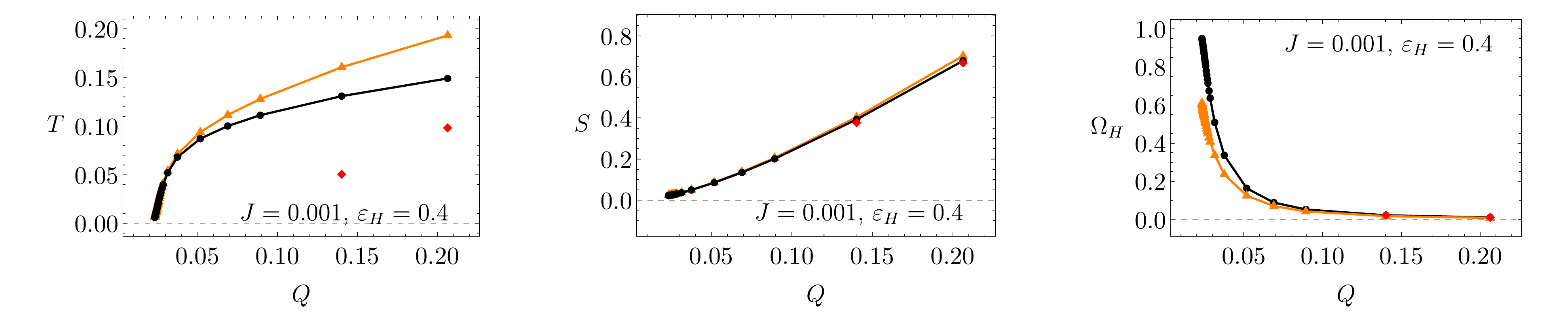}
  \end{minipage}
      \caption{\textit{First line}: The comparison with the approximation of~\cite{Bhattacharyya:2010yg} for the same charge, mass, and angular momentum $J=0.001$. Black disks are numerical results, and orange triangles are the predicted  curves. We expect the model to be valid for small $Q$, small $J$, and near extremality. For small horizon scalar $\varepsilon_H$, the hairy black holes are just above the extremality and coexist with CLP solutions. As $T\rightarrow 0$, they approach the Gutowski-Reall solution. For small charge the participating soliton contributes $Q_s/Q\simeq 0.5 \%$ of the total charge. Here the lowest  $T=0.00633$, and $S=0.0228$. The error of the approximation to $S$ close to the BPS bound is less than $1\%$.\\
     \textit{ Second line}: The constant $\varepsilon_H=0.4$ curve crosses the extremality line. The red rhombi are the corresponding CLPs where they coexist in micro-canonical ensemble. For small charge, $Q_s/Q\simeq 20 \%$. Here lowest $T=0.00597$, and $S=0.0224$. The error of the approximation to $S$ close to the BPS bound is $2-4\%$.}
        \label{fig:perturbative1}
\end{figure}	

\begin{figure}[t]
\centering
  \begin{minipage}[t]{1.0\textwidth}
    \includegraphics[width=\textwidth]{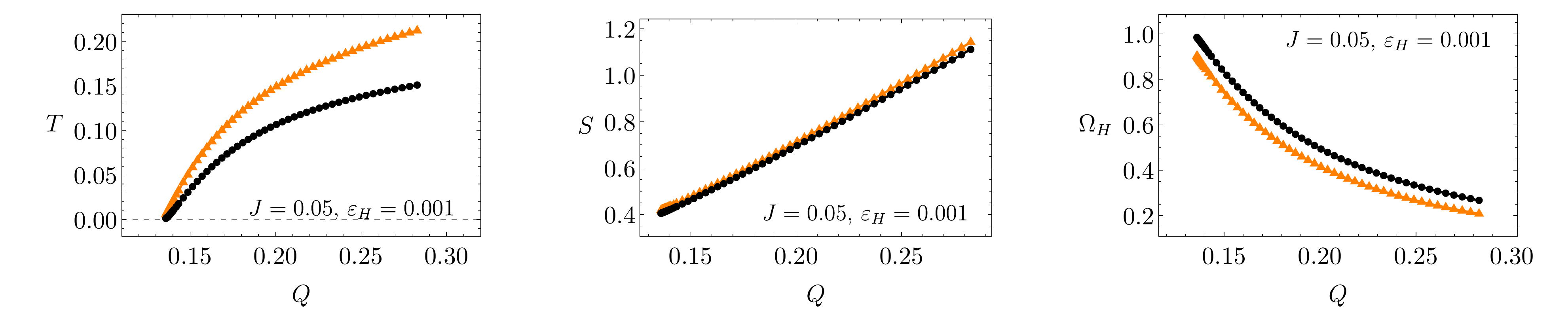}
  \end{minipage}
    \begin{minipage}[t]{1.0\textwidth}
    \includegraphics[width=\textwidth]{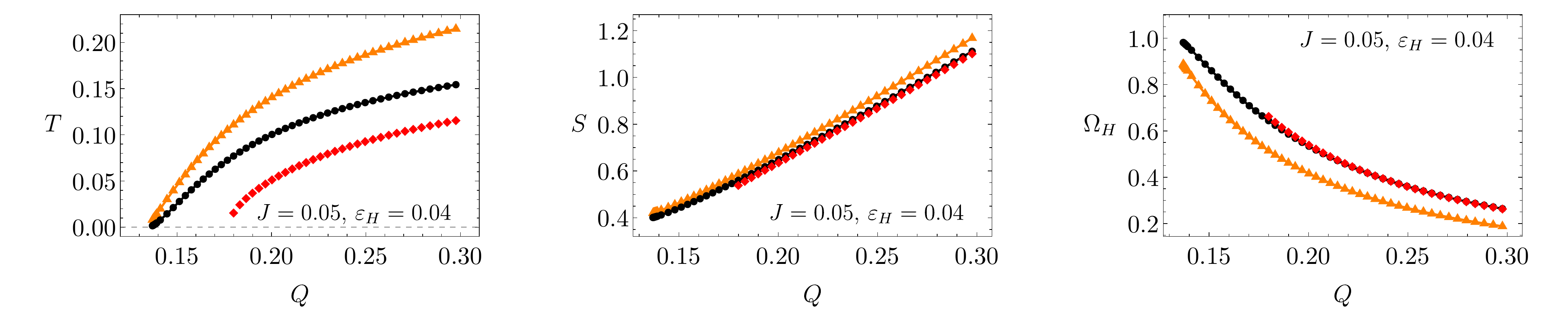}
  \end{minipage}
      \caption{\textit{First line}: The comparison with the approximation of~\cite{Bhattacharyya:2010yg} for the same charge, mass, and angular momentum $J=0.05$. Black disks are numerical results, and orange triangles are the predicted  curves. For small charge, $Q_s/Q\simeq 0.5 \%$. Here the lowest $T=0.00127$, and $S=0.405$. The error in $S$ close to the BPS bound is $2-4\%$. Note that in particular as $T\rightarrow 0$, $\Omega_H\rightarrow 1$.\\
     \textit{Second line}: Red rhombi show the coexisting CLP black holes. For small charge, $Q_s/Q\simeq 0.6 \%$. Here lowest $T=0.00164$, and $S=0.401$. The error in $S$ close to the BPS bound is~$5\%$.}
        \label{fig:perturbative2}
\end{figure}

One might wonder whether we can access the near extremal regime of the small rotating hairy black holes perturbatively. The small hairy non-rotating black holes were studied in~\cite{Bhattacharyya:2010yg}, where a matched asymptotic expansion was carried out\footnote{In~\cite{Markeviciute:2016ivy} we found a good agreement with fully non-linear numerical results, for small asymptotic charges.}. The hairy solutions were also modeled as a non interacting mix of the small RNAdS black hole and the supersymmetric ($\mu=1$, $M=3Q$) soliton, and the leading order thermodynamics of the matched expansion were reproduced. Such approach has also been succesfully employed in a similar context by~\cite{Basu:2010uz,Dias:2011tj,Dias:2016pma,Dias:2018zjg,Dias:2018yey}.

The authors of~\cite{Bhattacharyya:2010yg} also modeled small hairy rotating black holes as a non interacting mix of the CLP black hole with $\mu=1$ (but not necessarily $\Omega_H=1$, which is only equal at the supersymmetry) and the supersymmetric soliton, i.e. the system is in thermodynamic equilibrium. In $J=0$ plane, hairy black holes with fixed horizon scalar charge reduce to the smooth soliton as $T\rightarrow 0$. We find that the hairy rotating black holes exist arbitrarily close to the BPS bound, therefore it would be reasonable to expect that for small $J$ some mix of the soliton and the CLP black hole could approximate the hairy solutions, if we are sufficiently close to the Gutowski-Reall solution, \textit{i.e.} both to the extremal, and BPS limits.

We briefly describe the results of~\cite{Bhattacharyya:2010yg}. We assume that that\footnote{Here \textit{b} is the black hole index, and \textit{s} is the soliton index.}
\begin{align}
Q=Q_b+Q_s, \qquad jQ^2=j_bQ_b^2,\qquad M=M_b+M_s,
\end{align}
\noindent where angular momentum is given by $J=jQ^2$, charge $Q$ is small, and parameter $j$ is arbitrary. For CLP black holes $j_b\leq 3$, saturating at extremality. Requiring $Q_s>0$ implies $j_b> j$, so $j<3$. Leading order entropy, temperature and rotational charge are given by
\begin{align}
\begin{split}
&S=2\sqrt{2}\pi\left(\frac{1}{6}\Delta+\frac{1}{6}\sqrt{\Delta^2-4J^2}\right)^{3/4},\\
&T=\frac{\sqrt{2}}{\pi}\left(\frac{1}{6}\Delta+\frac{1}{6}\sqrt{\Delta^2-4J^2}\right)^{1/4}\left(1-4J^2\left(\Delta+\sqrt{\Delta^2-4J^2}\right)^{-2}\right),\\
&\Omega_H=\frac{1}{3}j\,,\\
&\Delta=M-3Q.
\end{split}
\end{align}
We present a comparison with the numerical results in Fig.~\ref{fig:perturbative1} and Fig.~\ref{fig:perturbative2}. For small $J$, $Q$ and the horizon scalar field $\varepsilon_H$ we find a good agreement, as the hairy black holes approach the supersymmetric black hole along the merger line (Fig.~\ref{fig:perturbative1}, \textit{first line}). If we increase the~$\varepsilon_H$, constant $\varepsilon_H$ curves will cross the extremality plane for the CLP black holes (Fig.~\ref{fig:perturbative1},~\textit{second line}), and no longer coexist with the non-hairy solutions. We still find a relatively good agreement in $T$ and $S$ (also see \cite{Markeviciute:2018yal}), however there is a clear difference in the behaviour of $\Omega_H$. As we have seen in \cite{Markeviciute:2016ivy}, intensive thermodynamic variables typically require higher order terms.

The hairy solutions exist for
\begin{equation}
3Q+2J \leq M \leq 3Q+\frac{1}{3}(j^2+9)Q^2-\frac{2}{81}j^2(j^2+9)Q^3+\mathcal{O}(Q^4),
\end{equation}
\noindent where at the upper range there is no contribution from the soliton, and the CLP black hole is recovered. Note, that in this regime the merger and extremality lines are very close to each other. In the limit $j\rightarrow 3$, the upper and lower limits coincide, and we obtain the Gutowski-Reall solution at the intersection of the extremal and BPS planes. 

At the lower range the hairy black holes approach the BPS bound where $T=0$. We also have that  $j_b=3$~\cite{Bhattacharyya:2010yg}, and the black hole is supersymmetric; the mix on the BPS can be seen as a weakly interacting black hole and a scalar cloud configuration, both of which are itself supersymmetric. We assume that $j<3$, and thus $\Omega_H<1$. Numerical results indicate that $\Omega_H\rightarrow 1$, which might be a non-linear effect. In~\cite{Bhattacharyya:2010yg} it was conjectured that these hairy black holes comprise a two parameter family of supersymmetric hairy black holes. On the BPS bound, the prediction for the entropy only matches when we are close to the Gutowski-Reall solution, and for a fixed $J$ is given by $S=2\sqrt{2}\pi J^{3/4}/3^{3/4}$. We find that for a fixed $J$, away from Gutowski-Reall black hole, the entropy is always decreasing. 

 
\bibliography{hairybh}{}
\bibliographystyle{JHEP}
\end{document}